\documentclass[11pt,epsfig]{article} 
\usepackage[utf8]{inputenc}
\usepackage[top=1in, left=0.95in, bottom=1.1in, right=0.95in]{geometry}
\usepackage[colorlinks=true,
linkcolor=blue,
urlcolor=red,
citecolor=red]{hyperref}
\usepackage[vcentermath,enableskew]{youngtab}
\usepackage{ytableau}
\usepackage{tikz}

\usepackage{tabularx}
\usepackage{diagbox}
\usepackage[toc]{appendix}
\usepackage{longtable}
\usepackage{enumerate}
\usepackage{float}
\usepackage{subfig}
\usepackage{xcolor}
\usepackage{setspace}

\usepackage{cite}

\let\counterwithin\relax
\usepackage{chngcntr}
\usepackage{amssymb, amsmath,mathrsfs}
\usepackage{multicol,multirow}

\usepackage{graphics}
\usepackage{graphicx}
\usepackage{epsf}
\usepackage{epsfig}
\usepackage{float}
\usepackage{makecell}

\usepackage{color}
\usepackage{xcolor}
\usepackage{simplewick}
\usepackage{amsmath}
\usepackage{amsfonts}
\usepackage{makeidx} 
\newcommand{\lra}[1]{\langle #1 \rangle}

\newcommand{\bin}[1]{(\overline{N} #1 N)}
\newcommand{\lrd}{\overleftrightarrow{D}}

\newcommand{\bea}{\begin{eqnarray}}
\newcommand{\eea}{\end{eqnarray}}
\newcommand{\calt}{\mathcal{T}}

\def\baselinestretch{1.16}

\begin{document}
\begin{center}

{\Large \textbf  {Chiral Effective Field Theories for Strong and Weak Dynamics}}\\[10mm]


Hao Sun$^{a, b}$\footnote{sunhao@itp.ac.cn}, Yi-Ning Wang$^{a, b}$\footnote{wangyining@itp.ac.cn}, Jiang-Hao Yu$^{a, b, c}$\footnote{jhyu@itp.ac.cn}\\[10mm]

\noindent 
$^a${\em \small CAS Key Laboratory of Theoretical Physics, Institute of Theoretical Physics, Chinese Academy of Sciences,    \\ Beijing 100190, P. R. China}  \\
$^b${\em \small School of Physical Sciences, University of Chinese Academy of Sciences,   Beijing 100049, P.R. China}   \\
$^c${\em \small School of Fundamental Physics and Mathematical Sciences, Hangzhou Institute for Advanced Study, UCAS, Hangzhou 310024, China} \\

\date{\today}   
          
\end{center}

\begin{abstract}

The chiral effective field theory (ChEFT) is an extension of the chiral perturbation theory that includes the nuclear forces and weak currents at the hadronic and nuclear scales. We propose a systematic framework of parametrising the pion-nucleon and nucleon-nucleon Lagrangian via the Weinberg power counting rules. We enumerate the operator bases of ChEFT by extending the Hilbert series of the pure meson sector to the nucleon sector with the CP symmetries. The Young tensor method is utilized to obtain the complete sets of the nucleon-nucleon, three-nucleon operators with/without pions ($\not\pi$-EFT). Then we use the spurion technique to reconstruct the ChEFT by taking the adjoint spurion and leptonic fields as the building blocks, without the need of external sources. These operators can be applied to both the strong dynamics, and the nucleon/nuclear weak current processes.

\end{abstract}

\newpage
\setcounter{tocdepth}{3}
\setcounter{secnumdepth}{3}

\tableofcontents

\setcounter{footnote}{0}

\def\baselinestretch{1.5}
\counterwithin{equation}{section}

\newpage

\section{Introduction}

Quantum chromodynamics (QCD) can not be perturbatively calculable at the low-energy regions since the coupling becomes so strong that the fundamental fields, the quarks, and the gluons, are confined in the colorless mesons and baryons. 
At the low energy, the QCD Lagrangian exhibits an approximate chiral symmetry $SU(N)_L \times SU(N)_R$ spontaneously broken down to the Gell-mann $SU(N)_V$ symmetry due to the quark condensation. 
After the chiral symmetry breaks, eight Nambu-Goldstone bosons (NGBs) appear in the hadron spectra, consisting of the eightfold mesons, such as the pion and the Kaon, etc.
These Goldstone modes are the quantum fluctuations around the vacuum background in the direction of the broken generators, with the following properties: they are gapless modes, they are weakly coupled in the infrared long-wave limit, and transform non-linearly under the coset group. 
These properties indicate a perturbatively derivative expansion on the NGB's interactions, and thus a generic effective field theory (EFT) for them can be constructed following this power counting rule.
Applying to the meson and baryon system, it is the chiral perturbative theory (ChPT)~\cite{Weinberg:1968de,Weinberg:1978kz,Gasser:1983yg,Gasser:1984gg,Gasser:1987rb}.

The ChPT describes the strong and weak interactions among the meson and baryon fields below the cutoff scale $ \Lambda_{\chi} \approx 1 \text{ GeV}$, identifying the chiral symmetry breaking $G = SU(N)_L\times SU(N)_R\rightarrow H = SU(N)_V$, where $N=2, 3$.  Using the $G/H$ coset construction formalism developed by Callan, Coleman, Wess, and Zumino (CCWZ)~\cite{Weinberg:1968de, Coleman:1969sm, Callan:1969sn}, 
the ChPT Lagrangian can be constructed systematically order by order following the power counting rule $p/(4\pi f)$, where $f$ is the pion decay constant. The power of $p/(4\pi f)$ is called the chiral dimension $\nu$~\cite{Weinberg:1968de,Weinberg:1978kz,Weinberg:1990rz, Weinberg:1991um}. After Weinberg constructed the leading-order (LO) $p^2$ Lagrangian for the pure meson system~\cite{Weinberg:1978kz}, many progresses have been made, including the $p^4$ order~\cite{Gasser:1983yg, Gasser:1984gg}, $p^6$ order~\cite{Fearing:1994ga,Bijnens:1999hw,Bijnens:1999sh,Ebertshauser:2001nj} and recently the $p^8$ order Lagrangian~\cite{Bijnens:2017wba,Bijnens:2018lez,Hermansson-Truedsson:2020rtj,Bijnens:2023hyv,Li:2024ghg}. On the other hand, the power counting breaking terms appear in the meson-baryon system due to the mass scale of the baryon around the cutoff, and thus the heavy baryon expansion~\cite{Jenkins:1990jv, Ecker:1995rk, Fettes:2000gb, Kobach:2018pie} was utilized to recover the power counting, and later the IR scheme~\cite{Becher:1999he,Becher:2001hv,Torikoshi:2002bt,Alarcon:2011kh}, and the extending-on-mass-shell (EOMS) scheme~\cite{Gegelia:1999gf, Fuchs:2003qc,Geng:2008mf,Alarcon:2011zs,Alarcon:2011px,MartinCamalich:2011py,Alarcon:2012kn} were developed to treat the baryon relativistically and recover the analytical behavior of the S-matrix. The effective Lagrangian on the meson-nucleon sector was first written by Ref.~\cite{Gasser:1987rb}, then the meson-nucleon interactions are constructed to $p^4$~\cite{Krause:1990xc,Ecker:1995rk,Fettes:1998ud,Fettes:2000gb}. Later extending to the $SU(3)$ flavor up to $p^3$ and $p^4$-orders~\cite{Krause:1990xc,Frink:2004ic,Oller:2006yh,Frink:2006hx,Jiang:2016vax}. 
Recently, the meson-nucleon operators up to $p^5$ for the $SU(2)$ case and the meson-baryon operators up to $p^4$ for  the $SU(3)$ case have been constructed in Ref.~\cite{Song:2024fae}, which are classified in terms of their transformations under the parity ($P$) and the charge conjugation ($C$).

Pioneered by Weinberg's works~\cite{Weinberg:1990rz,Weinberg:1991um}, the chiral symmetry has also been applied to describe the nuclear forces. The long- and intermediate-range forces are described by the pion exchange from the chiral Lagrangian, while the repulsive central forces at the short range are described by the hard-core nucleon-nucleon contact interactions, instead of the heavy $\rho, \omega, \sigma$ meson exchanges in the meson exchange model. In this way, the chiral Lagrangian is extended to the system containing more than single nucleon, the so-called chiral effective field theory (ChEFT)~\cite{Weinberg:1990rz, Weinberg:1991um, Weinberg:1992yk, vanKolck:1994yi, Ordonez:1995rz,Kaiser:1997mw,Epelbaum:1998ka,Epelbaum:1999dj,Epelbaum:2005bjv,Epelbaum:2008ga,Epelbaum:2009sd,Machleidt:2011zz, Hammer:2019poc}. In the nucleon-nucleon system, the nuclear potential can be constructed order-by-order using Weinberg's power counting rules~\cite{Weinberg:1968de,Weinberg:1978kz,Weinberg:1990rz,WEINBERG19913}, while the infrared enhancement in the two-particle reducible (2PR) amplitudes leads to the failure of the perturbation theory and thus the resummation over the 2PR contributions is necessarily needed to explain the bound states and the resonances of the nuclei. 
Adopting Weinberg's scheme, the nucleon-nucleon interactions and the three-nucleon interactions have been constructed up to $\nu=6$ order~\cite{Weinberg:1990rz, WEINBERG19913, Ordonez:1995rz, Kaiser:1997mw, Kaiser:1998wa, Girlanda:2010ya, Girlanda:2010zz, Xiao:2018jot, Filandri:2023qio, Nasoni:2023adf}, although the renormalization problem needs to be addressed. 
Later, Kaplan, Savage, and Wise (KSW)~\cite{Kaplan:1996xu,Kaplan:1998we,Kaplan:1998xi,Kaplan:1998tg,Kaplan:1998sz} proposed a different power-counting scheme concerning the unnaturally large nucleon-nucleon scattering length, which works well in the pion-less EFT ($\not\pi$EFT)~\cite{vanKolck:1998bw, Kaplan:1998we}, while cannot be converged in the  ChEFT~\cite{Fleming:1999ee}. Since then several modified Weinberg schemes have been proposed to address both the cutoff-dependence problem and the converge problem, but still no satisfactory scheme to address all yet.

Apart from describing the strong interactions among the nucleons and the mesons, the ChEFT can also be used to investigate the electroweak currents in the nucleon-nucleon processes, such as the beta decay~\cite{Cirigliano:2002ng,Cirigliano:2022hob,Cirigliano:2023fnz,Cirigliano:2024rfk}, the neutrinoless double beta decay~\cite{Savage:1998yh,Prezeau:2003xn,Graesser:2016bpz,Cirigliano:2017ymo,Cirigliano:2017djv,Cirigliano:2017tvr,Pastore:2017ofx,Cirigliano:2018yza,Cirigliano:2019vdj,Dekens:2024eae}, the coherent neutrino-nucleus scattering~\cite{Altmannshofer:2018xyo,Hoferichter:2020osn,Du:2021idh,Breso-Pla:2023tnz,Kouzakov:2024xnq}, and other weak processes. Traditionally the electroweak interactions are incorporated by adding some external sources to the Lagrangian.
However, the external source method encounters problems when constructing the High-dimension operators of the ChEFT. The ChEFT external sources originate from the leptonic currents in the QCD Lagrangian, 
\begin{equation}
    \mathcal{L}_{QCD} \supset \overline{q}_L \chi q_R + \overline{q}_R \chi^\dagger q_L + \overline{q}_L(\gamma^\mu l_\mu ) q_L + \overline{q}_R(\gamma^\mu r_\mu ) q_R\,,
\end{equation}
where $\chi\,,\chi^\dagger\,, l_\mu$ and $r_\mu$ are related to the leptonic currents, for example~\cite{Bishara:2016hek},
\begin{align}
    \chi = T_s (\overline{e}e)\,,\quad l_\mu= T_l (\overline{e}\gamma_\mu e)\,, \quad r_\mu= T_r (\overline{e}\gamma_\mu e)\,.
\end{align}
$T$'s are the Wilson coefficients in the $SU(2)_V$ space. Firstly, if there are more than 2 quarks in the QCD Lagrangian, which leads to more mesons and nucleons in the ChEFT Lagrangian, the currents above are inadequate. The tensor external sources and even more complicated ones could appear. Secondly, even for a specific current, it contains many different leptonic bilinears, for example, the scalar external source $\chi$ could contain $(\overline{e}e)\,,(\overline{e}\gamma^5 e)$, and so on, which are different and should be distinguished in the low-energy ChEFT Lagrangian more transparently.

In this paper, we propose a consistent description of the strong and weak interactions of the mesons and the nucleons by the ChEFT. In this framework, the effective Lagrangian takes the general form that
\begin{equation}
\label{eq:cheft_lagrangian_intro}
\mathcal{L}_\text{ChEFT} = \underbrace{\mathcal{L}_\pi+\mathcal{L}_{\pi N}}_{\text{ChPT}}+\mathcal{L}_{NN}+\mathcal{L}_{\pi NN} + \mathcal{L}_{NNN}+\mathcal{L}_{\text{leptonic}}\dots\,,     
\end{equation}
where the pure meson sector $\mathcal{L}_\pi$ and the meson-nucleon sector $\mathcal{L}_{\pi N}$ compose the ChPT Lagrangian, while the other sectors $\mathcal{L}_{NN}\,,\mathcal{L}_{\pi NN}\,,\mathcal{L}_{NNN}$ and $\mathcal{L}_{\text{leptonic}}$ are the 2-nucleon sector, the 2-nucleon-meson sector, the 3-nucleon sector, and the leptonic sector, respectively. Among them, the pionless EFT is also included by removing all the pion fields. To construct the most general effective operator basis, we adopt the Hilbert series~\cite{Feng:2007ur,Jenkins:2009dy,Hanany:2010vu,Hanany:2014dia,Lehman:2015via, Lehman:2015coa,Henning:2015daa,Henning:2015alf,Henning:2017fpj,Marinissen:2020jmb} and the Young tensor method~\cite{Li:2020gnx,Li:2022tec,Li:2020xlh},
\begin{itemize}
    \item The Hilbert series is a generating function counting the independent operators of EFTs. It utilizes the orthogonality of the group characters to eliminate the redundancies such as the equation of motion (EOM) and the integrating-by-part (IBP)~\cite{Lehman:2015via, Lehman:2015coa,Henning:2015daa,Henning:2015alf,Henning:2017fpj,Marinissen:2020jmb}. In addition, the discrete symmetries such as the $C$ and $P$ can also be incorporated in the Hilbert series~\cite{Graf:2020yxt,Sun:2022aag}, which has been applied to the pure-meson sector $\mathcal{L}_\pi$~\cite{Graf:2020yxt}. In this paper, we extend the Hilbert series of the pure meson sector to the other sectors containing nucleons for the first time and use it to count the independent operators of the ChEFT Lagrangian Eq.~\eqref{eq:cheft_lagrangian_intro}.
    \item The Young tensor method is another technique to explicitly present the independent and complete operators. It is general and can applied to the operators of arbitrary dimensions and any EFTs such as the standard model EFT (SMEFT)~\cite{Li:2020gnx,Li:2020xlh}, the low-energy EFT (LEFT)~\cite{Li:2020tsi,Li:2021tsq}, the Higgs EFT (HEFT)~\cite{Sun:2022ssa,Sun:2022snw}, the ChPT~\cite{Li:2024ghg,Song:2024fae,Low:2022iim}, and so on. In particular, the Young tensor method should be modified for the pions in the ChPT, since they satisfy the Adler's zero condition~\cite{Adler:1969gk}, and the external sources~\cite{Ren:2022tvi}, since they are of no EOM. We apply the Young tensor method to the ChEFT so that the independent operator basis can be obtained systematically.
\end{itemize}

As discussed before, the traditional external source method characterizing the interactions involving the leptons in the $\mathcal{L}_{\text{leptonic}}$ is not convenient for the constructions of the ChEFT operators from the bottom-up viewpoint since its building blocks are not well defined due to increasing number of external sources, and the mixed different leptonic currents in the chiral Lagrangian. In this paper, we adopt the spurion method, which regards the leptons as the building blocks of the ChEFT, and contains another two building blocks to maintain the $SU(2)_V$ symmetry-breaking effects~\cite{Song:2025snz}. In this work we address the spurion method in detail as follows
\begin{itemize}
    \item In the spurion method, the building blocks of the ChEFT are the mesons, the nucleons, the leptons, and two scalar fields, $\Sigma_\pm\,,Q_\pm$, where the two scalar fields $\Sigma_\pm\,,Q_\pm$ are formally covariant under the $SU(2)_V$ symmetry and are referred to as spurions.
    Given these building blocks, the complicated leptonic currents can be included naturally by the Young tensor method~\cite{Li:2020gnx,Li:2022tec,Li:2020xlh}, and the operators with different leptonic currents are treated independently in the ChEFT Lagrangian. 
\end{itemize}

Based on the tools and methods developed above, the complete and independent operators of the ChEFT can be obtained and classified by their sectors, detailed in the following 
\begin{itemize}
    \item In the pure-meson sector, we construct the effective operators conserving the parity and the charge conjugation up to $p^8$ order, where the external source method is used. In particular, for the $p^8$ order, only the operators with at least 4 mesons are listed. The complete operators of this sector can be found in Ref.~\cite{Li:2024ghg}.
    \item At the same time, in the meson-nucleon sector, we present the $C$-even and $P$-even operators up to 5 derivatives, which partly cover the operators of the $p^4$ and $p^5$ orders~\cite{Song:2024fae} due to the power-counting scheme. The external sources are taken into account.
    \item In the 2-nucleon sector $\mathcal{L}_{NN}$, the 2-nucleon-meson sector $\mathcal{L}_{\pi NN}$, and the 3-nucleon sector $\mathcal{L}_{NNN}$, we construct the complete and independent $C$-even and $P$-even operators up to $p^6$ order for the first time, following the modified power-counting formula proposed by Weinberg~\cite{Meissner:2014lgi,Epelbaum:2008ga,Hammer:2019poc,Machleidt:2011zz} . The operators are of the relativistic form without redundancy, and expanding to the non-relativistic form using the heavy nucleon expansion is straightforward. 
    \item For the operators with leptonic currents in the sector $\mathcal{L}_{\text{leptonic}}$ of the ChEFT, we utilize the spurion method~\cite{Song:2025snz}. We present the lepton-nucleon and the lepton-meson operators up to $p^4$ order. In particular, the leading-order operators contributing to the neutrinoless double beta decay~\cite{Savage:1998yh,Prezeau:2003xn,Graesser:2016bpz,Cirigliano:2017ymo,Cirigliano:2017djv,Cirigliano:2017tvr,Pastore:2017ofx,Cirigliano:2018yza,Cirigliano:2019vdj,Dekens:2024eae} are given by the spurion method.
\end{itemize}
The numbers of the operators above are consistent with the Hilbert series results.

This paper is organized as follows. In Sec.~\ref{sec:ccwz}, we briefly review the CCWZ formalism, the power-counting scheme, and the construction of the effective operators of the ChEFT. We outline the Hilbert series and the Young tensor method in Sec.~\ref{sec:hs} and Sec.~\ref{sec:young}.
Subsequently, we discuss the effective Lagrangian of the ChEFT by the different sectors respectively. (Tab.~\ref{tab:organization}) In Sec.~\ref{sec:operators}, we discuss the pure-meson sector $\mathcal{L}_\pi$ and present the operators up to $p^8$ order. In the Sec.~\ref{sec:NMoperators}, we discuss the meson-nucleon sector $\mathcal{L}_{\pi N}$ and present the operators up to $p^5$ order. Next, we present the multi-nucleon operators of the $\mathcal{L}_{NN}$, $\mathcal{L}_{\pi NN}$, and the $\mathcal{L}_{NNN}$ sectors up to $p^6$ order in Sec.~\ref{sec:NNoperators}, where we also discuss the modifications of the power-counting scheme for the muli-mucleon operators. In Sec.~\ref{sec:weakoperators}, we compare the external source method and the spurion method, and present the operators involving the leptons in the $\mathcal{L}_{\text{leptonic}}$ utilizing the spurion method.  Finally, we conclude in Sec.~\ref{sec:conclu}.

\section{Chiral Lagrangian for Meson and Baryon}
\label{sec:ccwz}



In this section, we review the CCWZ coset construction~\cite{Weinberg:1968de, Coleman:1969sm, Callan:1969sn} in general and apply it to the ChPT by identifying the building blocks. We discuss the extension to the ChEFT, including the modifications of the power-counting scheme and the construction of its effective Lagrangian.

\subsection{CCWZ Coset Construction}

Although we concentrate on the ChPT and the ChEFT of the chiral symmetry breaking $SU(2)_L\times SU(2)_R\rightarrow SU(2)_V$ in this paper, the CCWZ coset construction can be used for general $G\rightarrow H$ symmetry-breaking pattern~\cite{Goon:2012dy,Contino:2011np,Naegels:2021ivf}. In this section, we review the CCWZ construction generally and then apply it to the chiral case.

\subsubsection*{Coset representatives}

Consider a theory that is invariant under a generic Lie group $G$, spontaneously broken down to its subgroup $H\subset G$, according to the Goldstone theorem~\cite{Goldstone:1961eq, Goldstone:1962es}, dim$(G/H)$ NGBs appear corresponding to the breaking symmetries, which can be parameterized by the left coset space $G/H$ as a specific function $u(x)$.
We consider a dim$(G)$ field $\Phi(x)$ which takes value in the group $G$ and transforms linearly under it,
\begin{equation}
    \Phi(x) \rightarrow g \Phi(x)\,, \quad g \in G\,.
\end{equation}
The symmetry-breaking pattern $G\rightarrow H$ defines an equivalent relation $\Phi(x) \sim \Phi(x) h$ with $h \in H$, which implies that all the fields in the same coset are not independent. To implement such an equivalence we can prompt the subgroup $H$ to be the gauged one since the gauge symmetry is fake and implies redundant degrees of freedom. The $\text{dim}(H)$ gauge transformations mean the number of the physical NGBs would be dim$(G)$ $-$ dim$(H)$ = dim$(G/H)$ as expected. Thus the transformation of the field $\Phi(x)$ can be specifies as 
\begin{equation}
    \Phi(x)\rightarrow g \Phi(x)h(x)^{-1}\,,\quad g\in G\,, h(x)\in H\,.
\end{equation}
As $g\,, h(x)$ taking their values in $G$ and $H$ respectively, the field $\Phi(x)$ ranges over the group space of $G$.

\begin{figure}[ht]
    \centering
    \includegraphics[scale=0.5]{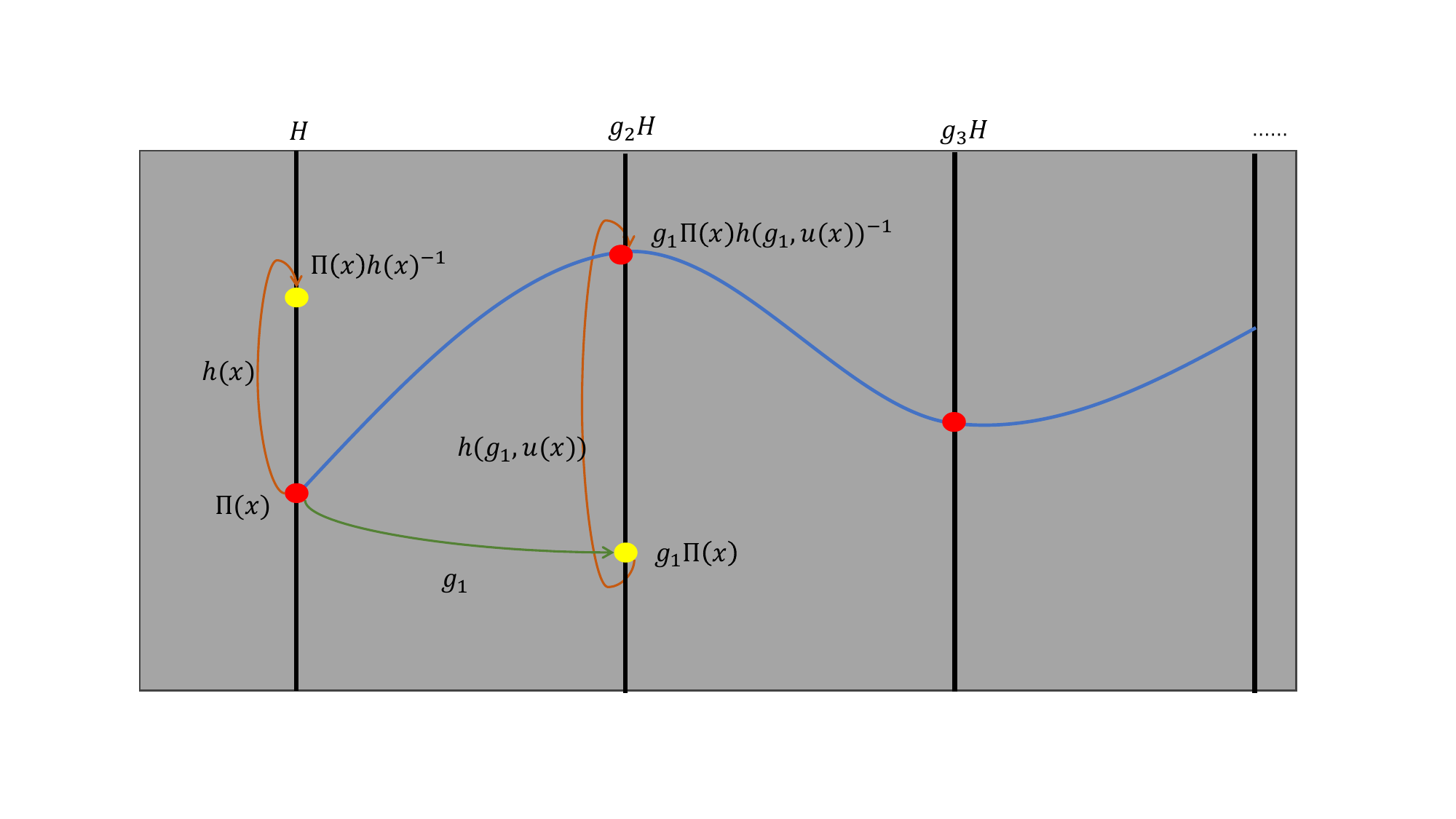}
    \caption{This is an illustrating diagram of the definition of $u(x)$. The gray box is the group space of $G$, on which the field $\Pi(x)$ takes values. The vertical black lines are the cosets $H\,, g_1H\,, g_2H$, and so on of $G/
    H$. The blue curve is the chosen $u(x)$, and the cross points of it and the cosets (the red circles) are the representatives of each coset. The equivalent fields are gathered in the same coset and are related by $h(x)\in H$, while the nonequivalent fields are related by $g_1,g_2,\dots \in G$. A chosen representative in a coset $\Pi(x)$ does not fall in the $u(x)$ after a transformation $g_1$, thus an element $h(g_1, u(x))$ in the subgroup $H$ is needed to ensure the transformed field is still on the $u(x)$, indicated by the brown arrows.}
    \label{fig:ccwz}
\end{figure}

According to the discussion above the independent NGBs are isomorphic to the cosets in $G/H$, 
thus we can fix the gauge transformation $h(x)$ by choosing a representative of each coset, and define the collection of these representatives of the cosets as $u(x)$, which takes value in $G/H$ and contains the $\text{dim}(G/H)$ NGBs. Since the image of $u(x)$ under the global transformation $g$ generally does not fall in the chosen representatives, a compensating transformation $h(x)\in H$ is needed to map it to the chosen ones, which fixes the gauge and depends on both the global transformation $g$ and the chosen representatives $u(x)$. So the transformation takes the form
\begin{equation}
\label{eq:uxtr}
    u(x)\rightarrow g u(x) h(g,u(x))^{-1}\,,\quad g \in G\,, h\in H\,,
\end{equation}
for a certain $u(x)$. An illustrating diagram of the definition and the transformation of $u(x)$ is shown in Fig.~\ref{fig:ccwz}.
Furthermore, if the commutator between any generator of $H$ and any broken generator in $G$ is again a broken generator, 
\begin{equation}
    [T_a\,,T_{\hat{a}}] \not\in H\,,
\end{equation}
for any $T_a\in \mathfrak{h}$ and $T_{\hat{a}} \in \mathfrak{g}\,,\not \in\mathfrak{h}$, where $\mathfrak{h}(\mathfrak{g})$ is the Lie algebra of the group $H (G)$, 
the field $u(x)$ can be parameterized in an exponential form
\begin{equation}
\label{eq:ux}
    u(x) = \exp\left(i\frac{\Pi(x)}{2f}\right)\,,
\end{equation}
where $\Pi(x)=\sum_{\hat{a}=1}^{\text{dim}(G/H)}\phi^{a}(x) T_{\hat{a}}$, with $T_{\hat{a}}\,,\hat{a}=1,2,\dots,\text{dim}(G/H)$ the broken generators, and $\phi^{a}$ are the NGBs, which transform non-linearly according to Eq.~\eqref{eq:uxtr} as
\begin{equation}
\label{eq:trans}
    \exp\left(i\frac{\Pi(x)}{2f}\right) \rightarrow \exp\left(i\frac{\Pi'(x)}{2f}\right) = g \exp\left(i\frac{\Pi(x)}{2f}\right) h(g,\Pi)^{-1}\,,\quad g \in G\,, h\in H\,.
\end{equation}
In particular, the field $\Pi(x)$ transforms linearly under the subgroup $H$, 
\begin{equation}
    \Pi(x)\rightarrow h \Pi(x) h^{-1}\,.
\end{equation}


We could apply these results to the ChPT, in which the group $G$ is the direct product group $SU(N)_L\times SU(N)_R$ with the Lie algebra $\mathfrak{g} = \mathfrak{g}_L \oplus \mathfrak{g}_R$, which is a linear space of $2\times\text{dim}(SU(N))$ dimension. Suppose the generators of the $SU(N)_{L/R}$ are $\mathbf{t}_a\,,a=1,2,\dots\,,\text{dim}(SU(N))$, a natural basis of $\mathfrak{g}$ is 
\begin{equation}
    \left\{ \begin{array}{l}
T_a = \mathbf{t}_a\otimes \mathbf{I} + \mathbf{I}\otimes \mathbf{t}_a \\
T_{\hat{a}} = \mathbf{t}_a\otimes \mathbf{I} - \mathbf{I}\otimes \mathbf{t}_a
    \end{array}\right.\,,\quad a=1,2\dots \text{dim}(SU(N))\,.
\end{equation}
The $T_a$ and $T_{\hat{a}}$ span two disjoint subspaces of $\mathfrak{g}$, but only the one spanned by $T_a$ is a subalgebra, which corresponds to the diagonal subgroup $SU(N)_V$. Thus it is $H=SU(N)_V$ the unbroken subgroup, and all the generators $T_{\hat{a}}$ are broken.
According to the definition of $u(x)$ in Eq.~\eqref{eq:ux}, there is 
\begin{equation}
\label{eq:ux2}
    u(x) = \exp\left(i\frac{\sum_a\phi^a(x)T_{\hat{a}}}{2f}\right) = 
    \xi_L\otimes \xi_R = \exp\left(i\frac{\Pi(x)}{2f}\right) \otimes \exp\left(-i\frac{\Pi(x)}{2f}\right)
    \,,
\end{equation}
where
\begin{equation}
    \xi_L = \xi_R^{-1} = \xi = \exp\left(i\frac{\Pi(x)}{2f}\right) \,.
\end{equation}
For the $N=3$ case, the meson octet is
\begin{equation}
\label{eq:field1}
    \Pi(x)=\sum_{A=1}^8\phi^A(x)\lambda^A = \left(\begin{array}{ccc}
    \pi^0 + \frac{1}{\sqrt{3}}\eta & \sqrt{2}\pi^+ & \sqrt{2}K^+ \\
    \sqrt{2}\pi^- & -\pi^0+\frac{1}{\sqrt{3}}\eta & \sqrt{2}K^0 \\
    \sqrt{2}K^- & \sqrt{2}~\overline{K}^0 & -\frac{2}{\sqrt{3}}\eta
    \end{array}\right)\,,
\end{equation}
and for the $N=2$ case, only the three pions are included,
\begin{equation} 
\label{eq:field2}
    \Pi(x)=\sum_{I=1}^3\phi^I(x)\sigma^I = \left(\begin{array}{cc}
    \pi^0 & \sqrt{2}\pi^+ \\ \sqrt{2}\pi^-&-\pi^0
    \end{array}\right)\,.
\end{equation}
In particular, the transformations of the $\xi_L$ and $\xi_R$ are
\begin{align}
    \xi_L(\Pi) &\rightarrow g_L \xi_L h(g_L,g_R,\Pi)^{-1}\,,\notag \\
    \xi_R(\Pi) &\rightarrow g_R \xi_R h(g_L,g_R,\Pi)^{-1}\,,
\end{align}
according to Eq.~\eqref{eq:uxtr}, where $g_L\in SU(N)_L$, $g_R\in SU(N)_R$ and $h\in SU(N)_V$.
For the chiral symmetry, the cosets are called the symmetric cosets, because the parity transformation keeps the unbroken generators unchanged and converts the broken generators to their minus,
\begin{equation}
\label{eq:pp}
    \mathcal{P}:\quad \left\{
    \begin{array}{l}
T_a \rightarrow T_a \\
T_{\hat{a}} \rightarrow -T_{\hat{a}} 
    \end{array}
    \right.\,,
\end{equation}
since the parity interchanges the left and right generators. 
For the symmetric cosets, we can construct a function that transforms globally under the group $G$.
Applying the parity transformation to the $u(x)$ transformation relation in Eq.~\eqref{eq:uxtr}, and taking the inverse, we can get $u(x)$ transforms as
\begin{equation}
\label{eq:uxtr2}
    u(x)\rightarrow h u(x)g_\mathcal{P}^{-1}\,,
\end{equation}
where $g_\mathcal{P}$ is the $g\in G$ applied by $\mathcal{P}$. Thus the $\xi_L$ and $\xi_R$ fields transform alternatively
\begin{align}
    \xi_L(\Pi) &\rightarrow h(g_L,g_R,\Pi) \xi_L g_R^{-1}\,,\notag \\
    \xi_R(\Pi) &\rightarrow h(g_L,g_R,\Pi) \xi_R g_L^{-1}\,,
\end{align}
thus we construct a new function 
\begin{equation}
    U(x)= \xi_R^{-1}\xi_L = \xi^2 = \exp\left(i\frac{\Pi(x)}{f}\right)\rightarrow g_L U(x) g_R^{-1}\,,
\end{equation}
which transforms globally under the $G= SU(N)_L\times SU(N)_R$ group. 
Thus for the symmetric cosets, once the transformation satisfying Eq.~\eqref{eq:pp} exists, the function $U(x)$ transforming globally is an equivalent parameterization of the mesons when constructing the invariant Lagrangian. For generality, we use the $u(x)$ in the subsequent discussion and present the $U$-parameterization for completeness. 

\subsubsection*{Invariant one-form}

When constructing the invariant Lagrangian, it is convenient to adopt the invariant one-form $-i u(x)^{-1}\partial_\mu u(x)$ of the group $G$, which can be projected into the broken and the unbroken generators as 
\begin{equation}
\label{eq:oneform}
    -i u(x)^{-1}\partial_\mu u(x) = u^{\hat{a}}_\mu(x) T_{\hat{a}} + \Gamma_\mu^a(x) T_a \equiv u_\mu(x)+\Gamma_\mu(x)\,,
\end{equation}
called the Maurer-Cartan form. According to Eq.~\eqref{eq:uxtr}, the transformations of $\Gamma_\mu $ and $u_\mu$ take the form
\begin{equation}
    \Gamma_\mu(x) \rightarrow h(g,\Pi) \Gamma_\mu(x) h(g,\Pi)^{-1} -i h(g,\Pi)\partial_\mu h(g,\Pi)^{-1}\,,\quad u_\mu(x)\rightarrow h(g,\Pi) u_\mu(x) h(g,\Pi)^{-1}\,,
\end{equation}
which implies that $u_\mu$ transforms covariantly under the $H$ and can be used to be the basic building block to construct the invariant Lagrangian, while $\Gamma_\mu$ transforms like a connection in the field space, thus can be used to define the covariant derivative
\begin{equation}
    D_\mu u_\nu = \partial_\mu u_\nu + i[\Gamma_\mu,u_\nu]\,, \quad \text{so that }\quad D_\mu u_\nu \rightarrow h D_\mu u_\nu h^{-1}\,,
\end{equation}
for $h\in H$. 
Furthermore, we can define the covariant field-strength tensor 
\begin{equation}
    \Gamma_{\mu\nu}= \partial_\mu \Gamma_\nu - \partial_\nu \Gamma_\mu + i [\Gamma_\mu,\Gamma_\nu]\,,
\end{equation}
which transforms under the local symmetry $H$ as $\Gamma_{\mu\nu} \rightarrow h \Gamma_{\mu\nu} h^{-1}$\,.
However, the field-strength tensor $\Gamma_{\mu\nu}$ is not independent in the case of symmetric coset because of the relation
\begin{equation}
\label{eq:gammamunu}
    \Gamma_{\mu\nu} = -i[u_\mu,u_\nu]\,.
\end{equation}

Considering the ChPT now, the one-form takes the form
\begin{align}
    -i u(x)^{-1}\partial_\mu u(x) &= -i(\xi^{-1}\otimes \xi)(\partial_\mu\xi\otimes \xi^{-1}+\xi\otimes\partial_\mu\xi^{-1}) \notag \\
    &= \frac{1}{2}\left[-i(\xi^{-1}\partial_\mu\xi-\xi\partial_\mu\xi^{-1})^a T_{\hat{a}} -i(\xi^{-1}\partial_\mu\xi+\xi\partial_\mu\xi^{-1})^a T_{a}\right]\,.
\end{align}
We define the covariant field $u_\mu$ and the connection $\Gamma_\mu$ as
\begin{align}
    u_\mu &= i(\xi^{-1}\partial_\mu\xi-\xi\partial_\mu\xi^{-1})^a T_{\hat{a}}\,, \notag \\
    \Gamma_\mu &= \frac{1}{2}(\xi^{-1}\partial_\mu\xi+\xi\partial_\mu\xi^{-1})^a T_{a}\,, \label{eq:chptu}
\end{align}
then 
\begin{equation}
    -i u(x)^{-1}\partial_\mu u(x) = -\frac{1}{2}u_\mu -i\Gamma_\mu\,.
\end{equation} 
In particular, they satisfy
\begin{equation}
    {u_\mu}^{-1} = u_\mu\,,\quad {\Gamma_\mu}^{-1} = -\Gamma_\mu\,.
\end{equation}
In terms of the $U(x)$, since it transforms under the global symmetry $G$, the covariant derivative is actually normal, and the building block corresponding to the $u_\mu$ is, 
\begin{equation}
    V_\mu = -i U(x)^{-1}\partial_\mu U(x) \rightarrow g_R V_\mu g_R^{-1}\,,
\end{equation}
which is a $SU(N)_L$-invariant one-form, and transforms under the global $SU(N)_R$.
Equivalently we can construct a $SU(N)_R$-invriant one-form $V'_\mu = -i U(x)\partial_\mu U(x)^{-1} \rightarrow g_L V_\mu g_L^{-1} $. 
The field-strength tensor $\Gamma_{\mu\nu}$ appears in the $u$-parameterization while does not appear in the $U$-parameterization, so it is not an independent building block as shown in Eq.~\eqref{eq:gammamunu} as expected.


\subsubsection*{External sources}

In addition to the field $u_\mu$ and the covariant derivative $ D_\mu$, we can introduce additional covariant building blocks in the ChPT called the external sources, which could be symmetry-breaking terms or the electroweak currents. As mentioned in the introduction, though the external source method is used frequently, it is not convenient when constructing the high-dimension operators, especially the operators with the leptons, because of the complicated leptonic currents,
which, however, can be managed naturally in the spurion method~\cite{Song:2025snz}.
We review the traditional external source method here and turn to the spurion method when we discuss the weak interactions between the hadrons and the leptons in Sec.~\ref{sec:weakoperators}, where it is shown that the spurion method is appropriate to construct the high-dimension operators.

Focusing on the external source method, the low-energy QCD Lagrangian may contain the external source terms such as
\begin{equation}
\label{eq:QCD_ext}
    \mathcal{L}_{QCD} \supset \overline{q}_L \chi q_R + \overline{q}_R \chi^\dagger q_L + \overline{q}_L(\gamma^\mu l_\mu ) q_L + \overline{q}_R(\gamma^\mu r_\mu ) q_R\,,
\end{equation}
where $\chi$ is the scalar external source, and $l_\mu\,,r_\mu$ are the left- and right-handed vector external sources, respectively.

For the vector external sources, they can be included by prompting the global group $G=SU(N)_L\times SU(N)_R$ to be local and regarding them as the corresponding gauge fields,
\begin{equation}
    l_\mu = l_\mu^a \, \mathbf{t}_a \otimes \mathbf{I}\,,\quad r_\mu = r^a_\mu \, \mathbf{I}\otimes \mathbf{t}_a\,,
\end{equation}
whose covariant field-strength tensors take the form
\begin{align}
    f^L_{\mu\nu} &= \partial_\mu l_\nu - \partial_\nu l_\mu + i[l_\mu,l_\nu] \rightarrow g_L f^L_{\mu\nu}g_L^{-1}\,, \label{eq:fl}\\
    f^R_{\mu\nu} &= \partial_\mu r_\nu - \partial_\nu r_\mu + i[r_\mu,r_\nu] \rightarrow g_R f^R_{\mu\nu} g_R^{-1}\,. \label{eq:fr}
\end{align}
For the $U$-parameterization,
the covariant derivative on $U(x)$ is
\begin{equation}
    \mathbb{D}_\mu U(x) = \partial_\mu U(x) -il_\mu U(x) + i U(x)r_\mu\,,
\end{equation}
satisfying $\mathbb{D}_\mu U(x)\rightarrow g_L \mathbb{D}_\mu U(x) g_R^{-1}$, so that the meson building block becomes
\begin{equation}
    V_\mu = -iU^{-1}\mathbb{D}_\mu U\,.
\end{equation}

At the same time, the partial derivative in the definition of the one-form in Eq.~\eqref{eq:oneform} becomes covariant
\begin{equation}
    \partial_\mu u(x) \rightarrow \mathbb{D}_\mu u(x) = (\partial_\mu-il_\mu)\xi(x) \otimes \xi(x)^{-1} + \xi(x)\otimes (\partial_\mu -ir_\mu) \xi(x)^{-1}\,,
\end{equation}
in terms of which the one-form becomes $-i u(x)^{-1}\mathbb{D}_\mu u(x) = -\frac{1}{2}u_\mu -i\Gamma_\mu$, where
\begin{align}
    u_\mu &= i\left[\xi^{-1}(\partial_\mu-il_\mu)\xi-\xi(\partial_\mu-ir_\mu)\xi^{-1}\right]^a T_{\hat{a}}\,, \\
    \Gamma_\mu &= \frac{1}{2}\left[\xi^{-1}(\partial_\mu-il_\mu)\xi+\xi(\partial_\mu-ir_\mu)\xi^{-1}\right]^a T_{a}\,.
\end{align}
In the $u$-parameterization, we can transform the left- and right-handed field-strength tensors to be covariant under the subgroup $H=SU(N)_V$,
\begin{align}
    & u^{-1}(f^L_{\mu\nu}{}^a \,\mathbf{t}_a\otimes \mathbf{I} +f^R_{\mu\nu}{}^a\, \mathbf{I}\otimes \mathbf{t}_a ) u \notag \\
    =& \left(\frac{\xi^{-1}f^L_{\mu\nu}\xi+\xi f^R_{\mu\nu}\xi^{-1}}{2}\right)^a T_a + \left(\frac{\xi^{-1}f^L_{\mu\nu}\xi-\xi f^R_{\mu\nu}\xi^{-1}}{2}\right)^a T_{\hat{a}} \notag \\
    =& f_+{}_{\mu\nu} + f_-{}_{\mu\nu}\,,
\end{align}
where
\begin{equation}
\label{eq:fpm}
    f_+{}_{\mu\nu} = \left(\frac{\xi^{-1}f^L_{\mu\nu}\xi+\xi f^R_{\mu\nu}\xi^{-1}}{2}\right)^a T_a\,,\quad 
    f_-{}_{\mu\nu} = \left(\frac{\xi^{-1}f^L_{\mu\nu}\xi-\xi f^R_{\mu\nu}\xi^{-1}}{2}\right)^a T_{\hat{a}}\,,
\end{equation}
and both of them are covariant under the $SU(N)_V$ group, 
\begin{equation}
    f_{\pm}{}_{\mu\nu} \rightarrow h f_{\pm}{}_{\mu\nu} h^{-1}\,,\quad h \in SU(N)_V\,.
\end{equation}
In particular, in the symmetric coset, even if the field-strength tensor $f_\pm$ is introduced, the $\Gamma_{\mu\nu}$ is still redundant,
\begin{equation}
\label{eq:cdc}
    \Gamma_{\mu\nu} = -i[u_\mu,u_\nu] + f_+{}_{\mu\nu}\,.
\end{equation}
In particular, the covariant derivative $\mathbb{D}_\mu$ corresponding to the $SU(N)_L\times SU(N)_R$ should be distinguished from the $D_\mu$ corresponding to the subgroup $SU(N)_V$, where the latter is needed to keep the gauge invariance, and the former is used to introduce the vector external sources. When constructing the effective operators in the $u$-parameterization, it is the $SU(N)_V$ symmetry to be respected, while the $SU(N)_L\times SU(N)_R$ invariance is realized trivially by the definitions of the building blocks, such as the definition of $u_\mu$ in Eq.~\eqref{eq:oneform}, and the definitions of the field strength tensors in Eq.~\eqref{eq:fpm}.

Similarly the scalar external sources $\chi\,,\chi^\dagger$ can be regarded as spurions whose transformations under the $SU(N)_L\times SU(N)_R$ symmetry are
\begin{equation}
    \chi\rightarrow g_L \chi g_R^{-1}\,, \quad \chi^\dagger \rightarrow g_R \chi^\dagger g_L^{-1}\,,
\end{equation}
and they can also be transferred in the $u$-parameterization since the cosets of the chiral symmetry breaking are symmetrical. To this end, we take the tensor product $\tilde{\chi} = \chi\otimes \chi^\dagger $
so that
\begin{equation}
    \tilde{\chi} \rightarrow g \tilde{\chi} g^{-1}_\mathcal{P}\,, \quad g = (g_L, g_R) \in SU(N)_L \times SU(N)_R\,.
\end{equation}
Because of the transformations of $u(x)$ in Eq.~\eqref{eq:uxtr} and Eq.~\eqref{eq:uxtr2}, the combination
\begin{equation}
    u(x)^{-1} \tilde{\chi} u(x)^{-1} = (\xi^{-1} \chi \xi^{-1} )\otimes (\xi \chi^\dagger \xi) \,,
\end{equation}
is covariant under the subgroup $SU(N)_V$
\begin{equation}
    u(x)^{-1} \tilde{\chi} u(x)^{-1} \rightarrow h(x)\left(u(x)^{-1} \tilde{\chi} u(x)^{-1}\right) h(x)^{-1} = (h(x)\xi^{-1} \chi \xi^{-1} h(x)^{-1})\otimes (h(x)\xi \chi^\dagger \xi h(x)^{-1})\,,
\end{equation}
where $h(x)\in SU(N)_V$. Furthermore, it can be expressed as
\begin{equation}
    u(x)^{-1} \tilde{\chi} u(x)^{-1} = \chi_+ + \chi_- = \chi_+^a T_a + \chi_-^a T_{\hat{a}}\,,
\end{equation}
where 
\begin{align}
    \chi_+^a = \left(\xi^\dagger \chi \xi^\dagger + \xi \chi^\dagger \xi\right)^a\,, \quad \chi_-^a = \left(\xi^\dagger \chi \xi^\dagger - \xi \chi^\dagger \xi\right)^a\,.
\end{align}

We summarize all the bosonic building blocks and the external sources of the ChPT in the two different parameterizations in Tab.~\ref{tab:building_blocks}, where the building blocks of the $u$-parameterization are covariant under the $SU(N)_V$ group, while the ones of the $U$-parameterization are covariant under the $SU(N)_L\times SU(N)_R$ group. It should be emphasized that different parameterizations are equivalent, and in this paper, we insist on the $ u$-parameterization.
\begin{table}
\renewcommand{\arraystretch}{1.5}
\begin{center}
    \begin{tabular}{|c|c|}
\hline
u-parameterization & U-parameterization \\
\hline
\multicolumn{2}{|c|}{Bosonic building blocks} \\
\hline
$u_\mu \rightarrow h u_\mu h^{-1}$ & $V_\mu \rightarrow g_R V_\mu g_R^{-1}$ \\
\hline
\multicolumn{2}{|c|}{external sources building blocks} \\
\hline
$f_+{}_{\mu\nu}\rightarrow hf_+{}_{\mu\nu}h^{-1} $ & $f^L_{\mu\nu}\rightarrow g_L f^L_{\mu\nu}g_L^{-1}$ \\ 
$f_-{}_{\mu\nu}\rightarrow hf_-{}_{\mu\nu}h^{-1} $ & $f^R_{\mu\nu}\rightarrow g_R f^R_{\mu\nu}g_R^{-1}$ \\ 
$\chi_+ \rightarrow h\chi_+ h^{-1}$ & $\chi \rightarrow g_L \chi g_R^{-1}$ \\
$\chi_- \rightarrow h\chi_- h^{-1}$ & $\chi^\dagger \rightarrow g_R \chi^\dagger g_L^{-1}$ \\
\hline
\multicolumn{2}{|c|}{Baryon building blocks}\\
\hline
\multicolumn{2}{|c|}{
    $SU(2): \quad N \rightarrow h N $
} \\
\multicolumn{2}{|c|}{
$ SU(3): \quad B \rightarrow h Bh^{-1}$
} \\
\hline
\multicolumn{2}{|c|}{Lepton building blocks}\\
\hline
\multicolumn{2}{|c|}{
    $\psi \rightarrow \psi\,,\quad \psi = e_L,e_R,\nu_L $
} \\
\hline
    \end{tabular}
\end{center}
\caption{The building blocks of two different parameterizations, including the bosonic field, the external sources, the baryon fields, and the leptons.}
\label{tab:building_blocks}
\end{table}

\subsubsection*{Baryons and Leptons}

In addition to the mesons and the external sources, the CCWZ formalism can be further extended to include the fermion fields $\psi$. The fermion fields transform under some linear representation $\mathbf{R}$ of the local group $H$, and the covariant derivative on the fermions takes the form that
\begin{equation}
    D_\mu \psi = \partial_\mu\psi + i \mathbf{R}(\Gamma_\mu) \cdot \psi\,.
\end{equation}
In the ChPT, the baryons are added as the light baryon octet $B$ for the $SU(3)$ case and the nucleon doublet $N$ for the $SU(2)$ case,
\begin{align}
    B&=\left(
    \begin{array}{ccc}
\frac{1}{\sqrt{2}}\Sigma^0+\frac{1}{\sqrt{6}}\Lambda & \Sigma^+ & p \\
\Sigma^- & -\frac{1}{\sqrt{2}}\Sigma^0+\frac{1}{\sqrt{6}}\lambda & n \\
\Xi^- & \Xi^0 & -\frac{2}{\sqrt{6}}\lambda \\
\end{array}
    \right)\,,\quad 
    N= \left(\begin{array}{c}
p\\n
    \end{array}\right)\,,
\end{align}
which transform linearly under the adjoint representation of the $SU(3)_V$, and the fundamental representation of the $SU(2)_V$,  respectively,
\begin{align}
    B &\rightarrow h Bh^{-1}\,,\quad h\in SU(3)_V\,,\\
    N &\rightarrow h N \,,\quad h\in SU(2)_V\,.
\end{align}
Thus the covariant derivatives on them are defined as
\begin{align}
    & SU(3):\quad D_\mu B = \partial_\mu B + [\Gamma_\mu,B]\,, \\
    & SU(2):\quad D_\mu N = \partial_\mu N + \Gamma_\mu N\,.
\end{align}

Besides, there could be other fermions such as the leptons from the electroweak interactions in the LEFT~\cite{Jenkins:2017jig,Jenkins:2017dyc,Liao:2020zyx,Li:2020tsi,Murphy:2020cly}, the right-handed neutrinos from the $\nu$LEFT~\cite{Li:2021tsq}, and the fermionic dark matter from new physics~\cite{Bishara:2016hek,Song:2023jqm}.
For example, the leptons $e_L\,, e_R$ and $\nu_L$ are of the trivial representation of the $SU(N)_V$, thus the covariant derivatives on them are just the partial derivatives
\begin{equation}
    D_\mu \psi = \partial_\mu \psi\,, \quad \psi = e_L\,, e_R\,, \nu_L\,.
\end{equation}
Treating these leptons as basic building blocks, we can avoid introducing various external sources, which will be illustrated further in Sec.~\ref{sec:weakoperators}. In summary, the baryons and the leptons of the ChPT are also listed in Tab.~\ref{tab:building_blocks}.

\subsubsection*{Discrete Symmetries}

Apart from the continuous symmetries such as the Lorentz symmetry and the chiral symmetry $SU(N)_L \times SU(N)_R$ or $SU(N)_V$, the discrete symmetries are important. For example, the charge conjugation ($C$) and the parity ($P$) are two symmetries of the strong interaction.

As mentioned before, the parity transformation makes the broken generators $T_{\hat{a}}$ transform to their minus, which means under the parity transformation, $u_\mu $ becomes its minus, and the external sources $f_\pm\,, \chi_\pm$ transform according to their subscripts.
The charge conjugation interchanges the particles and their anti-particles, which means the field matrices in Eq.~\eqref{eq:field1} and Eq.~\eqref{eq:field2}, transform into their transposed matrices. Thus the transformation of $u_\mu$ is determined,
\begin{align}
    C: \quad  u_\mu \rightarrow u_\mu^T\,.
\end{align}
At the same time, the scalar external sources transform as $\chi_\pm \rightarrow \chi_\pm ^T$,
while for the vector sources, we further assume that $l_\mu$ and $r_\mu$ are of negative intrinsic charge and obtain 
\begin{equation}
    C: \quad f_\pm {}_{\mu\nu} \rightarrow \mp f_\pm{}_{\mu\nu}^T\,.
\end{equation}

The extension to the fermions makes the $C\,,P$ properties of the effective operators more complicated since the Dirac matrices $\Gamma$ are included.
In particular, the derivatives applied on the fermion bilinears are defined as
\begin{equation}
    (\overline{\psi}\Gamma \overleftrightarrow{D}^\mu \psi) = (\overline{\psi}\Gamma D^\mu \psi) - (D^\mu \overline{\psi}\Gamma \psi)\,.
\end{equation}
The $C\,, P$ properties of all the bosonic and fermionic building blocks are summarized in Tab.~\ref{tab:buildingblocks}, including all the Dirac bilinear structures.

\begin{table}
    \centering
\renewcommand\arraystretch{1.5}
    \begin{tabular}{|c|c|c|c|c|c|}
\hline
\multicolumn{6}{|c|}{Bosonic building blocks} \\
\hline
fields & $u_\mu$ & $f_+$ & $f_-$ & $\chi_+$ & $\chi_-$ \\
\hline
$C$ & $u_\mu^T$ & $-f_+^T$ & $f_-^T$ & $\chi_+$ & $\chi_-$ \\
$P$ & $-u_\mu$ & $f_+$ & $-f_-$ & $\chi_+$ & $-\chi_-$ \\
$\mathcal{O}(p)$ & 1 & 2 & 2 & 2 & 2 \\
\hline
    \end{tabular}
    \\
    \vspace{0.5cm}
    \begin{tabular}{|c|c|c|c|c|c|c|c|}
\hline
\multicolumn{8}{|c|}{Fermionic building blocks} \\
\hline
& 1 & $\gamma^5$ & $\gamma^\mu$ & $\gamma^5\gamma^\mu$ & $\sigma^{\mu\nu}$ & $\epsilon_{\mu\nu\rho\lambda}$ & $\overleftrightarrow{D}^\mu$ \\
 \hline
 $C$ & $+$ & $+$ & $-$ & $+$ & $-$ & $+$ & $-$ \\  
 $P$ & $+$ & $-$ & $+$ & $-$ & $+$ & $-$ & $+$ \\
 $\mathcal{O}(p)$ & 0 & 1 & 0 & 0 & 0 & 0 & 0 \\ 
\hline
    \end{tabular}
    \caption{The $C\,, P$ properties and the power-countings of the ChEFT building blocks.}
    \label{tab:buildingblocks}
\end{table}

\subsection{Effecive Lagrangian and Power-Counting Scheme}

In this subsection, we will discuss the ChEFT effective Lagrangian invariant under the Lorentz symmetry, the chiral symmetry, and the discrete  $C\,,\,P$ symmetries. The building blocks are the mesons, nucleons, and the leptons as shown in Tab.~\ref{tab:building_blocks}, 
and generally, the ChEFT Lagrangian can be divided into different sectors,
\begin{equation}
\label{eq:la}
\mathcal{L}_\text{ChEFT} = \underbrace{\mathcal{L}_\pi+\mathcal{L}_{\pi N}}_{\text{ChPT}}+\mathcal{L}_{NN}+\mathcal{L}_{\pi NN} + \mathcal{L}_{NNN}+\mathcal{L}_{\text{leptonic}}\dots\,,     
\end{equation}
where the ellipsis contains the terms involving more nucleons. Each sector takes the form of an infinite expansion in terms of the chiral power-counting scheme.
Next, we will discuss all these sectors separately except the leptonic sector $\mathcal{L}_{\text{leptonic}}$, which is postponed to Sec.~\ref{sec:weakoperators}.

\subsubsection*{The ChPT Lagrangian and the Weinberg Power-Counting Scheme}

The sectors $\mathcal{L}_{\pi}$ and $\mathcal{L}_{\pi N}$ compose the ChPT Lagrangian~\cite{Weinberg:1968de,Weinberg:1978kz,Gasser:1983yg,Gasser:1984gg,Gasser:1987rb}. For the pure meson sector $\mathcal{L}_\pi$, its leading-order (LO) Lagrangian respects the $C\,,P$ symmetry contains the kinematic term and the mass term,  
\begin{equation}
\label{eq:leadboson}
        \mathcal{L}_\pi^{(2)} = \frac{{f}^2}{4}\text{Tr}(u_\mu u^\mu) + \frac{{f}^2}{4}\text{Tr}(\chi_+)\,,
\end{equation}
where the superscript $2$ is associated with the power-counting scheme and will be discussed later. As for the meson-nucleon sector $\mathcal{L}_{\pi N}$, its LO Lagrangian takes the form~\cite{Gasser:1987rb}
\begin{align}
    \mathcal{L}_{\pi N}^{(1)}&=\overline{N}(i\gamma^\mu D_\mu-M+\frac{g_A}{2}\gamma^5\gamma^\mu u_\mu)N\label{eq:lead}\,,
\end{align}
where $M$ is the nucleon mass matrix and $g_A$ is the axial vector coupling constants. The superscript $1$ is also related to the power-counting scheme and will be clarified subsequently. Besides, the LO meson-baryon interactions of the $SU(3)$ case is
\begin{equation}
    \mathcal{L}_{\pi B} = \lra{\overline{B}(i\gamma^\mu D_\mu - M)B} - \frac{D}{2}\lra{\overline{B}\gamma^\mu\gamma^5\{u_\mu,B\}}-\frac{f}{2}\lra{\overline{B}\gamma^\mu\gamma^5[u_\mu,B]}\,,
\end{equation}
where $M$ denotes the mass of the baryon octet, $f,D$ are the low-energy coefficients, and $\lra{...}$ denotes the trace over the $SU(3)_V$ space. Throughout this paper, we will use $\lra{...}$ to represent the traces of both the $SU(2)_V$ and $SU(3)_V$ cases.

The power of the effective operators in the $\mathcal{L}_{\pi}$ and the $\mathcal{L}_{\pi N}$ is determined by the powers of the ratios $p/\Lambda_\chi\sim p/(4\pi f)$ and $m/\Lambda_\chi$, where $p$ is the transferring momentum and $m$ is the mass of the pions. Thus we can set $u_\mu$ and the covariant derivative $D_\mu$ of order $p/\Lambda_\chi$, while the external sources are of order $(p/\Lambda_\chi)^2$. As for the nucleons, its propagator is of $p^{-1}$, thus all the effective interactions from these two sectors can be assessed consistently by the Weinberg power-counting formula~\cite{Weinberg:1990rz, WEINBERG19913}
\begin{equation}
\label{eq:nuw}
    \nu = 2-A+2L + \sum_i\Delta_i\,,
\end{equation}
where $A$ is the number of the nucleons, $L$ is the number of the loops and $\Delta_i$ is the so-called interaction index, defined for the $i$th vertex as
\begin{equation}
\label{eq:index}
    \Delta_i = d_i +\frac{n_i}{2}-2\,,
\end{equation}
by the number of the derivatives $d_i$ and the number of the nucleons $n_i$ of this vertex. 
The quantity $\nu$ assessing the interactions is called the chiral dimension sometimes.
The interaction index is non-negative, and the LO interactions must be composed of the interactions corresponding to the tree-level diagrams with the vertices of the minimal interaction indices.
For the pure meson sector $\mathcal{L}_\pi$, the minimum of $\Delta$ is 0 when there are only two derivatives, thus the LO Lagrangian takes the form of Eq.~\eqref{eq:leadboson}, with $\nu=2$, or of $p^2$ order. As for the pion-nucleon sector $\mathcal{L}_{\pi N}$, the minimal $\Delta$ is 0 as well, while in such case, these vertices contain two nucleons and one derivative, thus the LO Lagrangian corresponds to Eq.~\eqref{eq:lead}, with $\nu=1$, or of $p$ order. The chiral dimensions of the bosonic building blocks are listed in Tab.~\ref{tab:buildingblocks}.

However, the interaction index defined in Eq.~\eqref{eq:index} is not well defined for the relativistic nucleon operators because of the considerable nucleon mass. The nucleon mass $M$ does not vanish in the chiral limit and is comparable with the large energy scale $\Lambda_\chi$, so the time-derivative $D_0$ applying to the nucleon gives 
\begin{equation}
    D_0 \rightarrow E \approx M \sim \Lambda_\chi\,,
\end{equation} 
thus can not be considered as a small quantity. 
As a consequence, the nucleon mass $M$ appearing in the loop evaluation contributes to the divergent terms that break the power-counting scheme in Eq.~\eqref{eq:nuw}, the so-called power-counting break (PCB) terms. This problem is severe for the other sectors of multiple nucleons.

\subsubsection*{Heavy Baryon Form}

To remedy the problem above, the heavy baryon projection (HBP)~\cite{Jenkins:1990jv, Ecker:1995rk, Fettes:2000gb, Kobach:2018pie} is proposed to expand the relativistic Lagrangian to a non-relativistic form. 

In terms of the projection operator
\begin{equation}
    P_v^{\pm} = \frac{1\pm v\!\!\!\slash}{2}\,,
\end{equation}
where $v$ is a time-like vector satisfying $v^2=1$, the relativistic nucleon can be decomposed as
\begin{equation}
    N(x) = e^{-iMv\cdot x}\left[\underbrace{e^{iMv\cdot x} P_v^+N(x)}_{\equiv \mathcal{N}(x)} + \underbrace{e^{iMv\cdot x}P_v^- N(x)}_{\equiv \mathcal{H}(x)}\right]\,,
\end{equation}
then the ChEFT Lagrangian can be expressed by the new fields $\mathcal{N}(x)$ and $\mathcal{H}(x)$.

Consider the meson-baryon Lagrangian $\mathcal{L}_{\pi N}$ as an example, a general operator $\overline{N}\Gamma N$ takes the form
\begin{equation}
    \overline{N}\Gamma N = (\overline{\mathcal{H}} + \overline{\mathcal{N}})e^{iMv\cdot x} \Gamma e^{-iMv\cdot x}(\mathcal{N} + \mathcal{H} )\,,
\end{equation}
where $\Gamma = i\gamma^\mu D_\mu - M + \dots$, and the dots contain the structures composed of the pions and the external sources of higher order. Because $\overline{\mathcal{N}}v\!\!\!\slash \mathcal{H} = 0$, we can replace the term $\overline{\mathcal{H}}i\gamma^\mu D_\mu \mathcal{N}$ by $\overline{\mathcal{H}}i\gamma_\mu D^\mu_\bot \mathcal{N}$, and the Lagrangian becomes
\begin{equation}
    \mathcal{L}_{\pi N} = \overline{\mathcal{N}}(i\gamma^\mu D_\mu + \dots) \mathcal{N} + \overline{\mathcal{H}}(i\gamma_\mu D^\mu_\bot -2M) \mathcal{H} + \overline{\mathcal{N}}(i\gamma_\mu D^\mu_\bot + \dots) \mathcal{H} + h.c.\,,
\end{equation}
where $D^\mu_\bot = D^\mu - v^\mu(v\cdot D)$. Once the anti-field $\mathcal{H}$ is integrated out, the Lagrangian takes the heavy baryon form
\begin{equation}
    \mathcal{L}_{\pi N} = \overline{\mathcal{N}} \tilde{\Gamma} \mathcal{N} \,,
\end{equation}
where $\tilde{\Gamma}$ is composed of the mesons and the external sources, taking the expansion form such as
\begin{equation}
    \tilde{\Gamma} = \tilde{\Gamma}_{(1)} + \tilde{\Gamma}_{(2)} + \tilde{\Gamma}_{(3)} + \dots \,.
\end{equation}
The leading order is
\begin{equation}
    \tilde{\Gamma}_{(1)} = i v\cdot \nabla + g_A S\cdot u \,,
\end{equation}
where
\begin{equation}
    S^\mu = \frac{i\gamma^5 \sigma^{\mu\nu}v_\nu}{2}\,.
\end{equation}

By this method, the effective Lagrangian is expanded in terms of both $p/\Lambda_\chi$ and $p/M$, with an implicit assumption that the transferring momentum $p$ satisfies $p\ll M$. Nevertheless, the HBP form makes the Dirac matrices be of different dimensions since the remaining fields $\mathcal{H}$ and $\mathcal{N}$ satisfy the equation of motion (Up to LO)
\begin{equation}
    (-iv\cdot D -2M +\frac{g_A}{2} {u\!\!\slash}_\perp  \gamma^5) \mathcal{H} + (i{D\!\!\!\!\slash}_\perp - \frac{g_A}{2}v\cdot u \gamma^5) \mathcal{N}=0\,.
\end{equation}
Taking $v=(1,0,0,0)$ and recalling that $D^0 \rightarrow M \sim \Lambda_\chi \gg p$, there is an approximate solution that
\begin{equation}
    \mathcal{H} \approx \frac{\vec{p}\cdot \vec{\gamma}}{2M}\mathcal{N}\,,
\end{equation}
which means the component $\mathcal{H}$ is suppressed relative to the $\mathcal{N}$. Among the Dirac matrices under the Dirac representation, only $\gamma^5$ is not block-diagonal and mixes $\mathcal{H}$ and $\mathcal{N}$, thus should be counted as a higher-order term. The chiral dimension of the Dirac matrices and the other associated quantities are summarized in Tab.~\ref{tab:buildingblocks}.

We adopt the HBP power-counting scheme in this paper, but there are some other schemes since though the HBP scheme solves the PCB terms, the Lorentz symmetry is lost. Besides, the HBP method can not generate the correct analytic properties of the amplitudes. To fix that, covariant approaches should be adopted.
The infrared scheme (IR method)~\cite{Becher:1999he,Becher:2001hv,Torikoshi:2002bt,Alarcon:2011kh} was proposed so that it reserves the power-counting formula in Eq.~\eqref{eq:nuw} by the Lorentz-invariant form, as it performs a resummation over the positive-energy part of the kinematic terms. A drawback of the IR method is that the resummation of the kinematic terms completely omits the negative energy pole or the anti-nucleon contribution, resultant from which an unphysical cut will emerge so that another relativistic scheme called extended-on-mass-shell (EOMS) renormalization scheme was developed~\cite{Gegelia:1999gf, Fuchs:2003qc,Geng:2008mf,Alarcon:2011zs,Alarcon:2011px,MartinCamalich:2011py,Alarcon:2012kn}. Compared with the former two schemes, the EOMS scheme keeps the correct poles as well as the manifestly Lorentz invariance. 

\subsubsection*{Multi-Nucleon Lagrangian and Power-Counting Scheme}

The multi-nucleon sectors contain the 2-nucleon sector $\mathcal{L}_{NN}$, the 3-nucleon sector $\mathcal{L}_{NNN}$, the 2-nucleon-meson sector $\mathcal{L}_{\pi NN}$, and so on. many authors have been discussing the effective operator basis of these sectors~\cite{Ordonez:1995rz, Kaiser:1997mw, Kaiser:1998wa, Girlanda:2010ya, Xiao:2018jot}, but no systematic construction method has been developed. In this paper, we adopt the Young tensor method~\cite{Li:2020xlh, Li:2020gnx, Li:2022tec}, which will be discussed in Sec.~\ref{sec:young}, to efficiently establish the complete and independent operators of these sectors. In addition, we utilize the Hilbert series method~\cite{Feng:2007ur,Jenkins:2009dy,Hanany:2010vu,Hanany:2014dia,Lehman:2015via, Lehman:2015coa,Henning:2015daa,Henning:2015alf,Henning:2017fpj,Marinissen:2020jmb} to count the invariant operators conserving the $C$ and $P$ symmetries for complementation, which will be discussed in Sec.~\ref{sec:hs}.

Despite the modifications above, the formula in Eq.~\eqref{eq:nuw} is not applicable in the 2-nucleon sector $\mathcal{L}_{NN}$, since it excludes any bound states at LO, thus can not explain the shallow deuteron bound state. Thus the LO nucleon-nucleon scattering must be non-perturbative, and all the 2PR diagrams should be resumed. Weinberg attributed such perturbation break to the enhancement of the intermediate nucleon states~\cite{WEINBERG19913}. As a consequence, the LO interactions from $\mathcal{L}_{NN}$ contain not only the connected diagram and the one-pion-exchange diagram but also all their ladder diagrams. If we define the effective potential~\cite{Weinberg:1990rz} as the sum of all the connected diagrams without pure nucleon states, the power-counting formula in Eq.~\eqref{eq:nuw} is valid for it, since the momenta of nucleons and of the pions are of the same order due to the momentum conservation. Then the non-perturbative result can be obtained by the Lippman-Schwinger equation with the effective potential.

Although it is phenomenologically successful, Weinberg's approach is inadequate since the resultant LO potential is not renormalizable in the traditional sense, which means that it can generate divergent terms that are out of the LO structures and infinite counter terms should be involved to cancel them. Another method developed by Kaplan, Savage, and Wise (KSW) is to promote the one-pion-exchange diagram to be perturbative and next-to-leading order (NLO)~\cite{Kaplan:1996xu,Kaplan:1998we,Kaplan:1998xi,Kaplan:1998tg,Kaplan:1998sz}. However, the convergence of this method is not satisfactory~\cite{Mehen:1998tp,Mehen:1998zz,Fleming:1999bs,Fleming:1999ee}. An alternative way to manage such a non-perturbative problem is to introduce a finite cutoff $\Lambda_b$ as a regulator~\cite{Epelbaum:2009sd,Gasparyan:2021edy}, of the same order of the chiral scale $\Lambda_b\sim \Lambda_\chi$. The advantage of this method is that it can reserve both the Lorentz and the chiral symmetry, while the results would be cutoff-dependent.

In summary, although there are many power-counting schemes, the non-relativistic form of operators was used extensively in the early nuclear force literature~\cite{Weinberg:1990rz,WEINBERG19913,Kaplan:1996xu,Kaplan:1998tg,Kaplan:1998we}, because it is consistent with both the non-relativistic nature of the nucleons and the power-counting formula in Eq.~\eqref{eq:nuw}.
In this paper, we present the effective operators of the multi-nucleon sectors of the ChEFT in their relativistic form according to Weinberg's power-counting formula in Eq.~\eqref{eq:nuw} with certain modifications as shown in Sec.~\ref{sec:NNoperators}.

\subsection{Redundancies}

\label{sec:redundancies}

When constructing the higher-dimension effective operators, many redundancies should be eliminated. In the case of the ChEFT, these redundancies can be divided into two sectors, the Lorentz redundancies, and the chiral symmetry redundancies, where the former contains the equation of motion (EOM), the integrating-by-part (IBP), the covariant derivatives' commutation (CDC), and the Fierz identities, the latter is mainly the Cayley-Hamilton theorem. In addition, there are some extra relations about the external source $f_-$.
In this section, we will present all of these redundancies explicitly.

\subsubsection*{Lorentz Redundancies}

For the Lorentz redundancies, the EOM of the pions $u_\mu$ and the nucleons $N$ can be obtained from the LO Lagrangian in Eq.~\eqref{eq:leadboson} and Eq.~\eqref{eq:lead}, which are
\begin{equation}
    D^\mu u_\mu = 0\,,
\end{equation}
and 
\begin{align}
    &i\gamma^\mu D_\mu N  = (M-\frac{g_A}{2}\gamma^5\gamma^\mu u_\mu) N\,, \\
    &-iD_\mu\overline{N}\gamma^\mu = \overline{N}(M-\frac{g_A}{2}\gamma^5\gamma^\mu u_\mu)\,.
\end{align}
These EOMs imply $D^\mu u_\mu$, $i\gamma^\mu D_\mu N$ or $iD_\mu \overline{N}\gamma^\mu$ can be replaced by the terms without derivatives. As for the external sources $\chi_\pm$ and $f_\pm$, they possess no EOM since they are not the dynamic degrees of freedom of the ChEFT. This means the terms such as
\begin{equation}
    D^2 \chi_\pm\,,\quad  D_\mu f_\pm^{\mu\nu}\,, \quad D^2 f_\pm {}_{\mu\nu}\,,
\end{equation}
should be reserved.


The IBP redundancy originates from that the total derivative terms are vanishing when doing the integration over the whole space, which allows us to move the derivatives among the fields. For a general operator composed of several nucleons and several bosons, the total derivatives $D^\mu$ on the fermion bilinears can be moved to the bosons, 
\begin{equation}
    D^{\mu_1}D^{\mu_2}\dots D^{\mu_n}(\overline{N}\Gamma \Pi\overleftrightarrow{D}^{\nu_1}\dots\overleftrightarrow{D}^{\nu_m} N) \rightarrow (\overline{N}\Gamma (D^{\mu_1}D^{\mu_2}\dots D^{\mu_n}\Pi) \overleftrightarrow{D}^{\nu_1}\dots\overleftrightarrow{D}^{\nu_m}N)\,,
\end{equation}
thus all the derivatives on the nucleons are $\overleftrightarrow{D}$, 
where $\Pi$ could be composed of some bosons or leptons.

As shown in Eq.~\eqref{eq:cdc}, the covariant derivatives' commutation $[D_\mu, D_\nu]$ is not independent and is equivalent to the $[u_\mu,u_\nu]$ and the field strength tensor $f_+{}_{\mu\nu}$, thus we can treat all the covariant derivatives as they are commutable.
The Fierz identities between the fermion bilinears is an important redundancy, which allows us to consider only the operators with the spinor indices contracting within the bilinears,
\begin{equation}
    (\overline{N}^\alpha{\Gamma_1}_{\alpha\beta} N^\lambda)(\overline{N}^\rho{\Gamma_2}_{\rho\lambda} N^\beta) \rightarrow (\overline{N}^\alpha{\Gamma_3}_{\alpha\beta} N^\beta)(\overline{N}^\rho{\Gamma_4}_{\rho\lambda} N^\lambda) = (\overline{N}\Gamma_3N)(\overline{N}\Gamma_4N)\,,
\end{equation}
where $\alpha\,,\beta\,,\rho\,,\lambda$ are the spinor indices.

\subsubsection*{Chiral symmetry Redundancies}

There are two important redundancies, one is the Cayley-Hamilton relation and the other is the completeness relation of the $SU(2)$ group.

The Cayley-Hamilton theorem states that any $2\times 2$ matrices $A$ satisfies that
\begin{equation}
\label{eq:chr}
    A^2 = \lra{A} A + \frac{1}{2}(\lra{A^2}-\lra{A}^2)\mathbb{I}_{2\times 2}\,,
\end{equation}
where $\mathbb{I}_{2\times 2}$ is the identity matrix. Considering the adjoint representation, the fields are traceless, thus the relation becomes
\begin{equation}
    A^2 = \frac{1}{2}\lra{A^2}\mathbb{I}_{2\times 2}\,,
\end{equation}
Let $A=A+B$ and extract the terms proportional to $AB$, the Cayley-Hamilton relation implies
\begin{equation}
\label{eq:ch1}
    AB+BA = \lra{AB}\,,
\end{equation}
which constrains the traces of more than 2 fields. For example, in the simplest case of 3 fields, there are only two traces considering the cyclic property,
\begin{equation}
    T_1 = \lra{ABC}\,,\quad T_2 = \lra{BAC}\,,
\end{equation}
while the Cayley-Hamilton relation in Eq.~\eqref{eq:ch1} implies
\begin{equation}
    T_1 + T_1 = \lra{ABC} + \lra{BAC} = \lra{AB}\lra{C}=0\,,
\end{equation}
thus only one trace is independent, which can be chosen to be $T_1-T_2 = \lra{[A,B]C}$. 

In the case of 4 fields, there are 9 traces,
\begin{align}
    & T_1 = \lra{AB}\lra{CD}\,, \quad T_2=\lra{AC}\lra{BD}\,, \quad T_3=\lra{AD}\lra{BC}\,, \notag \\
    & T_4 = \lra{ABCD}\,, \quad T_5 =\lra{ABDC} \,,\quad T_6= \lra{ACBD} \,, \notag \\
    & T_7 = \lra{ACDB}\,,\quad T_8 = \lra{ADBC}\,, \quad T_9 = \lra{ADCB}\,,
\end{align}
and there exit 6 independent Cayley-Hamilton relations
\begin{align}
    & \lra{ABCD}+\lra{BACD} = \lra{AB}\lra{CD}\,, \\
    & \lra{ABDC}+\lra{BADC} = \lra{AB}\lra{CD}\,, \\
    & \lra{ACBD}+\lra{CABD} = \lra{AC}\lra{BD} \,, \\
    & \lra{ACDB}+\lra{CADB} = \lra{AC}\lra{BD}\,, \\
    & \lra{ADBC}+\lra{DABC} = \lra{AD}\lra{BC}\,, \\
    & \lra{ADCB}+\lra{DACB} = \lra{AD}\lra{BC}\,,
\end{align}
thus there are only 3 independent traces, and they can be chosen as
\begin{equation}
    T_1 = \lra{AB}\lra{CD}\,, \quad T_2=\lra{AC}\lra{BD}\,, \quad T_3=\lra{AD}\lra{BC}\,.
\end{equation}

In addition, in the multi-nucleon sectors, the fermions can be rearranged via the completeness relation of the $SU(2)$ group 
\begin{equation}
    \delta_{ij}\delta_{kl} = \delta_{il}\delta_{kj} + \frac{1}{2} \sum_{I=1}^3 {\tau^I}_{il}{\tau^I}_{kj}\,,
\end{equation}
where $\tau^I$ are the $SU(2)$ generators. This relation implies any operators containing $\tau^I$ can be eliminated, 
\begin{equation}
    (\overline{N}^i \Gamma_1 {\tau^I}_{il} N^l)(\overline{N}^k \Gamma_2 {\tau^I}_{kj} N^j) \rightarrow (\overline{N}^i \Gamma_1  N_j)(\overline{N}^j \Gamma_2  N_i) - (\overline{N}^i \Gamma_1  N_i)(\overline{N}^j \Gamma_2  N_j)\,.
\end{equation}

\subsubsection*{Constraint on $f_-$}

The field-strength $f_-{}_{\mu\nu}$ satisfies the relation that
\begin{equation}
    D_\mu u_\nu - D_\nu u_\mu - f_-{}_{\mu\nu}=0\,,
\end{equation}
which means all the Lorentz indices of the object $D_\mu u_\nu$ should be symmetric since we have adopted $f_-$ as an independent building block.

In this paper, we focus on the ChEFT of the $SU(2)$ case, and in Sec.~\ref{sec:young}, we will show that all the relevant redundancies mentioned here can be eliminated systematically via the Young tensor method~\cite{Li:2020gnx,Li:2022tec,Li:2020xlh}. Especially, the trace basis obtained by the Cayley-Hamilton relations is equivalent to the tensor basis presented there. 

\section{Operator Counting via the Hilbert Series}
\label{sec:hs}

Before writing down all the independent operators in the Lagrangian, the Hilbert series~\cite{Feng:2007ur,Jenkins:2009dy,Hanany:2010vu,Hanany:2014dia,Lehman:2015via, Lehman:2015coa,Henning:2015daa,Henning:2015alf,Henning:2017fpj,Marinissen:2020jmb} can be used to count the numbers of them. To reach this goal, let us consider the general structure of the operators. The most general form of an operator is a polynomial of some building blocks $\{\phi_{i}\},i=1,\dots,m$ (whether boson or fermion) and the derivative $D$. 
The polynomial is invariant under the symmetry group of the theory, which we note as $G$ generally. 

Suppose the group $G$ is a tensor product of some $SU(N)$ groups, which is the case for the ChEFT considered in this paper since the symmetry group of the ChEFT is the tensor product of the Lorentz group $SO(3,1)\cong SU(2)_l\times SU(2)_r$ and the internal group $SU(2)_V$. The double-cover group of the Lorentz group $SU(2)_l\times SU(2)_r$ should no be confused with the chiral group $SU(2)_L\times SU(2)_R$. The character of a specific representation of the group $G$ can be expressed by the variables of its maximal torus,
\begin{equation}
    \chi(g) \equiv \chi(\{x\})\,,\quad g\in G\,,
\end{equation}
and $\{x\} = x_1\,,x_2\,,\dots\,,x_n$ are the maximal torus variables satisfying $|x_i| = 1$. In this section, we use $\chi$ to represent the group characters, which should not be confused with the external source introduced in Sec.~\ref{sec:ccwz}. As $\{x\}$ ranges on the maximal torus, the character $\chi(g)$ runs over all the group elements $g\in G$. The symmetric product of the characters is realized by the plethystic exponential (PE)~\cite{Feng:2007ur},
\begin{equation}\label{eq:PEdef}
    \text{PE}(\chi(x_1,\cdots,x_n))\equiv \exp{\left(\sum_{k\geq1} \frac{\chi(x_{1}^{k},\cdots,x_{n}^{k})}{k}\right)}.
\end{equation}
When applying to the effective operators, there exists a PE for each field $\phi_i$, and the product of the fermion characters should be antisymmetric, thus we extend the PE as
\begin{equation}
    \text{PE}(\phi_i,\chi_{\phi_i}(\{x\}))\equiv \left\{
    \begin{aligned}
    &\exp{\left(\sum_{k\geq1} \frac{\phi_{i}^{k}}{k} \chi_{\phi_i}(x_1^k\,,\dots x_n^k)\right)} &,\phi_i\;\text{is boson},
    \\
    &\exp{\left(\sum_{k\geq1} \frac{(-1)^{k-1}\phi_{i}^{k}}{k}\chi_{\phi_i}(x_1^k\,,\dots\,,x_n^k)\right)} &,\phi_i\;\text{is fermion\,,}
    \end{aligned}
        \right.
\end{equation}
where we have multiplied a variable $\phi_i$ with the associated character $\chi_{\phi_i}$ to distinguish them. Then we can obtain the total character of all the fields by the product of their PEs,
\begin{equation}
    Z(\{\phi_i\},g) = Z(\{\phi_i\},\{x\})\equiv \prod_{i}^{m} \text{PE}(\phi_i,\chi_{\phi_i}(\{x\})).
\end{equation}

Having gotten the character, we can get the generating function. Due to Schur's lemma, we can use the orthonormality of the character to project out the trivial representation:
\begin{equation}\label{eq:hswithoutredun}
\mathcal{H}\equiv \int_{G} Z(\{\phi_i\},g) d\mu(g).
\end{equation}
Here $d\mu(g)$ is the Haar measure of the group, which can also be expressed by the maximal torus variables.
Let us list these measures in Tab.~\ref{tab:measures}. Note that the Lorentz group is considered by using the double-cover group: $SO(3,1)\cong SU(2)_l\times SU(2)_r$. After this preparation, we just work out the integral in Eq.~\eqref{eq:hswithoutredun} and get a formal series,
\begin{table}[]
\renewcommand{\arraystretch}{1.8}
    \centering
    \begin{tabular}{|c|c|}
    \hline
          $d\mu_{SU(2)}/\frac{dx}{2\pi xi}$&$d\mu_{SU(3)}/\frac{dz_1dz_2}{(2\pi i)^2z_1z_2}$ \\
          \hline
        $\frac{1}{2}(1-x^{2})(1-x^{-2})$ &$\frac{1}{6}\Pi_{3\geq i>j\geq1,z_1z_2z_3=1}(1-z_{i}z_{j})(1-z_{i}^{-1}z_{j}^{-1})$ \\
        \hline
        $d\mu_{U(1)}/\frac{dx}{2\pi xi}$&$d\mu_{SO(4)}/\frac{dx_1dx_2}{(2\pi i)^2x_1x_2}$ \\
        \hline
        $1$&$\frac{1}{4}(1-x_{1}^{2}x_{2}^{2})(1-x_{1}^{-2}x_{2}^{-2})(1-x_{1}^{-2}x_{2}^{2})(1-x_{1}^{2}x_{2}^{-2})$ \\
        \hline
    \end{tabular}
    \caption{The Haar measures of the groups we used in this work. 
    }
    \label{tab:measures}
\end{table}
\begin{equation}
\label{eq:hs_form}
    \mathcal{H}= \sum_{k_1,\cdots,k_m}\left(\#_{k_1,k_2,\cdots,k_m}\right)\phi_1^{k_1} \phi_2^{k_2} \cdots \phi_m^{k_m}.
\end{equation}
We can find that $\#_{k_1,k_2,\cdots,k_m}$ gives the number of invariants of the type $\phi_1^{k_1} \phi_2^{k_2} \cdots \phi_m^{k_m}$.

Eq.~\eqref{eq:hswithoutredun} is the Hilbert series without removing the redundancies of the EOM and the IBP. Thus the expression in Eq.~\eqref{eq:hswithoutredun} should be modified.   
The two redundancies can be naturally removed by using the unitary irreducible representations of the conformal group in the Hilbert series method~\cite{Minwalla1997RestrictionsIB,Ferrara:2000nu,Dolan:2005wy}. But we will not follow their steps and will only outline the important results.



\subsection{Counting with EOM and IBP}

The derivative $D$ is not an independent building block because of the redundancies such as the EOM and the IBP. To eliminate these redundancies, the Lorentz group should be enlarged to the conformal group. 

In the conformal group, the derivative is a lowing operator that extends the irreducible fields $\phi_i$ under the Lorentz group to a tower structure. For example, an irreducible field $\phi_i$ of the representation $(a,b)$ of the Lorentz group is extended as
\begin{equation}
\label{eq:spm}
    \phi_i \longrightarrow M_{\phi_i} = \left(\begin{array}{c}
\phi_i \\ D\phi_i \\D^2\phi \\\vdots
    \end{array}\right)\,,
\end{equation}
where $M_{\phi_i}$ is called a single particle module (SPM). The SPM is a direct sum of many subspaces, each of which is a generally reducible representation of the Lorentz group, for example,
\begin{equation}
    \phi_i \in (a,b)\,,\quad D\phi_i \in (a,b)\otimes (\frac{1}{2},\frac{1}{2})\,,\quad D^2\phi_i \in (a,b)\otimes (\frac{1}{2},\frac{1}{2})\otimes (\frac{1}{2},\frac{1}{2})\,,\quad \dots
\end{equation}
Considering the addition and the multiplicity properties of the characters, it is not difficult to obtain the SPM characters of different fields, which take the universal form that
\begin{equation}
    \chi_{SPM}(\phi_i,g) = P(D,g)\chi_{\phi_i}(g)\,,
\end{equation}
where $P(D,g)$ is a factor appearing frequently, and its definition is
\begin{equation}
    P(D,g)\equiv\text{PE}(D) = \exp\left(\sum_{k\geq 1}\frac{D^k \chi_{(\frac{1}{2}\,,\frac{1}{2})}(x_1^k\,,\dots\,,x_n^k)}{k}\right).
\end{equation}
At the same time, the operators are also of module form similar to the fields in Eq.~\eqref{eq:spm},
\begin{equation}
\label{eq:spm_o}
    M_{\mathcal{O}} = \left(\begin{array}{c}
\mathcal{O} \\D\mathcal{O}\\ D^2\mathcal{O} \\ \vdots
    \end{array}\right)\,,
\end{equation}
where the first subspace $\mathcal{O}$ is of trivial representation of the Lorentz group, $\chi_\mathcal{O}=1$, and all the secondary subspaces are obtained from the first one by applying derivatives, thus are redundant due to the IBP. The SPM character of the operators is just 
\begin{equation}
    \chi_{SPM}(\mathcal{O},g) = P(D,g)\,.
\end{equation}
To eliminate the IBP redundancy, we should keep only the first subspace in Eq.~\eqref{eq:spm_o}, which means the total character $Z$ in the integral should be divided by $P(D,g)$, 
\begin{equation}
    Z \rightarrow \frac{Z}{P(D,g)}\,.
\end{equation}

As for the EOM redundancy, since the EOM changes the SPMs, its elimination is equivalent to the subtraction of the SPM characters that are related. For example, the SPM of a scalar field $\phi$ and its EOM $D^2\phi$ are the same, except there is an extra factor $D^2$ in the latter, thus its character with the EOM removed is the subtraction that   
\begin{equation}
    \chi(\phi,g) = \chi_{\phi}(D,g) - \chi_{D^2\phi}(D,g) = (1-D^2)P(D,g)\,,
\end{equation}
where $\chi_{(0,0)}(g)=1$ has been used. Similarly, the characters of the left-handed and right-handed spinors are
\begin{equation}\label{eq:psiLcha}
    \chi(\psi_L,g)=(\chi_{(\frac{1}{2},0)}(g)-D\chi_{(0,\frac{1}{2})}(g))P(D,g)\,,
\end{equation}
\begin{equation}\label{eq:psiRcha}
    \chi(\psi_R,g)=(\chi_{(0,\frac{1}{2})}(g)-D\chi_{(\frac{1}{2},0)}(g))P(D,g)\,,
\end{equation}
and the characters of the left-handed and right-handed field-strength tensors are 
\begin{equation}\label{eq:FLcha}
    \chi(F_L,g)=(\chi_{(1,0)}(g)-D\chi_{(\frac{1}{2},\frac{1}{2})}(g)+D^2\chi_{(0,0)}(g))P(D,g)\,,
\end{equation}
\begin{equation}\label{eq:FRcha}
    \chi(F_R,g)=(\chi_{(0,1)}(g)-D\chi_{(\frac{1}{2},\frac{1}{2})}(g)+D^2\chi_{(0,0)}(g))P(D,g)\,.
\end{equation}
Thus the character of field strength tensor $F=F_L\oplus F_R$ is
\begin{equation}\label{eq:Fcha}
    \chi(F,g) = \chi(F_L,g) + \chi(F_R,g)\,.
\end{equation}
In particular, the mesons $u_\mu$ in the ChEFT are NGBs, which means their operators should satisfy the Adler's zero condition~\cite{Adler:1969gk}, thus the first subspace in the SPM in Eq.~\eqref{eq:spm} should be removed, so that the character of $u_\mu$ is 
\begin{equation}
    \chi(u,g) = \chi(\phi,g)-1 = (1-D^2)P(D,g)-1\,.
\end{equation}

Defining the new character with the EOM removed as
\begin{equation}
    Z(\{\phi_i\},D,g) = \prod_i^m \text{PE}(\phi_i,\chi(\phi_i,g))\,,
\end{equation}
the Hilbetr series without both the EOM and the IBP redundancies are obtained by the integral that
\begin{equation}
\label{eq:hs_2}
\mathcal{H}= \int_{G} \frac{Z(\{\phi_i\},D,g)}{P(D,g)} d\mu(g).
\end{equation}
\subsection{Charge Conjugation and Parity}

The parity and the charge conjugation are two important discrete symmetries of the ChEFT. To calculate the Hilbert series under this condition, we should contain two discrete group elements $\mathcal{C}$ (charge conjugation) and $\mathcal{P}$ (parity), and enlarge the group $G$ to $G\rtimes \{1,\mathcal{C},\mathcal{P},\mathcal{CP}\}$. 
Then the group manifold is divided into four disconnected branches, and the Hilbert series also becomes
\begin{equation}
      \mathcal{H}=\frac{1}{4}(\mathcal{H}_{++}+\mathcal{H}_{+-}+\mathcal{H}_{-+}+\mathcal{H}_{--}),
\end{equation}
where the Hilbert series on the left-hand counts the operators conserving both the parity and the charge conjugation. On the right hand, the first subscript indicates the two branches related by $\mathcal{C}$, and the second subscript indicates the two branches related by $\mathcal{P}$. The Hilbert series in these four branches are obtained by the integrals
\begin{equation}
\begin{split}
   & {\mathcal{H}_{++} \equiv \int_{G} \frac{Z(\{\phi_i\},D,g)}{P(D,g)} d\mu(g)}\,,\\
    &{\mathcal{H}_{+-} \equiv \int_{G} \frac{Z(\{\phi_i\},D,g\mathcal{P})}{P(D,g\mathcal{P})} d\mu(g)}\,,\\
   & \mathcal{H}_{-+} \equiv \int_{G} \frac{Z(\{\phi_i\},D,g\mathcal{C})}{P(D,g\mathcal{C})} d\mu(g)\,,\\
   & \mathcal{H}_{--} \equiv \int_{G} \frac{Z(\{\phi_i\},D,g\mathcal{CP})}{P(D,g\mathcal{CP})} d\mu(g)\,.
    \end{split}
\end{equation}
Considering the two branches related by $\mathcal{P}$, according to the Lie group theory, the integral on the $\mathcal{P}+$ branch can be reduced on the maximal torus,
so that the integral over the Lorentz group can be calculated as usual by the integral in Eq.~\eqref{eq:hs_2}, The factor $P(D,g)$ of this branch takes the form
\begin{equation}
    P(D,g)=P(D,x_1,x_2)\equiv\frac{1}{(1-Dx_1^2)(1-Dx_1^{-2})(1-Dx_2^2)(1-Dx_2^{-2})}\,,
\end{equation}
because a Lorentz group element $g$ of $(\frac{1}{2}\,,\frac{1}{2})$ representation can be reduced to the maximal torus as 
\begin{equation}
    g \sim \text{diag}(x_1^2\,,x_1^{-2}\,,x_2^2\,,x_2^{-2})\,.
\end{equation}

However, in the $\mathcal{P}-$ branch we have~\cite{Graf:2020yxt,Sun:2022aag} 
\begin{equation}
    g\mathcal{P}\sim \text{diag}(x_1,x_1^{-1},1,-1)\,,
\end{equation}
which is dependent on only $x_1$.
Because we have $(g\mathcal{P})^{2}=\text{diag}(x_1^{2},x_1^{-2},1,1)$, the even power contribution in the generating function is the same as that in the $\mathcal{P}+$ branch by taking $x_2\rightarrow1 $: 
\begin{equation}
    \chi((g_\mathcal{P})^{2n})=x_1^{2n} + x_1^{-2n} + 2\,.
\end{equation}
For the odd powers, we consider a field $\phi$ with the representation $(a,b)$. The parity $\mathcal{P}$ must change $\phi$ into another field $\phi^{\mathcal{P}}$ with representation $(b,a)$. Thus the group action is
\begin{equation}
\begin{pmatrix}
\phi \\
\phi^{\mathcal{P}}
\end{pmatrix}
\xrightarrow[]{g\mathcal{P}}
\begin{pmatrix}
0 & \lambda \\
\lambda' &0
\end{pmatrix}
\begin{pmatrix}
\phi \\
\phi^{\mathcal{P}}
\end{pmatrix}.
\end{equation}
Here $\lambda\,,\lambda'$ are the representation matrices of $g$ with the representations $(a,b)\,,(b,a)$ respectively. Obviously, the character of the odd-power terms would vanish if $\phi\neq \phi^{\mathcal{P}}$ since the representation matrices are off-diagonal.
 If we have $\phi=\phi^{\mathcal{P}}$, then $\mathcal{P}$ is block-diagonalized. In fact, $\mathcal{P}$ presents an intrinsic parity phase $\eta_p$. In this case, we must have $a=b= j$. 
For the general $j$ in this branch, one will get~\cite{Henning:2017fpj}
\begin{equation}\label{eq:characterinC-}
    \chi_{(j,j)}(g^{2n-1})=\eta_p\chi_{j}^{SU(2)}(x_1^{2n-1})\,,
\end{equation}
where the Lorentz group has been reduced to a $SU(2)$ group.
After introducing the character in this branch, the factor $P(D,g)$ is
\begin{equation}\label{eq:kinematic-branch}
    P(D,g\mathcal{P})=P_{-}(D,x_1)\equiv\frac{1}{(1-Dx_1)(1-Dx_1^{-1})(1-D^2)}.
\end{equation}
Based on the discussion above, we can get the odd power terms in the generating functions of the $\mathcal{P}-$ branch for various fields.
\begin{align}
    \chi(\phi,g^{2n-1})&=P_{-}(D,x_1^{2n-1})\,,\\
    \chi(\psi_L,g^{2n-1}) &= \chi(\psi_L,g^{2n-1})  =0\,, \\
    \chi(F,g^{2n-1}) &=D(\chi_{\frac{1}{2}}^{SU(2)}(x_1^{2n-1})-D)P_-(D,x_1^{2n-1})\,.
\end{align}
For practice, the intrinsic parity $\eta_p$ should be determined by the theory, which of the ChEFT in the paper has been presented in Tab.~\ref{tab:buildingblocks}. Besides the character, the Haar measure of this branch should also be replaced by the measure $d\mu_{SU(2)}$ of the $SU(2)$ group.

In the case of $\mathcal{C}$, 
We consider the $SU(2)$ group specifically. First, we should define the action of the charge conjugation $\mathcal{C}$. In fact two different kinds of conjugation are allowed, noted as $\mathcal{C}_1$ and $\mathcal{C}_2$. The definitions are
\begin{equation}
\begin{pmatrix}
N^{c i} \\
N_i
\end{pmatrix}
\xrightarrow[]{\mathcal{C}_1}
\begin{pmatrix}
N_i \\
N^{c i}
\end{pmatrix}
=
\begin{pmatrix}
0 & 1 \\
1 & 0 
\end{pmatrix}
\begin{pmatrix}
N^{c i} \\
N_i
\end{pmatrix},
\end{equation}
\begin{equation}
\begin{pmatrix}
N^{c}_i \\
N_i
\end{pmatrix}
\xrightarrow[]{\mathcal{C}_2}
\begin{pmatrix}
N_i \\
N^{c}_i
\end{pmatrix}
=
\begin{pmatrix}
0 & 1 \\
1 & 0 
\end{pmatrix}
\begin{pmatrix}
N^{c}_i \\
N_i
\end{pmatrix}.
\end{equation}
Here $N_i$ and $N^{c i}$ are $\frac{1}{2}$-isospin fermion and its anti-particle with the $SU(2)$ doublet indices with $N^{c}_i=\epsilon_{ij} N^{c j}$. $\epsilon_{ij}$ is the Levi-Civita symbol. The $\mathcal{C}_1$ is what we used in this work. The difference between these two definitions is reflected by a minus sign in the character function. Let us show how this happened by ignoring the rest group structure for convenience. When a charge conjugation and a general $SU(2)$ transformation $g$ is acted on the fermion particle, the representation matrix $\rho(\mathcal{C}_{1}g)$ is
\begin{equation}
    \rho(\mathcal{C}_{1}g)=\begin{pmatrix}
0 & 1 \\
1 & 0 
\end{pmatrix}
\begin{pmatrix}
\epsilon g \epsilon^{-1} & 0 \\
0 & g 
\end{pmatrix}\,.
\end{equation}
The odd-power terms in the character function are zero, but the even-power terms are not. Hence the character is
\begin{equation}
    \chi((\mathcal{C}_{1}g)^{2n})=(-1)^n \text{tr}\left(
    \begin{pmatrix}
(\epsilon g \epsilon g)^n& 0 \\
0 & (g\epsilon g \epsilon)^n 
\end{pmatrix}
    \right).
\end{equation}
The equality have used $\epsilon^{-1}=-\epsilon$. By using the group variable redefinition $g\rightarrow g\epsilon^{-1}$ and the invariance of the Haar measure $d\mu(g)=d\mu(g\epsilon^{-1})$ and reducing the group integral onto the maximal torus, the character function changes with a factor $(-1)^n$. As a conclusion, the $SU(2)$ group character on the $\mathcal{C}-$ branch 
is 
\begin{equation}
    \chi_{N}^{SU(2)C-}(g^{2n})=\eta_c(-1)^n(y^{2n}+y^{-2n}).
\end{equation}
where $y$ is the torus variable of $SU(2)$, and $\eta_c$ is the intrinsic charge conjugation phase. In Tab.~\ref{tab:summarycha} we list the Haar measures and the characters of the two $\mathcal{C}$ branches for some groups.

At last, we give the character function used in this work in Tab.~\ref{tab:eachcha} for convenience. In the rest of this paper, we give the Hilbert series of the $C$-even and $P$-even operators for different sectors, and in Appendix~\ref{sec:app1}, some complementary material is presented.

\begin{table}
\renewcommand{\arraystretch}{1.5}
\centering
\begin{tabular}{|c|clcccl|}
\hline
Group Branch                                      & \multicolumn{3}{c|}{$g\in U(1)$}                                                                        & \multicolumn{3}{c|}{$g=g \mathcal{C}$}                                                                               \\ \hline
Measure                                      & \multicolumn{3}{c|}{$d\mu_{U(1)}(x)$}                                                                 & \multicolumn{3}{c|}{$d\mu_{U(1)}(x)$}                                                                \\ \hline
$tr(g^{2n})$\; for\;charge q\; rep              & \multicolumn{3}{c|}{$\chi_{q}^{U(1)}(x^{2n})+\chi_{-q}^{U(1)}(x^{2n})$}                                                 & \multicolumn{3}{c|}{$\chi_{q}^{U(1)}(1)+\chi_{-q}^{U(1)}(1)$}                                                                       \\ \hline
 \multicolumn{7}{|c|}{}
\\
\hline
Group Branch                                      & \multicolumn{3}{c|}{$g\in SU(2)$}                                                                       & \multicolumn{3}{c|}{$g=g \mathcal{C}$}                                                                              \\ \hline
Measure                                      & \multicolumn{3}{c|}{$d\mu_{SU(2)}(y)$}                                                                 & \multicolumn{3}{c|}{$d\mu_{SU(2)}(y)$}                                                                        \\  \hline
$tr(g^{n})$\; for\; $j=1$                 & \multicolumn{3}{c|}{$\chi_{1}^{SU(2)}(y^n)$}                                                            & \multicolumn{3}{c|}{$ \chi_{1}^{SU(2)}(y^n)$}                                                        \\ \hline
$tr(g^{2n})$\; for\; fund-anti\; rep               & \multicolumn{3}{c|}{$\chi_{\frac{1}{2}}^{SU(2)}(y^{2n})+\chi_{\frac{1}{2}}^{SU(2)}(y^{2n})$}                        & \multicolumn{3}{c|}{$(-1)^n( \chi_{\frac{1}{2}}^{SU(2)}(y^{2n})+\chi_{\frac{1}{2}}^{SU(2)}(y^{2n}))$}                 \\ \hline
 \multicolumn{7}{|c|}{}
\\
\hline
Group Branch                                      & \multicolumn{3}{c|}{$g\in SU(3)$}                                                                       & \multicolumn{3}{c|}{$g=g \mathcal{C}$}                                                                              \\ \hline
Measure                                      & \multicolumn{3}{c|}{$d\mu_{SU(3)}(z_1,z_2)$}                                                                 & \multicolumn{3}{c|}{$d\mu_{SU(2)}(z_1)$}                                                                        \\  \hline
$tr(g^{2n})$\; for\; adjoint\; rep                 & \multicolumn{3}{c|}{$\chi_{\text{ad}}^{SU(3)}(z_{1}^{2n},z_{2}^{2n})$}                                                            & \multicolumn{3}{c|}{$\chi_{\text{ad}}^{SU(3)}(z_{1}^{n},1)$}            \\      
\hline
$tr(g^{2n-1})$\; for\; adjoint\; rep                 & \multicolumn{3}{c|}{$\chi_{\text{ad}}^{SU(3)}(z_{1}^{2n-1},z_{2}^{2n-1})$}                                                            & \multicolumn{3}{c|}{$ \chi_{1}^{SU(2)}(z_{1}^{2n-1})$}                   
\\ \hline
$tr(g^{2n})$\; for\; fund-anti\; rep               & \multicolumn{3}{c|}{$\chi_{\text{fund}}^{SU(3)}(z_{1}^{2n},z_{2}^{2n})+\chi_{\text{anti}}^{SU(3)}(z_{1}^{2n},z_{2}^{2n})$}                        & \multicolumn{3}{c|}{$\chi_{\text{fund}}^{SU(3)}(z_{1}^{n},1)+\chi_{\text{anti}}^{SU(3)}(z_{1}^{n},1)$}                 \\ \hline
\end{tabular}
\caption{Here we list the results of the character $\chi_{\phi_i}(D^n,g^n)$ and measures in different group branches. 
In this table, we haven't added any intrinsic phase so $\eta_{c}$ and $\eta_p$ should be multiplied by hand. In each branch, we also ignore the other groups' structures so the character may be simply the sum of the two character functions. In general, the character is the sum of the characters of the particle and the anti-particle. Of course, the characters of the odd-power on the $C$-branches we haven't listed are zero.}
\label{tab:summarycha}
\end{table}

\begin{table}[htb]
\renewcommand{\arraystretch}{1.5}
\centering
\begin{tabular}{|c|c|}
\hline
$\chi_{q}^{U(1)}(x)$                            & $x^q$           \\ \hline
$\chi^{SU(2)}_{j}(x)$                                & $x^{2j}+x^{2j-2}+\cdots+x^{-2j}$\\ \hline
$\chi_{(j_1,j_2)}^{SO(4)}(x_1,x_2)$                                              & $\chi^{SU(2)}_{j_1}(\frac{x_1}{x_2})\chi^{SU(2)}_{j_2}(x_1 x_2)$ \\ \hline
$\chi^{SU(3)}_{\text{fund}}(z_1,z_2)$&$z_1+z_2+\frac{1}{z_1z_2}$\\ \hline
$\chi^{SU(3)}_{\text{anti}}(z_1,z_2)$&$\frac{1}{z_1}+\frac{1}{z_2}+z_1z_2$\\ \hline
$\chi^{SU(3)}_{\text{ad}}(z_1,z_2)$&$2+\frac{z_1}{z_2}+\frac{z_2}{z_1}+\frac{1}{z_{1}^{2}z_2}+z_{1}^{2}z_2+\frac{1}{z_{1}z_{2}^{2}}+z_{1}z_{2}^{2}$\\ \hline
\end{tabular}
\caption{The detailed expressions of the character corresponding to the symmetry group involved in the ChEFT.}
\label{tab:eachcha}
\end{table}




\section{Operator Bases via Spinor Young Tensor}

\label{sec:young}

In this section, we will briefly review the Young tensor method developed in Ref.~\cite{Li:2020gnx, Li:2020xlh, Li:2022tec}.  
In this method, both the Lorentz structures and the internal structures are obtained via the Young tableaux. 

\subsection{Lorentz Structures}

A general operator is composed of some fields and derivatives, all of which are irreducible representations of the Lorentz group,

\begin{align}
    \text{Scalar field: }&\phi\in (0,0)\,, \notag \\
    \text{Left-handed spinor field: } & \psi \in (\frac{1}{2},0)\,, \notag \\
    \text{Right-handed spinor field: } & \psi^\dagger \in (0,\frac{1}{2})\,, \notag \\
    \text{Left-handed field strength tensor: } & F_L=\frac{F-i\tilde{F}}{2} \in (1,0)\,, \notag \\
    \text{Right-handed field strength tensor: } & F_R=\frac{F+i\tilde{F}}{2} \in (0,1)\,, \notag \\
    \text{Derivative: } & D \in (1,1) \,,
\end{align}
and correspond to the spinor expressions that

\begin{align}
    & \phi \sim 1 \,,\notag \\
    & \psi\sim \lambda\,,\quad \psi^\dagger\sim \tilde{\lambda}\,, \notag \\
    & {F_L} \sim \lambda\lambda ,,\quad {F_R} \sim \tilde{\lambda}\tilde{\lambda} \,, \notag\\
    & p\sim \lambda\tilde{\lambda}\,. \label{eq:corres}
\end{align}
The two-component spinors $\lambda/\tilde{\lambda}$ are of the irreducible representations of the Lorentz group $SO(3,1)\sim SU(2)_l\times SU(2)_r$,
\begin{equation}
    \lambda \in (\frac{1}{2},0)\,,\quad \tilde{\lambda} \in (0,\frac{1}{2})\,.
\end{equation}
Thus the general Lorentz structure of an operator is represented by the on-shell amplitude, which takes the form that
\begin{equation}
    \mathcal{A}= \prod^{n} \langle ij\rangle\times \prod^{\tilde{n}}[kl]\,,\label{eq:amp}
\end{equation}
where 
\begin{align}
    \langle ij\rangle &= \lambda_{i}^\alpha \lambda_{j\alpha} = -\lambda_{j}^\alpha \lambda_{i\alpha} = -\langle ji\rangle\,, \label{eq:contr1}\\
    [ij] &= \tilde{\lambda}^i_{\dot{\alpha}}\tilde{\lambda}^{j\dot{\alpha}} = -\tilde{\lambda}^i_{\dot{\alpha}}\tilde{\lambda}^{j\dot{\alpha}} = -[ji]\,.
\end{align}
Since the spinor contractions such as $\langle ii\rangle\,, [ii]$ vanish, the EOM redundancy is eliminated automatically. 
For a specific operator, whose field contents and the derivatives are fixed, thus the numbers $n$ and $\tilde{n}$ are determined,
\begin{equation}
    n=\frac{k}{2}-\sum_{h_i<0}h_i\,,\quad \tilde{n}=\frac{k}{2}+\sum_{h_i>0}h_i\,,
\end{equation}
where $k$ is the number of the derivatives and $h_i$'s are the helicities of the fields. 

With the spinor forms of the fields, we can construct the Lorentz structures by contracting them to form the Lorentz invariants. However, there are some redundancies, such as the Fierz identities (or the Schouten identity in the spinor form), the IBP, and so on, which make the construction non-trivial. To eliminate these redundancies, we notice that the contractions $\langle i j \rangle$ and $[kl]$ are all anti-symmetric, which means they can be related to the Young tableaux that
\begin{equation}
    \langle i j \rangle \sim \begin{ytableau}
        i \\j
    \end{ytableau}\,,\quad [kl] \sim \begin{ytableau}
        *(cyan) k \\ *(cyan) l 
    \end{ytableau}\sim \epsilon^{klm\dots n}\,\begin{ytableau}
            *(cyan) m \\ \none[\vdots] \\ *(cyan) n
        \end{ytableau}\,,
\end{equation}
where $\epsilon^{klm\dots n}$ is the anti-symmetric tenor of the $U(N)$ group, and $N$ is the number of the fields of the type. We have used different colors to distinguish the left- and right-handed spinors. Thus the Lorentz structures are obtained from the tensor product of these Young tableaux. It has been proved that all the tensor products are redundant because of the IBP except the one of the shape~\cite{Henning:2019enq, Henning:2019mcv},

\begin{equation}
    \begin{tikzpicture}
\filldraw [draw = black, fill = cyan] (10pt,10pt) rectangle (22pt,22pt);
\filldraw [draw = black, fill = cyan] (10pt,22pt) rectangle (22pt,34pt);
\draw [densely dotted] (24pt,22pt) -- (32pt,22pt);
\filldraw [draw = black, fill = cyan] (10pt,46pt) rectangle (22pt,58pt);
\filldraw [draw = black, fill = cyan] (10pt,58pt) rectangle (22pt,70pt);
\draw [densely dotted] (24pt,58pt) -- (32pt,58pt);
\draw [densely dotted] (16pt,36pt) -- (16pt,46pt);
\filldraw [draw = black, fill = cyan] (34pt,10pt) rectangle (46pt,22pt);
\filldraw [draw = black, fill = cyan] (34pt,22pt) rectangle (46pt,34pt);
\filldraw [draw = black, fill = cyan] (34pt,46pt) rectangle (46pt,58pt);
\filldraw [draw = black, fill = cyan] (34pt,58pt) rectangle (46pt,70pt);
\draw [densely dotted] (40pt,36pt) -- (40pt,46pt);
\draw [|<-] (0pt,10pt)--(0pt,34pt);
\draw [|<-] (0pt,70pt)--(0pt,46pt);
\node (n2) at (-5pt,40pt) {\small $N-2$};

\draw [|<-] (10pt,80pt)--(22pt,80pt);
\draw [|<-] (46pt,80pt)--(34pt,80pt);
\node (nt) at (28pt,80pt) {\small $\tilde{n}$};


\draw (46pt,58pt) rectangle (58pt,70pt);
\draw (46pt,46pt) rectangle (58pt,58pt);
\draw [densely dotted] (60pt,58pt) -- (70pt,58pt);
\draw (70pt,58pt) rectangle (82pt,70pt);
\draw (70pt,46pt) rectangle (82pt,58pt);
\draw [|<-] (46pt,36pt)--(58pt,36pt);
\draw [|<-] (82pt,36pt)--(70pt,36pt);
\node (n) at (64pt,36pt) {\small $n$};
    \end{tikzpicture}\,,
\end{equation}
which is called the primary Young diagram. According to the group theory, all the semi-standard Young tableaux (SSYTs) compose a basis of this representation, thus we can construct the independent and complete Lorentz structures by finding all the SSYTs of the primary Young diagram, by which the redundancies due to the Schouten identities are also removed. The numbers that can be filled range from $1$ to $N$, and the repetition of each of them is determined by the operator type
\begin{equation}
\label{eq:numi}
    \# i = \sum_{h_j>0}h_j -2h_i + \frac{k}{2}\,.   
\end{equation}
If the external sources are included, the SSYTs are not adequate. As discussed in Ref.~\cite{Ren:2022tvi}, we modify the correspondence relations in Eq.~\eqref{eq:corres} to 
\begin{align}
    \phi & \sim 1 \notag\\
    \psi & \sim \lambda_{0}\,, \quad \psi^\dagger \sim \tilde{\lambda}_{0} \notag \\
    F_L & \sim \lambda_{0}\lambda_{0}\,, \quad F_R \sim \tilde{\lambda}_{0}\tilde{\lambda}_{0} \notag \\
    p & \sim \lambda_{d}\tilde{\lambda}_{d}\,,
\end{align}
where each spinor is labeled by $0$ or $d$ distinguishing the fields and the derivatives. For each operator type, we can use the new correspondence relations to obtain all the spinor contractions, and then reduce them to a minimal set, which reserves the EOMs of the external sources. 

The Lorentz basis obtained from the SSYTs or the reduction with the external sources is called the y-basis, which are the products of brackets and generally are not monomials when they are recovered to the fields. However, we can transform the y-basis to another one in which all the Lorentz bases are monomials, called the m-basis. 

When applying to the ChEFT, the pions $u_\mu$ need extra management in the Young tensor method~\cite{Sun:2022ssa,Sun:2022aag,Low:2022iim}. Because the pions satisfy the Adler's zero condition~\cite{Adler:1969gk}, the amplitudes with pions vanish when the momenta of the pions go to zero, which presents new constraints on the Lorentz y-basis. Suppose a general amplitude $\mathcal{B}(p)$ satisfying the Adler's zero condition, $\lim_{p\rightarrow 0} \mathcal{B}(p)=0$, it can be expanded on a set of bases $\{\mathcal{B}_1,\mathcal{B}_2,\dots \mathcal{B}_n\}$ as
\begin{equation}
    \mathcal{B} = \sum_{i=1}^n c_i \mathcal{B}_i\,,
\end{equation}
with $c_i$ the expansion coefficients. The limit implies that
\begin{equation}
    \lim_{p\rightarrow 0} \mathcal{B}(p) = \sum_{i=1}^nc_i\lim_{p\rightarrow 0} \mathcal{B}_i = \sum_{i=1}^nc_i \mathcal{B}'_i \,,
\end{equation}
where the resultant amplitudes $B'_i$ are generally not independent and can be expanded again on the basis $\{\mathcal{B}_1,\mathcal{B}_2,\dots \mathcal{B}_n\}$ as
$\mathcal{B}'_i = \sum_j \mathcal{K}_{ij} \mathcal{B}_j$. Thus the limit becomes
\begin{equation}
    0 = \lim_{p\rightarrow0} \mathcal{B} = \sum_{i,j=1}^n c_i \mathcal{K}_{ij} \mathcal{B}_j \,,
\end{equation}
which means all the coefficients are zero, thus we obtain a system of linear equations
\begin{equation}
\label{eq:linearequations}
    \sum_{i,j=1}^n c_i \mathcal{K}_{ij}=0 \,,\quad j=1,2,\dots,n\,.
\end{equation}
The solution space is the amplitude satisfying the Adler's zero condition, which contributes to the effective operators with the pions. 

Next, we consider two typical examples. The first one is the type $N^2u^2$, which is composed of the operators of $N_LN_{Rc}u^2$ and its hermitian conjugation. In the Young tensor method, the type $N_LN_{Rc}u^2$ is equivalent to the one $N_L N_{Rc}\phi^2 D^2$ with the constraints that the two pions $\phi$ satisfy the Adler's zero condition. The SSYTs of $N_L N_{Rc}\phi^2 D^2$ are simple,
\begin{align}
    \mathcal{B}_1 &= \begin{ytableau}
*(cyan)1 & 1 & 3 \\
*(cyan)2 & 2 & 4 
    \end{ytableau} \sim \lra{12}\lra{34}[34]\,, \notag \\
    \mathcal{B}_2 &= \begin{ytableau}
*(cyan)1 & 1 & 2 \\
*(cyan)2 & 3 & 4 
    \end{ytableau} \sim \lra{13}\lra{24}[34] \,,
\end{align}
and both of them satisfy the Adler's zero condition, which means the system of the linear equations in Eq.~\eqref{eq:linearequations} is trivial. The corresponding m-basis can be taken as
\begin{align}
    \mathcal{B}_1^m &= (N_L N_{Rc})D^\mu\phi D_\mu\phi\,,\notag \\
    \mathcal{B}_2^m &= (N_L \sigma^{\mu\nu}N_{Rc}) D_\mu \phi D_\nu\phi \,.\label{eq:example1}
\end{align}
If there exist tensor sources $f_\pm$, which means we are regarding the type $N^2 u^2 f_\pm$, the Lorentz structures become complicated. Considering the subtype $N_L N_{Rc} \phi^2 f_\pm{}_L D^2$, 
the involving fields are $\{f_\pm,N_L,N_{Rc},\phi,\phi\}$, and there are 2 derivatives. The primary Young diagram is of the shape that
\begin{equation}
    \begin{ytableau}
*(cyan)  & & & & \\
*(cyan)  & & & & \\
*(cyan) 
    \end{ytableau}\,,
\end{equation}
and the numbers to fill are determined by Eq.~\eqref{eq:numi},
\begin{align}
    \#1 &= 3\,,\notag \\
    \#2 &= \# 3 = 2\,. \notag \\
    \#4 &- \# 5 = 1\,.
\end{align}
There are 7 SSYTs in this type
\begin{align}
    \mathcal{B}_1 &= \begin{ytableau}
*(cyan)1 & 1 & 1 & 4 \\
*(cyan)2 & 2 & 4 & 5 \\
*(cyan)3
    \end{ytableau}\sim -s_{45} \langle 12\rangle  \langle 13\rangle\,,\notag \\
    \mathcal{B}_2 &= \begin{ytableau}
*(cyan)1 & 1 & 1 & 3 \\
*(cyan)2 & 2 & 4 & 5 \\
*(cyan)3 
    \end{ytableau}\sim [45]\langle 12\rangle  \langle 14\rangle  \langle 35\rangle\,, \notag \\
    \mathcal{B}_3 &= \begin{ytableau}
*(cyan)1 & 1 & 1 & 3 \\
*(cyan)2 & 2 & 3 & 5 \\
*(cyan)4
    \end{ytableau}\sim s_{35} \langle 12\rangle  \langle 13\rangle\,,\notag \\
    \mathcal{B}_4 &= \begin{ytableau}
*(cyan)1 & 1 & 1 & 3 \\
*(cyan)2 & 2 & 3 & 4 \\
*(cyan)5
    \end{ytableau}\sim -s_{34} \langle 12\rangle  \langle 13\rangle \,,\notag 
\end{align}
\begin{align}
    \mathcal{B}_5 &= \begin{ytableau}
*(cyan)1 & 1 & 1 & 2 \\
*(cyan)2 & 3 & 3 & 5 \\
*(cyan)4
    \end{ytableau}\sim -[35] \langle 13\rangle ^2 \langle 25\rangle \,,\notag \\
    \mathcal{B}_6 &= \begin{ytableau}
*(cyan)1 & 1 & 1 & 2 \\
*(cyan)2 & 3 & 3 & 4 \\
*(cyan)5
    \end{ytableau}\sim [34] \langle 13\rangle ^2 \langle 24\rangle \,,\notag \\
    \mathcal{B}_7 &= \begin{ytableau}
*(cyan)1 & 1 & 1 & 2 \\
*(cyan)2 & 3 & 4 & 5 \\
*(cyan)3
    \end{ytableau}\sim [45] \langle 13\rangle  \langle 14\rangle  \langle 25\rangle \,, \label{eq:examlor}
\end{align}
thus we obtain the 7-dimension Lorentz y-basis $\{\mathcal{B}_1,\mathcal{B}_2,\dots \mathcal{B}_7\}$ of this type. Because the field $\phi$ is pion, not all the 7 y-bases satisfy the Adler's zero condition. Take the limit $p_\phi \rightarrow 0$, then
\begin{align}
    \lim_{p_\phi\rightarrow 0} \mathcal{B}_{1,2,7} &= 0 \,,\notag \\
    \lim_{p_\phi\rightarrow 0} \mathcal{B}_{3,4,5,6} &= \mathcal{B}_{3,4,5,6} \,,
\end{align}
thus the matrix $\mathcal{K}$ in Eq.~\ref{eq:linearequations} takes the form
\begin{equation}
    \mathcal{K} = \left(
\begin{array}{ccccccc}
 0 & 0 & 0 & 0 & 0 & 0 & 0 \\
 0 & 0 & 0 & 0 & 0 & 0 & 0 \\
 0 & 0 & 1 & 0 & 0 & 0 & 0 \\
 0 & 0 & 0 & 1 & 0 & 0 & 0 \\
 0 & 0 & 0 & 0 & 1 & 0 & 0 \\
 0 & 0 & 0 & 0 & 0 & 1 & 0 \\
 0 & 0 & 0 & 0 & 0 & 0 & 0 \\
\end{array}
\right)\,,
\end{equation}
and the system of linear equations is
\begin{equation}
    c_i = 0\,,\quad i=3,4,5,6\,.
\end{equation}
Thus the solution space is spanned by
\begin{align}
    \mathcal{B}'_1 = \mathcal{B}_1 &= -s_{45} \langle 12\rangle  \langle 13\rangle \,, \notag \\
    \mathcal{B}'_2 = \mathcal{B}_2 &= [45]\langle 12\rangle  \langle 14\rangle  \langle 35\rangle \,, \notag \\
    \mathcal{B}'_3 = \mathcal{B}_7 &= [45] \langle 13\rangle  \langle 14\rangle  \langle 25\rangle \,.
\end{align}

Recovering the spinor form of the y-basis to the field form, we need to apply the correspondence relation in Eq.~\eqref{eq:corres} conversely and utilize the relation
\begin{equation}
    {F_L}_{\alpha\beta} = \frac{i}{2}{F_L}_{\mu\nu}\sigma^{\mu\nu}_{\alpha\beta}\,,\quad {F_R}_{\dot{\alpha}\dot{\beta}} = -\frac{i}{2}{F_R}_{\mu\nu}\bar{\sigma}^{\mu\nu}_{\dot{\alpha}\dot{\beta}}\,,\quad D_{\alpha\dot{\alpha}} = D_\mu\sigma^\mu_{\alpha\dot{\alpha}}\,.
\end{equation}
Then we can obtain the m-basis that
\begin{align}
    \mathcal{B}^m_1 &= (N_L N_{Rc})f_\pm{}_L{}_{\mu\nu} D^\mu\phi D^\nu \phi\,, \notag \\
    \mathcal{B}^m_2 &= (N_L\sigma^{\nu\lambda}N_{Rc})f_\pm{}_L{}_{\nu\lambda} D^\mu\phi D_{\mu}\phi \,, \notag \\
    \mathcal{B}^m_3 &= (N_L\sigma^{\nu\lambda}N_{Rc})f_\pm{}_L{}_{\lambda\mu}D^\mu\phi D_\nu\phi \,, \label{eq:example2}
\end{align}
and the corresponding transforming matrix is 
\begin{equation}
    \mathcal{K}^{my}=\left(
\begin{array}{ccc}
0 & \frac{1}{4} & -\frac{1}{4} \\
i & 0 & 0 \\
0 & \frac{i}{4} & \frac{i}{4}
\end{array}
\right)\,.
\end{equation}

To summarize, the amplitude-operator correspondence gives an efficient method to construct the independent and complete Lorentz structures. In the language of the spinor, the redundancies such as the Schouten identity and the IBP can be eliminated via the SSYTs of the primary Young diagram, of which the shape and the numbers to fill are completely determined by the field contents.

In particular, the fermions used in the Young tensor method are the Weyl spinors, which are related by the Dirac spinors as
\begin{equation}
    N = \left(\begin{array}{c}
        {N_{L}}_\alpha\\
        {N^\dagger_{Rc}}^{\dot{\beta}}
    \end{array}\right)\,,
\end{equation}
where both $N_L$ and $N_{Rc}$ are left-handed spinors. As a consequence, we need to combine the effective operators composed of the Weyl spinors to form the $P$-even operators. For example, the m-basis obtained in Eq.~\eqref{eq:example1} and their hermitian conjugation can be combined as follows
\begin{align}
    \mathcal{B}_1^m + \mathcal{B}_1^m{}^\dagger &= (\overline{N}N)u^\mu u_\mu \,,\notag \\
    \mathcal{B}_2^m + \mathcal{B}_2^m{}^\dagger &= (\overline{N}\sigma^{\mu\nu} N)u_\mu u_\nu \,,
\end{align}
which are $P$-even operators, and we have replaced $D_\mu\phi$ by $u_\mu$.

\subsection{Internal Structures}

A complete operator takes the form that
\begin{equation}
    \mathcal{O} = \mathcal{T}\times \mathcal{A}\,,
\end{equation}
where $\mathcal{A}$ is the Lorentz structures discussed in the previous subsection, and $\mathcal{T}$ is the internal structures,  which is the chiral symmetry $SU(2)_V$ for the ChEFT. We have discussed that by the Young diagram technique, the independent Lorentz structures can be obtained, and in this section, we will show that the internal group structures can be obtained by manipulating the Young diagrams as well.

According to the group theory, each irreducible representation of the $SU(N)$ group is labeled by a Young diagram, and the corresponding SSYTs form a set of bases. For example, the first several irreducible representations of the $SU(2)$ and $SU(3)$ groups are listed in Tab.~\ref{tab:youngdiagrams}. All the fields of the irreducible representations should form the trivial representation $\mathbf{1}$ by the tensor product. In the language of the Young diagram, such a tensor product can be accomplished by the outer product according to the Littlewood-Richardson rule (LR rule).
\begin{table}
\renewcommand{\arraystretch}{1.5}
\ytableausetup
{boxsize=1em}
\ytableausetup
{aligntableaux=center}
    \centering
    \begin{tabular}{|c|c|c|c|c|c|c|c|c|c|}
    \hline
\multirow{2}{*}{$SU(2)$} & $\mathbf{1}$ & $\mathbf{2}$ & $\overline{\mathbf{2}}$ & $\mathbf{3}$ & \multirow{2}{*}{$SU(3)$} & $\mathbf{1}$ & $\mathbf{3}$ & $\overline{\mathbf{3}}$ & $\mathbf{8}$ \\
\cline{2-5} \cline{7-10}
                         & \ydiagram{1,1} & \ydiagram{1} & \ydiagram{1} & \ydiagram{2} & & \ydiagram{1,1,1} & \ydiagram{1} & \ydiagram{1,1} & \ydiagram{2,1} \\
    \hline
    \end{tabular}
    \caption{The corresponding Young diagrams of the first several irreducible representations of the $SU(2)$ and $SU(3)$ groups.}
    \label{tab:youngdiagrams}
\end{table}

Next, we will give two examples of the $SU(2)$ and $SU(3)$ symmetries respectively. Suppose the type $N^2u^2f_\pm$ is of the $SU(2)$ case, which means all the fermions are of the $SU(2)$ fundamental representation and all the bosons are of the $SU(2)$ adjoint representation.
\begin{align}
    &f^K_\pm {}_L{\tau^K}^t_r\epsilon_{ts} \in \mathbf{3} \sim \young(rs)\,, \notag \\
    &{N_L}_i \in \mathbf{2} \sim \young(i)\,,\notag \\
    &{N_{Rc}}_j \in \mathbf{2} \sim \young(j)\,, \notag \\
    &{\phi^I}{\tau^I}^p_k\epsilon_{pl} \in \mathbf{3} \sim \young(kl)\,, \notag \\
    &{\phi^J}{\tau^J}^q_m\epsilon_{qn} \in \mathbf{3} \sim \young(mn)\,,
\end{align}
the $SU(2)$ singlet representation corresponds to the Yong diagram
\begin{equation}
    \yng(4,4)\,,
\end{equation}
in this type, and it can be constructed by the Young diagram outer product. They are
\begin{align}
    & \young(i) \times \young(j) \times \young(kl) \times \young (mn) \times \young(rs) \notag \\
    &= \left(\young(ij)+\young(i,j)\right) \times \young(kl) \times \young (mn) \times \young(rs) \notag \\
    &= \left(\young(ijkl)+\young(ij,kl)+\young(ijk,l)+\young(ikl,j)\right)\times \young (mn) \times \young(rs) \notag \\
    &= \young(ijmn,klrs) + \young(ijkl,mnrs)+\young(ijkm,lnrs)+\young(iklm,jnrs) \notag \\
    &= \epsilon^{ik}\epsilon^{jl}\epsilon^{mr}\epsilon^{ns} + \epsilon^{im}\epsilon^{jn}\epsilon^{kr}\epsilon^{ls} + \epsilon^{il}\epsilon^{jn}\epsilon^{kr}\epsilon^{ms} + \epsilon^{ij}\epsilon^{kn}\epsilon^{lr}\epsilon^{ms} \notag \\
    &\rightarrow \calt_1 + \calt_2 + \calt_3 + \calt_4\,,\label{eq:examgaug}
\end{align}
thus there are 4 independent $SU(2)$ structures in this type, all of which are expressed as the product of the anti-symmetric tensors, and the tensors $\calt^{1\sim 4}$ are the m-basis,
\begin{equation}
\label{eq:tensor2}
    \calt_1 = \delta_i^j \epsilon^{IJK}\,, \quad \calt_2 = \delta^{IJ}\tau^K{}_i^j\,,\quad \calt_3=\delta^{IK}\tau^J{}_i^j\,,\quad \calt_4=\delta^{JK}\tau^I{}_i^j\,.
\end{equation} 
The complete effective operators are the tensor products of the Lorentz bases in Eq.~\eqref{eq:example2} and the tensor bases above, which means the 3-dimension Lorentz basis and the 4-dimension $SU(2)$ internal basis give 12 independent effective operators.
They are usually called the operator m-basis or just m-basis.

For the $SU(3)$ case, we suppose the type $N^2u^2$ is of the $SU(3)$ case, thus all the fields (both bosons and fermions) are of the $SU(3)$ adjoint representation,
\begin{equation}
    \Psi_x^a \lambda^a{}_{i_x}^{l_x}\epsilon_{l_xj_xk_x} \in \mathbf{8} \sim \begin{ytableau}
        i_x & j_x \\ k_x
    \end{ytableau}\,,\quad x=1,2,3,4\,,
\end{equation}
where $\{\Psi_1,\Psi_2,\Psi_3,\Psi_4\} = \{\overline{N},N,u,u\}$. The $SU(3)$ singlet representation of this type corresponds to the Young diagram that
\begin{equation}
    \yng(4,4,4)\,,
\end{equation}
and eight Young tableaux emerge when doing the outer product, which correspond to 8 tensors,
\begin{align}
  & \calt_1 = d^{abe}d^{cde} \,,\quad \calt_2 = d^{abe}f^{cde} \,,\quad \calt_3 = f^{abe}f^{cde} \,,\quad \calt_3 = \delta^{ab}\delta^{cd}\,,\notag \\
  & \calt_5 = f^{abe}d^{cde} \,,\quad \calt_6 = \delta^{ac}\delta^{bd} \,,\quad \calt_7 = d^{ace}d^{bde} \,,\quad \calt_8 = d^{ace}f^{bde} \,.\label{eq:su3tensor}
\end{align}

In summary, since every field is of an irreducible representation of $SU(N)$, corresponding to a unique Young diagram, the invariants they can form are the trivial representation in the decomposition of their tensor product. In terms of the Young diagrams, such decomposition can be done directly by the outer product according to the LR rule. With the Lorentz and the internal basis obtained, the operator m-basis is just the product of them. Although such operators are all independent, it is usually the case not all of them are physical since we have not considered the repeated fields in the flavor space up to now, thus such operators are also called the flavor-blind operators.

\subsection{Repeated Fields}

In the construction of the high-dimension operators, it is usually the case there are repeated fields. If the repeated $N$ fields have flavor number 1, they should be symmetric under the permutations. If the repeated fields are of flavor number $n_f$, the flavor structures should be the irreducible representations of the $SU(n_f)$ group, which at the same time are the irreducible representations of the symmetric group $S_N$. 

It is difficult to identify the $S_N$ irreducible representation from the m-basis, thus we adopt the Young operators to project the flavor-blind operators to the ones of specific permutation symmetries.
Considering the example of type  $N^2u^2f_\pm$ again, there are $3\times 4=12$ flavor-blind operators as discussed before. 
Because the two pions $\phi$ are repeated, they should be symmetric, since they are bosons. To obtain such operators, we need to construct the corresponding Young operator
\begin{equation}
    \mathcal{Y}[\young(45)] = \frac{1}{2}(1+\sigma_{45})\label{eq:young}\,,
\end{equation}
then apply on the 12-dimension flavor-blind operator basis. The representative of the group element $\sigma_{34}\in S_2$ is the tensor product of its representatives in the Lorentz and the internal basis respectively, 
\begin{equation}
    \sigma_{45} = \sigma^L_{45}\otimes \sigma^{SU(2)}_{45}\,,
\end{equation}
both of which are not difficult to get, since the permutations on the SSYTs are straightforward. These representatives are
\begin{equation}
    \sigma^L_{45}= \left(
\begin{array}{ccc}
 -1 & 0 & 0 \\
 0 & 1 & 0 \\
 0 & \frac{1}{2} & -1 \\
\end{array}
\right) \,,\quad 
\sigma^{SU(2)}_{45}= \left(
\begin{array}{cccc}
 -1 & 0 & 0 & 0 \\
 0 & 0 & 1 & 0 \\
 0 & 1 & 0 & 0 \\
 0 & 0 & 0 & 1 \\
\end{array}
\right)\,,
\end{equation}
respectively. 

If we require the $SU(2)$ group tensor to be symmetric, the corresponding Young operator is 
\begin{equation}
\label{eq:youngoperator1}
    \mathcal{Y}^{SU(2)}[\young(45)] = 
    \left(
\begin{array}{cccc}
 0 & 0 & 0 & 0 \\
 0 & \frac{1}{2} & \frac{1}{2} & 0 \\
 0 & \frac{1}{2} & \frac{1}{2} & 0 \\
 0 & 0 & 0 & 1 \\
\end{array}
\right)\,,
\end{equation}
which is a rank-2 matrix and implies there are only two symmetric tensors among the four. According to Eq.~\eqref{eq:tensor2}, these tensors are 
\begin{equation}
\label{eq:tensor1}
    \frac{1}{2}(\calt_2 + \calt_3) =  \frac{1}{2}(\delta^{IJ}\tau^K{}_i^j +\delta^{IK}\tau^J{}_i^j ) \,, \quad \calt_4 = \delta^{JK}\tau^I{}^i_j\,.
\end{equation}
At this stage, we can illustrate the equivalence between the Young tensor method and the Cayley-Hamilton relation for the $SU(2)$ structures. For the type $N^2u^2f_\pm$, the fields $u,f_\pm$ are of the adjoint representation, while the nucleon $N$ is of the fundamental representation. However, we can take the tensor product of the two nucleons to form a $2\times 2$ matrix, 
\begin{equation}
    \overline{N}^i \otimes N_j \rightarrow A^i_j\,,
\end{equation}
which, in particular, is not traceless, $\lra{A}\neq 0$. Thus we assign
\begin{equation}
    \overline{N} \otimes N \rightarrow A\,, \quad u\rightarrow B\,,\quad f_\pm \rightarrow C\,,
\end{equation}
and find all the distinct traces
\begin{equation}
    \lra{A}\lra{BBC}\,,\quad \lra{AB}\lra{BC} \quad\,, \lra{AC}\lra{BB}\,, \quad \lra{ABBC}\,, \quad \lra{ACBB}\,, \quad \lra{ABCB}\,.
\end{equation}
At the same time, the Cayley-Hamilton theorem discussed in Sec.~\ref{sec:ccwz} presents 4 independent relations,
\begin{align}
   & \lra{A}\lra{BBC} = 0 \,, \\
   & \lra{ABBC} + \lra{ABCB} = \lra{AB}\lra{BC} \,, \\
   & \lra{ABCB} + \lra{ACBB} = \lra{AB}\lra{BC} \,, \\
   & \lra{ACBB} + \lra{ABBC} = \lra{AC}\lra{BB} \,,
\end{align}
thus there are only 2 independent traces, which can be chosen as
\begin{equation}
    T_1 = \lra{AB}\lra{BC}\,, \quad T_2 = \lra{AC}\lra{BB}\,.
\end{equation}
This is consistent with the Young tensor result in Eq.~\eqref{eq:tensor1}. Actually, they can be related to each other explicitly by the identification
\begin{equation}
\label{eq:iden1}
    A = (\overline{N}N) \mathbb{I}_{2\times 2} + (\overline{N}\tau^L N) \tau^L{}_i^j\,, \quad B = u_\mu^J \tau^J \,, \quad f_\pm = f_\pm^I\tau^I\,.
\end{equation}
Besides the resultant tensors and traces, the constraints of the Cayley-Hamilton relations and the Young tensor method can be related as well, and the detailed discussion can be found in Ref.~\cite{Song:2024fae}.

Furthermore, the permutation symmetry of the repeated fields is determined by the permutations of the Lorentz and the internal structures simultaneously. Thus we need to take their tensor product, and substitute it in the expression of the Young operator in Eq.~\eqref{eq:young}, we can get
\begin{equation}
    \mathcal{Y}[\young(45)] =\left(
\begin{array}{cccccccccccc}
 1 & 0 & 0 & 0 & 0 & 0 & 0 & 0 & 0 & 0 & 0 & 0 \\
 0 & \frac{1}{2} & -\frac{1}{2} & 0 & 0 & 0 & 0 & 0 & 0 & 0 & 0 & 0 \\
 0 & -\frac{1}{2} & \frac{1}{2} & 0 & 0 & 0 & 0 & 0 & 0 & 0 & 0 & 0 \\
 0 & 0 & 0 & 0 & 0 & 0 & 0 & 0 & 0 & 0 & 0 & 0 \\
 0 & 0 & 0 & 0 & 0 & 0 & 0 & 0 & 0 & 0 & 0 & 0 \\
 0 & 0 & 0 & 0 & 0 & \frac{1}{2} & \frac{1}{2} & 0 & 0 & 0 & 0 & 0 \\
 0 & 0 & 0 & 0 & 0 & \frac{1}{2} & \frac{1}{2} & 0 & 0 & 0 & 0 & 0 \\
 0 & 0 & 0 & 0 & 0 & 0 & 0 & 1 & 0 & 0 & 0 & 0 \\
 0 & 0 & 0 & 0 & -\frac{1}{4} & 0 & 0 & 0 & 1 & 0 & 0 & 0 \\
 0 & 0 & 0 & 0 & 0 & 0 & \frac{1}{4} & 0 & 0 & \frac{1}{2} & -\frac{1}{2} & 0 \\
 0 & 0 & 0 & 0 & 0 & \frac{1}{4} & 0 & 0 & 0 & -\frac{1}{2} & \frac{1}{2} & 0 \\
 0 & 0 & 0 & 0 & 0 & 0 & 0 & \frac{1}{4} & 0 & 0 & 0 & 0 \\
\end{array}
\right)\,,
\end{equation}
which is a dimension-12 but rank-6 matrix. Applying it to the flavor-blind operators, we can see that there are only 6 independent operators, and they are
\begin{align}
    & (N_L u_\nu N_{Rc})\lra{f_\pm{}_L^{\mu\nu} u_\mu } \notag \\
    & (N_L \sigma^{\nu\lambda} u^\mu N_{Rc})\lra{f_\pm{}_L{}_{\nu\lambda} u_\mu} \notag \\
    & (N_L \sigma^{\lambda\nu} u_\nu N_{Rc})\lra{f_\pm{}_L{}_{\lambda\mu} u^\mu} \notag \\
    & (N_L \sigma^{\nu\lambda}f_\pm{}_L{}_{\nu\lambda} N_{Rc}) \lra{u_\mu u^\mu} \notag \\
    & (N_L N_{Rc}) \lra{f_\pm{}_L{}^{\mu\nu}[u_\mu,u_\nu]} \notag \\
    & (N_L \sigma^{\lambda\nu} N_{Rc})\lra{f_\pm{}_L{}_{\lambda\mu}[u^\mu,u_\nu]} \,,
\end{align}
where all the $SU(2)$ indices contracted within the fermion bilinears, $\lra{...}$ denotes the $SU(2)$ trace, and the $D_\mu\phi$ has been replaced by $u_\mu$. 

In the ChEFT Lagrangian, the flavor indices of both the fermions and bosons are contracted to form the $SU(2)_V$ invariant. Consequently, all the operators with repeated fields should be symmetrized.

\newcommand{\tra}[1]{\langle#1 \rangle}

\section{Pure Meson Sector}
\label{sec:operators}

The interactions of the pure meson sector $\mathcal{L}_{\pi}$ are organized by the Weinberg power-counting formula in Eq.~\eqref{eq:nuw}, which can be simplified by setting $L=A=n_i = 0$ to,
\begin{equation}
    \nu = d_i\,,
\end{equation}
where $d_i$ is the power of the small momentum $p$.
When $d_i=2$, the interactions are of the leading order $p^2$, or $\nu = 2$, in Eq.~\eqref{eq:leadboson}. 
In this section, we construct the complete and independent $C\,, P$-even interactions containing the pions up to $p^8$ in the $\mathcal{L}_\pi$ sector. In particular, only the effective operators involving 4 or more $u_\mu$ at the $p^8$ order are considered. The complete operators up to $p^8$ can be found in Ref.~\cite{Li:2024ghg}.

The Hilbert series gives the result of the $C\,, P$-even operators of the wanted types that
\begin{align}
    \mathcal{H}_{\text{C-even}}^{\text{P-even}} &= fu^2 + 2u^4 + u^2\chi & \mathit{\nu=4} \notag \\
    & + 2Df^2u + D^2fu + 10f^2 u^2 + 3Dfu^3 + 2D^2u^4 + 2fu^4 + 3u^6 & \mathit{\nu=6}\notag \\
    & + 2Dfu\chi + D^2 u^2 \chi + 4fu^2\chi + 2Du^3\chi + 2u^4\chi + 2Du\chi^2 + 6u^2\chi^2 \notag \\
    & + 3D^4u^4 + 14D^2fu^4 + 28 f^2u^4 + 12 Dfu^5 + 9D^2u^6 + 4fu^6 + 7 D^2 u^4\chi & \mathit{\nu=8} \notag \\
    &+ 10fu^4 + 7Du^5 \chi + 3 u^6\chi + 12 u^4 \chi^2 \,,
\end{align}
where the chiral dimensions are presented on the right.
Next, we present the effective operators according to their dimensions and types.

\begin{center}
    \large \textbf 1. $\mathcal{O}(p^4)$
\end{center}

$ {f}u^2$:
\begin{align}
    & \tra{{f_+}^{\mu\nu}[u_\mu,u_\nu]}
\end{align}

$u^4$:
\begin{align}
    & \tra{u_\mu u^\mu}\tra{u_\nu u^\nu} & \tra{u_\mu u_\nu}\tra{u^\mu u^\nu} 
\end{align}

$u^2\chi$:
\begin{align}
    & \tra{u_\mu u^\mu}\tra{\chi_+}
\end{align}

\begin{center}
    \large \textbf 2. $\mathcal{O}(p^6)$
\end{center}

$Df^2u$:
\begin{align}
    & \tra{u^\mu [D_\mu {f_+}_{\nu\lambda},{f_-}^{\nu\lambda}]} & \tra{u_\mu [D^\nu{f_+}_{\nu\lambda},{f_-}^{\mu\lambda}]} 
\end{align}

$f^2u^2$:
\begin{align}
    & \tra{{f_-}{}^{\nu\lambda}u^\mu}\tra{{f_-}{}_{\nu\lambda}u_\mu} & \tra{{f_-}{}^{\mu\lambda}u_\mu}\tra{{f_-}{}_{\nu\lambda}u^\nu} \notag \\
    & \tra{{f_-}{}_{\nu\lambda}{f_-}{}^{\nu\lambda}}\tra{u_\mu u^\mu}  & \tra{{f_-}{}^{\mu\lambda}u_\nu}\tra{{f_-}{}_{\nu\lambda}u^\mu} \notag \\
    & \tra{{f_-}{}^{\mu\lambda}{f_-}{}_{\lambda\nu}}\tra{u_\mu u^\nu} & \tra{{f_+}{}^{\mu\lambda}{f_+}{}_{\lambda\nu}}\tra{u_\mu u^\nu} \notag \\
    & \tra{{f_+}{}^{\nu\lambda}u^\mu}\tra{{f_+}{}_{\nu\lambda}u_\mu} & \tra{{f_+}{}^{\mu\lambda}u_\mu}\tra{{f_+}{}_{\nu\lambda}u^\nu} \notag \\
    & \tra{{f_+}{}_{\nu\lambda}{f_+}{}^{\nu\lambda}}\tra{u_\mu u^\mu}  & \tra{{f_+}{}^{\mu\lambda}u_\nu}\tra{{f_+}{}_{\nu\lambda}u^\mu} 
\end{align}

$D^2fu^2$:
\begin{align}
    \tra{D^\mu {f_+}^{\lambda\nu}[D_\lambda u_\mu, u_\nu]}
\end{align}

$Dfu^3$:
\begin{align}
    & \tra{D^\mu {f_-}^{\lambda\nu} u_\nu}\tra{u_\lambda u_\mu} & \tra{D^\mu {f_-}_{\mu\nu}u_\lambda}\tra{u^\lambda u^\nu} \notag \\
    & \tra{{f_-}_{\mu\nu} D^\mu u^\lambda}\tra{u_\lambda u^\nu}
\end{align}

$D^2u^4$:
\begin{align}
    & \tra{u_\nu D_\mu D^\lambda u^\nu}\tra{u_\lambda u^\mu} & \tra{D_\lambda u_\mu u_\nu}\tra{D^\lambda u^\mu u^\nu}
\end{align}

$fu^4$:
\begin{align}
    & \tra{{f_+}^{\lambda\nu}[u_\lambda,u_\nu]}\tra{u^\mu u_\mu} & \tra{{f_+}^{\mu\nu}[u_\nu,u_\lambda]}\tra{u_\mu u^\lambda}
\end{align}

$u^6$:
\begin{align}
    & \tra{u_\mu u^\nu}\tra{u_\lambda u^\mu}\tra{u_\nu u^\lambda} & \tra{u^\mu u_\mu}\tra{u_\lambda u_\nu}\tra{u^\lambda u^\nu} \notag \\
    & \tra{u^\mu u_\mu}\tra{u^\nu u_\nu}\tra{u^\lambda u_\lambda} 
\end{align}

$Dfu\chi$:
\begin{align}
    & \tra{D_\mu {f_+}^{\mu\nu}[u_\nu,\chi_-]} & \tra{D_\mu {f_-}^{\mu\nu} u_\nu}\tra{\chi_+}
\end{align}

$fu^2\chi$:
\begin{align}
    & \tra{{f_-}^{\mu\nu} u_\mu}\tra{\chi_- u_\nu} & \tra{{\tilde{f}_-}^{\mu\nu} u_\mu}\tra{\chi_+ u_\nu} \notag \\
    & \tra{{\tilde{f}_+}^{\mu\nu}[u_\mu,u_\nu]}\tra{\chi_-} & \tra{{f_+}^{\mu\nu}[u_\mu,u_\nu]}\tra{\chi_+}
\end{align}

$Du^3\chi$:
\begin{align}
    & \tra{u_\nu D_\mu u^\nu}\tra{u^\mu \chi_-} & \tra{u_\mu u_\nu}\tra{D^\mu u^\nu\chi_-}
\end{align}

$u^2\chi^2$:
\begin{align}
    & \tra{\chi_- u_\mu }\tra{\chi_- u^\mu} & \tra{u_\mu u^\mu}\tra{\chi_-\chi_-} \notag \\
    & \tra{\chi_+ u_\mu }\tra{\chi_+ u^\mu} & \tra{u_\mu u^\mu}\tra{\chi_+\chi_+} \notag \\
    & \tra{u_\mu u^\mu}\tra{\chi_+}\tra{\chi_+} & \tra{u_\mu u^\mu}\tra{\chi_-}\tra{\chi_-}
\end{align}

$Du\chi^2$:
\begin{align}
    & \tra{u_\mu D^\mu\chi_+}\tra{\chi_-} & \tra{u_\mu D^\mu\chi_-}\tra{\chi_+}  
\end{align}

$D^2 u^2\chi$:
\begin{align}
    & \tra{u^\mu u_\mu}D^2\tra{\chi_+} 
\end{align}

$u^4\chi$:
\begin{align}
    & \tra{u_\mu u^\mu}\tra{u_\nu u^\nu}\tra{\chi_+} & \tra{u_\mu u_\nu}\tra{u^\mu u^\nu}\tra{\chi_+} 
\end{align}

\begin{center}
    \large \textbf 3. $\mathcal{O}(p^8)$
\end{center}

$u^6\chi$:
\begin{align}
    & \tra{u_\mu u^\nu}\tra{u_\lambda u^\mu}\tra{u_\nu u^\lambda}\tra{\chi_+} & \tra{u^\mu u_\mu}\tra{u_\lambda u_\nu}\tra{u^\lambda u^\nu}\tra{\chi_+} \notag \\
    & \tra{u^\mu u_\mu}\tra{u^\nu u_\nu}\tra{u^\lambda u_\lambda}\tra{\chi_+} 
\end{align}

$D^2u^4\chi$:
\begin{align}
    & \tra{u_\nu D_\mu u_\rho}\tra{u_\lambda D_\eta u^\mu}\tra{\chi_-}\epsilon^{\eta\lambda\nu\rho} & \tra{u_\nu D_\mu D^\lambda u^\nu}\tra{u_\lambda u^\mu}\tra{\chi_+} \notag \\
    & \tra{D_\lambda u_\mu u_\nu}\tra{D^\lambda u^\mu u^\nu}\tra{\chi_+} & \tra{D_\lambda u_\mu u^\mu}\tra{D_\nu u^\lambda u^\nu}\tra{\chi_+} \notag \\
    & \tra{u_\nu D^\lambda u^\nu}\tra{u_\lambda u^\mu}D_\mu\tra{\chi_+} & \tra{u_\mu u_\nu}\tra{D^\lambda u^\mu u^\nu}D_\lambda\tra{\chi_+} \notag \\
    & \tra{u_\mu u^\mu}\tra{D_\nu u^\lambda u^\nu}D_\lambda\tra{\chi_+} 
\end{align}

$u^4\chi^2$:
\begin{align}
    & \tra{u_\mu u^\nu}\tra{u^\mu \chi_-}\tra{u_\nu \chi_-} & \tra{u_\mu u^\mu}\tra{u_\nu\chi_-}\tra{u^\nu \chi_-} \notag \\
    & \tra{u_\mu u_\nu}\tra{u^\mu u^\nu}\tra{\chi_-\chi_-} & \tra{u_\mu u^\mu}\tra{u_\nu u^\nu}\tra{\chi_-\chi_-} \notag \\
    & \tra{u_\mu u^\nu}\tra{u^\mu \chi_+}\tra{u_\nu \chi_+} & \tra{u_\mu u^\mu}\tra{u_\nu\chi_+}\tra{u^\nu \chi_+} \notag \\
    & \tra{u_\mu u_\nu}\tra{u^\mu u^\nu}\tra{\chi_+\chi_+} & \tra{u_\mu u^\mu}\tra{u_\nu u^\nu}\tra{\chi_+\chi_+} \notag \\
    & \tra{u_\mu u_\nu}\tra{u^\mu u^\nu}\tra{\chi_-}\tra{\chi_-} & \tra{u_\mu u^\mu}\tra{u_\nu u^\nu}\tra{\chi_-}\tra{\chi_-} \notag \\
    & \tra{u_\mu u_\nu}\tra{u^\mu u^\nu}\tra{\chi_+}\tra{\chi_+} & \tra{u_\mu u^\mu}\tra{u_\nu u^\nu}\tra{\chi_+}\tra{\chi_+}
\end{align}

$fu^6$:
\begin{align}
    & \tra{f_+^{\rho\lambda}u_\nu}\tra{u^\mu u^\nu}\tra{u_\rho[u_\mu,u_\lambda]} & \tra{f_+^{\rho\lambda}u^\nu}\tra{u^\mu u_\mu}\tra{u_\rho[u_\nu,u_\lambda]} \notag \\
    & \tra{u_\mu u^\mu}\tra{u_\nu u^\nu}\tra{f_+^{\rho\lambda}[u_\rho,u_\lambda]} & \tra{u_\mu u_\nu}\tra{u^\mu u^\nu}\tra{f_+^{\rho\lambda}[u_\rho,u_\lambda]}
\end{align}

$Du^5\chi$:
\begin{align}
    & \tra{D_\mu u^\lambda u^\nu}\tra{u_\lambda u^\mu}\tra{u_\nu \chi_-} & \tra{D_\mu u^\lambda u^\nu}\tra{u_\lambda u_\nu}\tra{u^\mu \chi_-} \notag \\
    & \tra{D_\mu u^\lambda u^\mu}\tra{u_\lambda u_\nu}\tra{u^\nu \chi_-} & \tra{u^\lambda D_\mu u_\lambda}\tra{u_\nu u^\nu}\tra{u^\mu \chi_-} \notag \\
    & \tra{u_\mu u^\nu}\tra{u_\lambda u^\mu}\tra{D_\nu u^\lambda \chi_-} & \tra{u_\mu u^\mu}\tra{u_\lambda u_\nu}\tra{D^\lambda u^\nu \chi_-} \notag \\
    & \tra{u_\lambda D_\eta u_\mu} \tra{u_\rho u^\mu}\tra{u_\nu \chi_+}\epsilon^{\eta\lambda\nu\rho} 
\end{align}

$fu^4\chi$:
\begin{align}
    & \tra{f_-^{\lambda\nu} u_\nu}\tra{u_\lambda u_\mu}\tra{u^\mu \chi_-} & \tra{f_-^{\lambda\nu} u^\mu}\tra{u_\lambda u_\mu}\tra{u_\nu\chi_-} \notag \\
    & \tra{f_-^{\lambda\nu} u_\lambda}\tra{u_\mu u^\mu}\tra{u_\nu\chi_-} &  \tra{\tilde{f}_-^{\lambda\nu} u_\lambda}\tra{u_\mu u^\mu}\tra{u_\nu\chi_+} \notag \\
    & \tra{\tilde{f}_-^{\lambda\nu} u_\nu}\tra{u_\lambda u_\mu}\tra{u^\mu \chi_+} & \tra{\tilde{f}_-^{\lambda\nu} u^\mu}\tra{u_\lambda u_\mu}\tra{u_\nu\chi_+} \notag \\
    & \tra{u_\lambda u^\mu}\tra{\tilde{f}_+^{\lambda\nu}[u_\mu,u_\nu]}\tra{\chi_-} & \tra{u_\lambda u^\lambda}\tra{\tilde{f}_-^{\nu\mu}[u_\mu,u_\nu]}\tra{\chi_-}  \notag \\
    & \tra{u_\lambda u^\mu}\tra{f_+^{\lambda\nu}[u_\mu,u_\nu]}\tra{\chi_+} & \tra{u_\lambda u^\lambda}\tra{f_+^{\nu\mu}[u_\mu,u_\nu]}\tra{\chi_+} 
\end{align}

$D^4u^4$:
\begin{align}
    & \tra{u_\nu D_\mu D_\rho D_\lambda u^\nu}\tra{u^\lambda D^\mu u^\rho} & \tra{u_\nu D_\lambda D_\mu u_\rho}\tra{u^\nu D^\lambda D^\mu u^\rho} \notag \\
    & \tra{u^\mu D_\lambda D_\mu u_\rho}\tra{u^\nu D_\nu D^\lambda u^\rho} 
\end{align}

$D^2u^6$:
\begin{align}
    & \tra{D_\lambda u_\mu D^\lambda u^\mu} \tra{u_\nu u^\nu} \tra{u_\rho u^{\rho}} & \tra{D_\lambda u_\mu D^\lambda u^\mu}\tra{u_\nu u_\rho} \tra{u^\nu u^\rho} \notag \\
    & \tra{D_\lambda u_\mu u^\mu}\tra{D^\lambda u_\nu u_\rho}\tra{u^\nu u^\rho} & \tra{u_\mu u^\mu}\tra{D_\lambda u_\nu D^\lambda u_\rho}\tra{u^\nu u^\rho} \notag \\
    & \tra{u_\mu u^\mu}\tra{D_\lambda u_\nu u_\rho}\tra{D^\lambda u^\nu u^\rho} & \tra{D_\lambda u_\mu D^\lambda u_\nu} \tra{u^\mu u^\rho}\tra{u_\rho u^\nu} \notag \\
    & \tra{D_\lambda u_\mu u_\nu} \tra{D^\lambda u^\mu u^\rho}\tra{u_\rho u^\nu} & \tra{D_\lambda u_\mu u_\nu} \tra{u^\mu D^\lambda u^\rho}\tra{u_\rho u^\nu} \notag \\
    & \tra{D_\lambda u_\mu u_\nu} \tra{u^\mu u^\rho}\tra{u_\rho D^\lambda u^\nu}
\end{align}

$D^2fu^4$:
\begin{align}
    & \tra{D^\lambda D^\mu f_+^{\rho\nu}[u_\mu,u_\nu]}\tra{u_\lambda u_\rho} & \tra{D^\nu f_+^{\rho\lambda}[D_\mu u_\rho,u^\mu]}\tra{u_\lambda u_\nu} \notag \\
    & \tra{D^\nu f_+^{\rho\lambda}[D_\mu u_\rho,u_\nu]}\tra{u_\lambda u^\mu} & \tra{D^\mu f_+^{\rho\nu}[D_\mu u_\rho,u_\nu]}\tra{u_\lambda u^\lambda} \notag \\
    & \tra{f_+^{\rho\lambda}[D_\mu D_\rho u^\nu,u^\mu]}\tra{u_\lambda u_\nu} & \tra{f_+^{\rho\nu}[D_\mu D_\rho u^\lambda,u_\nu]}\tra{u_\lambda u^\mu} \notag \\
    & \tra{f_+^{\rho\lambda}[D_\lambda u_\mu,D_\rho u^\mu]}\tra{u_\nu u^\mu} & \tra{f_+^{\rho\lambda}[D_\lambda u_\mu,D_\nu u_\rho]}\tra{u^\mu u^\nu} \notag \\
    & \tra{f_+^{\rho\lambda}[D_\mu u^\nu,D_\rho u^\mu]}\tra{u_\lambda u_\nu} & \tra{D^\nu f_+^{\rho\lambda}[u_\mu,u_\nu]}\tra{u_\lambda D_\rho u^\mu} \notag \\
    & \tra{D^\lambda f_+^{\rho\nu}[u_\mu,u_\nu]}\tra{u_\lambda D_\rho u^\mu} & \tra{f_+^{\rho\lambda}[D_\lambda u_\mu,u^\mu]}\tra{D_\nu u_\rho u^\nu} \notag \\
    & \tra{f_+^{\rho\lambda}[D_\lambda u_\mu,u_\nu]}\tra{D_\rho u^\mu u^\nu} & \tra{f_+^{\lambda\nu}[D_\mu u_\rho,u_\nu]}\tra{u_\lambda D^\mu u^\rho} 
\end{align}

$Dfu^5$:
\begin{align}
    & \tra{D^\nu f_-^{\rho\lambda}u^\mu}\tra{u_\lambda u_\mu}\tra{u_\nu u_\rho} & \tra{D^\lambda f_-^{\rho\nu}u_\nu}\tra{u_\lambda u_\mu}\tra{u_\rho u^\mu} \notag \\
    & \tra{D^\nu f_-^{\rho\lambda} u_\lambda} \tra{u_\mu u^\mu}\tra{u_\nu u_\rho} & \tra{f_-^{\rho\lambda} u_\nu}\tra{D_\mu u_\rho u^\nu}\tra{u_\lambda u^\mu} \notag \\
    & \tra{f_-^{\rho\lambda} u^\mu}\tra{D_\mu u_\rho u^\nu}\tra{u_\lambda u_\nu} & \tra{f_-^{\rho\lambda} u^\nu}\tra{D_\mu u_\rho u^\mu}\tra{u_\lambda u_\nu} \notag \\
    & \tra{f_-^{\rho\lambda} u_\nu}\tra{u_\lambda D_\mu u_\rho}\tra{u^\mu u^\nu} & \tra{f_-^{\rho\lambda} u^\mu}\tra{u_\lambda D_\mu u_\rho}\tra{u_\nu u^\nu} \notag \\
    & \tra{D_\rho f_-^{\rho\lambda} u_\lambda}\tra{u_\mu u^\mu}\tra{u_\nu u^\nu} & \tra{D_\rho f_-^{\rho\lambda} u_\lambda}\tra{u_\mu u_\nu}\tra{u^\mu u^\nu} \notag \\
    & \tra{D_\rho f_-^{\rho\lambda} u_\nu}\tra{u_\mu u^\nu}\tra{u_\lambda u^\mu} & \tra{D_\rho f_-^{\rho\lambda}u^\mu}\tra{u_\nu u^\nu}\tra{u_\mu u_\lambda} 
\end{align}

$f^2u^4$:
\begin{align}
    & \tra{{f_-}_{\lambda\rho} u^\mu}\tra{f_-^{\lambda\rho}u_\nu}\tra{u_\mu u^\nu} & \tra{{f_-}_{\lambda\rho}u_\nu}\tra{f_-^{\lambda\rho}u^\nu}\tra{u_\mu u^\mu} \notag \\
    & \tra{{f_-^\nu}_\rho u^\mu}\tra{f_-^{\rho\lambda} u_\nu}\tra{u_\lambda u_\mu} & \tra{{f_-^\nu}_\rho u_\nu}\tra{f_-^{\rho\lambda} u_\lambda}\tra{u^\mu u_\mu} \notag \\
    & \tra{f_-^{\mu\nu}u_\nu}\tra{f_-^{\rho\lambda}u_\lambda}\tra{u_\mu u_\rho} & \tra{{f_-}_{\rho\lambda}f_-^{\rho\lambda}}\tra{u_\mu u^\mu}\tra{u_\nu u^\nu} \notag \\
    & \tra{{f_-}_{\rho\lambda}f_-^{\rho\lambda}}\tra{u_\mu u_\nu}\tra{u^\mu u^\nu} & \tra{f_-^{\lambda\mu}f_-^{\rho\nu}}\tra{u_\mu u_\nu}\tra{u_\lambda u_\rho} \notag \\
    & \tra{\tilde{f}{}_-^{\rho\lambda}u^\mu}\tra{{\tilde{f}{}_-^\nu}_\rho u_\nu}\tra{u_\lambda u_\mu} & \tra{\tilde{f}{}_-^{\rho\lambda} u_\nu}\tra{{\tilde{f}{}^\nu_-}_\rho u^\mu}\tra{u_\lambda u_\mu} \notag \\
    & \tra{{\tilde{f}{}_-^\nu}_\rho u_\lambda}\tra{\tilde{f}{}_-^{\rho\lambda} u_\nu}\tra{u_\mu u^\mu} & \tra{\tilde{f}{}_-^{\mu\nu} u_\lambda}\tra{\tilde{f}{}_-^{\rho\lambda} u_\nu}\tra{u_\mu u_\rho} \notag \\
    & \tra{{\tilde{f}{}_-^\nu}_\rho\tilde{f}{}_-^{\rho\lambda}}\tra{u_\mu u^\mu}\tra{u_\lambda u_\nu} & \tra{{\tilde{f}{}_-^\nu}_\rho\tilde{f}{}_-^{\rho\lambda}}\tra{u_\mu u_\nu}\tra{u_\lambda u^\mu}\notag \\
    & \tra{{f_+}_{\lambda\rho} u^\mu}\tra{f_+^{\lambda\rho}u_\nu}\tra{u_\mu u^\nu} & \tra{{f_+}_{\lambda\rho}u_\nu}\tra{f_+^{\lambda\rho}u^\nu}\tra{u_\mu u^\mu} \notag \\
    & \tra{{f_+^\nu}_\rho u^\mu}\tra{f_+^{\rho\lambda} u_\nu}\tra{u_\lambda u_\mu} & \tra{{f_+^\nu}_\rho u_\nu}\tra{f_+^{\rho\lambda} u_\lambda}\tra{u^\mu u_\mu} \notag \\
    & \tra{f_+^{\mu\nu}u_\nu}\tra{f_+^{\rho\lambda}u_\lambda}\tra{u_\mu u_\rho} & \tra{{f_+}_{\rho\lambda}f_+^{\rho\lambda}}\tra{u_\mu u^\mu}\tra{u_\nu u^\nu} \notag \\
    & \tra{{f_+}_{\rho\lambda}f_+^{\rho\lambda}}\tra{u_\mu u_\nu}\tra{u^\mu u^\nu} & \tra{f_+^{\lambda\mu}f_+^{\rho\nu}}\tra{u_\mu u_\nu}\tra{u_\lambda u_\rho} \notag \\
    & \tra{\tilde{f}{}_+^{\rho\lambda}u^\mu}\tra{{\tilde{f}{}_+^\nu}_\rho u_\nu}\tra{u_\lambda u_\mu} & \tra{\tilde{f}{}_+^{\rho\lambda} u_\nu}\tra{{\tilde{f}{}^\nu_+}_\rho u^\mu}\tra{u_\lambda u_\mu} \notag \\
    & \tra{{\tilde{f}{}_+^\nu}_\rho u_\lambda}\tra{\tilde{f}{}_+^{\rho\lambda} u_\nu}\tra{u_\mu u^\mu} & \tra{\tilde{f}{}_+^{\mu\nu} u_\lambda}\tra{\tilde{f}{}_+^{\rho\lambda} u_\nu}\tra{u_\mu u_\rho} \notag \\
    & \tra{{\tilde{f}{}_+^\nu}_\rho\tilde{f}{}_+^{\rho\lambda}}\tra{u_\mu u^\mu}\tra{u_\lambda u_\nu} & \tra{{\tilde{f}{}_+^\nu}_\rho\tilde{f}{}_+^{\rho\lambda}}\tra{u_\mu u_\nu}\tra{u_\lambda u^\mu}
\end{align}

\section{Meson-Nucleon Sector}
\label{sec:NMoperators}

In the meson-nucleon sector $\mathcal{L}_{\pi N}$, the Weinberg power-counting formula in Eq.~\eqref{eq:nuw} can also be simplified by setting $L=0,A=1,n_i = 2$ as
\begin{equation}
    \nu = d_i \,.
\end{equation}
The leading order Lagrangian in Eq.~\eqref{eq:lead} is of $\nu=1$ or $p$ order, and the higher order interactions correspond to $\nu=2,3\dots$.
In this section, we construct the $C\,, P$-even effective operators in $\mathcal{L}_{\pi N}$ up to 5 derivatives. In particular, the operators up to 5 derivatives are not equivalent to the ones of $p^5$ order. Actually, the operators presented here cover the complete $C\,, P$-even operators of order $p^2$ and $p^3$, and partly cover the complete $C\,,P$-even operators of order $p^4$ and $p^5$, of which the complete operators can be found in Ref.~\cite{Song:2024fae}. As complementation, the Hilbert series result is presented in App.~\ref{sec:app1}.

As mentioned previously the derivatives applying on the relativistic nucleon are comparable with $\Lambda_\chi$, thus do not contribute to $d_i$ in the power-counting formula. However, the effective operators are organized in terms of the types in the Young tensor method, thus the operators of a specific type could be of different dimensions.
In this sector, we organize the effective operators in terms of their chiral dimension strictly.
In a specific type, the operators that are more dominated are listed in gray color and underlined, and the ones that are less dominated are listed in gray color solely.
In summary, we write down the meson-nucleon interactions following the notations that in a specific type, 
\begin{itemize}
    \item The operators with derivatives on the nucleons are of lower dimension, and are written in gray and underlined.
    \item The operators with pseudoscalar fermion bilinears are of higher dimension, and are written in gray (not underlined).
\end{itemize}
 The operators more dominant are also presented at the chiral dimension as they should be, with a * in front of their types.


\begin{center}
    \large \textbf{1. $\mathcal{O}(p^2)$}
\end{center}

$f N^2$:
\begin{align}
    & \bin{\sigma^{\mu\nu}f_+{}_{\mu\nu}} 
\end{align}

$N^2u^2$:
\begin{align}
    & \bin{}\lra{u_\mu u^\mu} & \bin{\sigma^{\mu\nu}[u_\mu,u_\nu]}
\end{align}

$N^2\chi$:
\begin{align}
    & {\color{gray}\bin{\gamma^5}\lra{\chi_-}} & \bin{}\lra{\chi_+} \notag \\
    & {\color{gray}\bin{\gamma^5\chi_-}} & \bin{\chi_+}
\end{align}

\textbf{Operators from next order:}
\vspace{2ex}

*$DN^2u^2$:
\begin{align}
    & \bin{\gamma^\mu\lrd^\nu}\lra{u_\mu u_\nu}
\end{align}

\begin{center}
    \large \textbf{2. $\mathcal{O}(p^3)$}
\end{center}

$DfN^2$:
\begin{align}
    & \bin{\gamma^5\gamma^\mu (D^\nu f_-{}_{\mu\nu})} & \bin{\gamma^\mu (D^\nu f_+{}_{\mu\nu})} 
\end{align}

$fN^2u$:
\begin{align}
    & \bin{\gamma^5\gamma^\nu[u^\mu,f_+{}_{\mu\nu}]} & \bin{\gamma^\mu}\lra{u^\nu\tilde{f}_+{}_{\mu\nu}} \notag \\
    & \bin{\gamma^\nu[u^\mu,f_-{}_{\mu\nu}]} & \bin{\gamma^5\gamma^\mu}\lra{u^\nu\tilde{f}_-{}_{\mu\nu}} 
\end{align}

$DN^2u^2$:
\begin{align}
    & \bin{\gamma^\mu[D^\nu u_\mu,u_\nu]} & {\color{gray} \underline{\bin{\gamma^\mu\lrd^\nu}\lra{u_\mu u_\nu}}}
\end{align}

$N^2u\chi$:
\begin{align}
    & \bin{\gamma^\mu[u_\mu,\chi_-]} & \bin{\gamma^5\gamma^\mu}\lra{u_\mu\chi_+} \notag \\
    & \bin{\gamma^5\gamma^\mu u_\mu}\lra{\chi_+} 
\end{align}

$N^2u^3$:
\begin{align}
    & \bin{\gamma_\rho}\lra{[u_\mu,u_\nu]u_\sigma}\epsilon^{\mu\nu\rho\sigma} & \bin{\gamma^5\gamma^\nu u^\mu}\lra{u_\nu u_\mu}\notag \\
    & \bin{\gamma^5\gamma^\mu u_\mu}\lra{u_\nu u^\nu}
\end{align}

\textbf{Operators from next order}

\vspace{2ex}

*$D^2N^2u^2$:
\begin{align}
    & \bin{\sigma^{\mu\nu}\lrd_\lambda}\tra{u_\mu D^\lambda u_\nu}
\end{align}

*$DN^2u^3$:
\begin{align}
    & \bin{\sigma^{\rho\sigma}\lrd^\nu u^\mu} \lra{u^\nu u^\lambda} \epsilon_{\mu\lambda\rho\sigma}
\end{align}

*$DufN^2$:
\begin{align}
    &  \bin{\sigma_{\mu\nu}\lrd_\lambda}\lra{f_-^{\nu\lambda}u^\mu} & \bin{\sigma_{\mu\nu}\lrd_\lambda [\tilde{f}_+^{\mu\lambda},u^\nu]}  
\end{align}

\begin{center}
    \large \textbf{3. $\mathcal{O}(p^4)$}
\end{center}

${f}^2N^2$:
\begin{align}
    & \bin{}\lra{{f_-}_{\mu\nu}{f_-}^{\mu\nu}} & {\color{gray}\bin{\gamma^5}\lra{{\tilde{f}{}_-}_{\mu\nu}{f_-}^{\mu\nu}}} \notag \\
    & \bin{\sigma^{\mu\lambda}[{f_-}_{\mu\nu},{f_-{}_\lambda}^\nu]} & \bin{\sigma^{\mu\lambda}}\lra{\tilde{f}{}_-{}_{\mu\nu}{f_+{}_\lambda}^\nu}\notag \\
    & {\color{gray}\bin{\gamma^5[f_-{}_{\mu\nu},{f_+}^{\mu\nu}]}} & \bin{[\tilde{f}{}_-{}_{\mu\nu},{f_+}^{\mu\nu}]} \notag \\
    & \bin{}\lra{{f_+}_{\mu\nu}{f_+}^{\mu\nu}} & {\color{gray}\bin{\gamma^5}\lra{{\tilde{f}{}_+}_{\mu\nu}{f_+}^{\mu\nu}}} \notag \\
    & \bin{\sigma^{\mu\lambda}[{f_+}_{\mu\nu},{f_+{}_\lambda}^\nu]}
\end{align}

$D^2 f N^2$:
\begin{align}
    & \bin{\sigma^{\mu\nu}(D^2{f_+{}_{\mu\nu}})} 
\end{align}

$DfN^2 u$:
\begin{align}
    & \bin{}\lra{D_\mu f_-^{\mu\nu}u_\nu} & \bin{\sigma_{\mu\nu}[D^\rho f_-^{\mu\nu},u_\rho]} \notag \\
    & \bin{\sigma_{\mu\nu}[f_-^{\mu\lambda},D_\lambda u^\nu]} & \bin{\sigma_{\mu\nu}[D_\lambda f_-^{\mu\lambda},u^\nu]} \notag \\
    & \bin{\sigma_{\mu\nu}}\lra{D^\rho\tilde{f}_+^{\mu\nu} u_\rho} & \bin{\sigma_{\mu\nu}}\lra{\tilde{f}_+^{\mu\lambda}D_\lambda u^\nu} \notag \\
    & \epsilon_{\mu\nu\rho\lambda}\bin{\sigma^{\rho\lambda}}\lra{D_\sigma f_+^{\mu\sigma} u^\nu} & {\color{gray} \bin{\gamma^5[D_\mu f_+^{\mu\nu},u_\nu]} } \notag \\
    & {\color{gray}\underline{ \bin{\sigma_{\mu\nu}\lrd_\lambda}\lra{f_-^{\nu\lambda}u^\mu} }} & {\color{gray}\underline{\bin{\sigma_{\mu\nu}\lrd_\lambda [\tilde{f}_+^{\mu\lambda},u^\nu]} }} 
\end{align}


$D^2N^2u^2$:
\begin{align}
    & \bin{}\lra{D_\nu u_\mu D^\nu u^\mu} & \bin{[D_\lambda u_\mu, D^\lambda u_\nu]\sigma^{\mu\nu}} \notag \\
    & {\color{gray}\underline{\bin{\sigma^{\mu\nu}\lrd_\lambda}\tra{u_\mu D^\lambda u_\nu}}}
\end{align}

$fN^2u^2$:
\begin{align}
    & {\color{gray}\bin{\gamma^5u^\mu}\lra{f_-{}_{\mu\nu}u^\nu}} & \bin{\sigma^{\lambda\nu}}\lra{[u_\mu,u_\nu]{\tilde{f}_-{}_\lambda}^\mu} \notag \\
    & \bin{u^\mu}\lra{\tilde{f}_-{}_{\mu\nu}u^\nu} & \bin{\sigma^{\nu\lambda}u^\mu}\lra{f_+{}_{\nu\lambda}u_\mu} \notag \\
    & \bin{\sigma^{\lambda\nu}u_\nu}\lra{{f_+{}_\lambda}^\mu u_\mu} & \bin{\sigma^{\nu\lambda}f_+{}_{\nu\lambda}}\lra{u^\mu u_\mu} \notag \\
    & \bin{}\lra{[u^\mu,u^\nu]f_+{}_{\mu\nu}} & {\color{gray}\bin{\gamma^5}\lra{[u^\mu,u^\nu]\tilde{f}_+{}_{\mu\nu}}} \notag \\
    & \bin{\sigma^{\lambda\nu}{f_+{}_\lambda}^\mu}\lra{u_\nu u_\mu} & \bin{\sigma^{\lambda\nu}u_\mu}\lra{{f_+{}_\lambda}^\mu u_\nu}
\end{align}

$DN^2u^3$:
\begin{align}
    & {\color{gray}\bin{\gamma^5 (D_\nu u_\mu)}\tra{u^\mu u^\nu}} & {\color{gray}\bin{\gamma^5 u_\mu}\tra{D_\nu u^\mu u^\nu}} \notag \\
    & \bin{\sigma^{\alpha\beta}}\tra{u_\mu [D^\mu u_\nu,u_\lambda]}\epsilon_{\lambda\nu\alpha\beta} & {\color{gray}\underline{\bin{\sigma^{\rho\sigma}\lrd^\nu u^\mu} \lra{u^\nu u^\lambda} \epsilon_{\mu\lambda\rho\sigma}} }
\end{align}

$N^2u^4$:
\begin{align}
    & \bin{\sigma^{\lambda\nu}u^\mu}\lra{[u_\lambda,u_\nu]u_\mu} & \bin{\sigma^{\lambda\nu}[u_\lambda,u_\nu]}\lra{u^\mu u_\mu} \notag \\
    & \bin{}\lra{u^\mu u_\mu} \lra{u^\nu u_\nu} & \bin{}\lra{u^\mu u^\nu}\lra{u_\mu u_\nu} 
\end{align}

$D^2N^2\chi$:
\begin{align}
    & {\color{gray}\bin{\gamma^5(D^2\chi_-)}} & \bin{(D^2\chi_+)} \notag \\
    & {\color{gray}\bin{\gamma^5}D^2\lra{\chi_-}} & \bin{}D^2\lra{\chi_+}
\end{align}

$fN^2\chi$:
\begin{align}
    & \bin{\sigma^{\mu\nu}[f_-{}_{\mu\nu},\chi_-]} & \bin{\sigma^{\mu\nu}}\lra{\tilde{f}_+{}_{\mu\nu}\chi_-} \notag \\
    & \bin{\sigma^{\mu\nu}[\tilde{f}_-{}_{\mu\nu},\chi_+]} & \bin{\sigma^{\mu\nu}}\lra{f_+{}_{\mu\nu}\chi_+} \notag \\
    & \bin{\sigma^{\mu\nu}\tilde{f}_+{}_{\mu\nu}}\lra{\chi_-} & \bin{\sigma^{\mu\nu}f_+{}_{\mu\nu}}\lra{\chi_+}
\end{align}

$DN^2u\chi$:
\begin{align}
    & \bin{}\lra{D^\mu\chi_- u_\mu} & \bin{\sigma^{\mu\nu}[D_\nu \chi_-,u_\mu]}  \notag \\
    & {\color{gray}\bin{\gamma^5}\lra{u_\mu D^\mu\chi_+}} & \bin{\sigma_{\rho\lambda}[D_\nu \chi_+,u_\mu]}\epsilon^{\mu\nu\rho\lambda}  \notag \\
    & \bin{u_\mu}D^\mu\lra{\chi_-} & {\color{gray}\bin{\gamma^5u_\mu}D^\mu\lra{\chi_+}}
\end{align}

$N^2u^2\chi$:
\begin{align}
    & \bin{\sigma_{\mu\nu}}\lra{[u_\alpha,u_\beta]\chi_-}\epsilon^{\mu\nu\alpha\beta} & {\color{gray}\bin{\gamma^5 u_\mu}\lra{u^\mu\chi_-}} \notag \\
    & {\color{gray}\bin{\gamma^5\chi_-}\lra{u^\mu u_\mu}} & \bin{\sigma^{\mu\nu}}\lra{[u_\mu,u_\nu]\chi_+} \notag \\
    & \bin{u_\mu}\lra{u^\mu\chi_+} & \bin{\chi_+}\lra{u^\mu u_\mu} \notag \\
    & \bin{\sigma_{\mu\nu}[u_\alpha,u_\beta]}\lra{\chi_-}\epsilon^{\mu\nu\alpha\beta} & {\color{gray}\bin{\gamma^5}\lra{u^\mu u_\mu}\lra{\chi_-}} \notag \\
    & \bin{\sigma^{\mu\nu}[u_\mu,u_\nu]}\lra{\chi_+} & \bin{}\lra{u^\mu u_\mu}\lra{\chi_+}
\end{align}

$N^2{\chi}^2$:
\begin{align}
    & \bin{}\lra{\chi_-\chi_-} & {\color{gray}\bin{\gamma^5}\lra{\chi_+\chi_-}} \notag \\
    & \bin{}\lra{\chi_+\chi_+} & \bin{\chi_-}\lra{\chi_-} \notag \\
    & {\color{gray}\bin{\gamma^5\chi_+}\lra{\chi_-}} & \bin{}\lra{\chi_-}\lra{\chi_-} \notag \\
    & {\color{gray}\bin{\gamma^5\chi_-}\lra{\chi_+}} & \bin{\chi_+}\lra{\chi_+} \notag \\
    & {\color{gray}\bin{\gamma^5}\lra{\chi_-}\lra{\chi_+}} & \bin{}\lra{\chi_+}\lra{\chi_+} 
\end{align}

\textbf{Operators from next order}
\vspace{2ex}

*$DN^2u^2{\chi}$:
\begin{align}
    & \bin{\gamma^5\gamma^\mu[u_\mu u_\nu]\lrd^\nu }\tra{\chi_-} & \bin{\gamma^\nu \lrd^\mu}\tra{u_\mu u_\nu}\tra{\chi_+} \notag \\
    & \bin{\gamma^5\gamma^\nu\lrd^\mu}\tra{\chi_-[u_\mu,u_\nu]}
\end{align}

*$D^2N^2u\chi$:
\begin{align}
    & \bin{\gamma^\mu u_\mu \lrd^\nu} D_\nu\tra{\chi_-} & \bin{\gamma^5\gamma^\mu[u_\mu D_\nu\chi_+]\lrd^\nu} \notag \\
    & \bin{\gamma^\mu\lrd^\nu}\tra{u_\mu D_\nu\chi_-}
\end{align}

*$DN^2f\chi$:
\begin{align}
    & \bin{\gamma^\mu f_-{}_{\mu\nu}\lrd^\nu}\tra{\chi_-} & \bin{\gamma^5\gamma^\mu f_+{}_{\mu\nu}\lrd^\nu}\tra{\chi_-} \notag \\
    & \bin{\gamma^5\gamma^\mu [f_-{}_{\mu\nu},\chi_+] \lrd^\nu} & \bin{\gamma^5\gamma^\mu\lrd^\nu}\tra{\tilde{f}_+{}_{\mu\nu}\chi_+} \notag \\
    & \bin{\gamma^5\gamma^\mu [\tilde{f}_-{}_{\mu\nu},\chi_-]\lrd^\nu} & \bin{\gamma^5\gamma^\mu \lrd^\nu}\tra{f_+{}_{\mu\nu}\chi_-}
 \end{align}

*$D^2N^2u^3$:
\begin{align}
     & \bin{\gamma^5\gamma^\lambda \lrd^\nu}\tra{D_\lambda u_\mu [u^\mu,u_\nu]} & \bin{\gamma^5\gamma^\nu\lrd_\lambda}\tra{D_\mu u^\lambda [u^\mu,u^\nu]}
\end{align}

*$DN^2u^2f$:
\begin{align}
    & {\bin{\gamma^5\gamma^\lambda\lrd_\mu u^\nu}\lra{\tilde{f}_+{}_{\nu\lambda} u^\mu}} & {\bin{\gamma^5\gamma^\lambda\lrd_\mu u^\mu}\lra{\tilde{f}_+{}_{\nu\lambda} u^\nu}} \notag \\
    & {\bin{\gamma^5\gamma^\lambda \lrd^\mu\tilde{f}_+{}_{\nu\lambda}}\lra{u_\mu u^\nu}} & {\bin{\gamma^\lambda\lrd^\mu}\lra{f_+{}_{\nu\lambda}[u_\mu,u^\nu]}} \notag \\
    & {\bin{\gamma^\lambda\lrd_\mu u^\nu}\lra{\tilde{f}_-{}_{\nu\lambda} u^\mu}} & {\bin{\gamma^\lambda\lrd_\mu u^\mu}\lra{\tilde{f}_-{}_{\nu\lambda} u^\nu}} \notag \\
    & {\bin{\gamma^\lambda \lrd^\mu\tilde{f}_-{}_{\nu\lambda}}\lra{u_\mu u^\nu}} & {\bin{\gamma^5\gamma^\lambda\lrd^\mu}\lra{f_-{}_{\nu\lambda}[u_\mu,u^\nu]}}
\end{align}
*$D^2N^2uf$:
\begin{align}
    & \bin{\gamma^5\gamma^\mu \lrd^\nu}\lra{f_+{}_{\mu\lambda} D^\lambda u_\nu} & \bin{\gamma^\mu \lrd^\nu [\Tilde{f}_+{}_{\mu\lambda} ,D^\lambda u_\nu]} \notag \\
    & \bin{\gamma^5\gamma^\mu \lrd^\nu}\lra{D^\lambda f_+{}_{\mu\lambda}u_\nu} & \bin{\gamma^\mu \lrd^\nu}\lra{f_-{}_{\mu\lambda} D^\lambda u_\nu} \notag \\
    & \bin{\gamma^5\gamma^\mu \lrd^\nu [\Tilde{f}_-{}_{\mu\lambda} ,D^\lambda u_\nu]} & \bin{\gamma^\mu \lrd^\nu}\lra{D^\lambda f_-{}_{\mu\lambda}u_\nu]} 
\end{align}

*$Df^2N^2$:
\begin{align}
    & \bin{\gamma^\mu\lrd_\nu}\lra{f_+{}_{\nu\lambda}f_+^{\lambda\mu}} & \bin{\gamma^5\gamma^\mu\lrd_\nu[\Tilde{f}_+{}_{\nu\lambda},f_+^{\lambda\mu}]} \notag \\
    & \bin{\gamma^\mu\lrd_\nu}\lra{f_-{}_{\nu\lambda}f_-^{\lambda\mu}} & \bin{\gamma^5\gamma^\mu\lrd_\nu[\Tilde{f}_-{}_{\nu\lambda},f_-^{\lambda\mu}]} \notag \\
    & \bin{\gamma^\mu\lrd_\nu [\tilde{f}_+{}_{\nu\lambda},f_-^\lambda{}_\mu]} & \bin{\gamma^5\gamma^\mu\lrd_\nu}\lra{f_-{}_{\nu\lambda}f_+^{\lambda\mu}} 
\end{align}

\begin{center}
    \large \textbf{4. $\mathcal{O}(p^5)$}
\end{center}

$D^3 N^2 u^2$:
\begin{align}
    & \bin{\gamma^\lambda[D_\lambda D_\mu u_\nu,D^\mu u^\nu]} \notag \\
    & {\color{gray}\underline{\bin{\gamma^5\gamma^\lambda \lrd^\nu}\tra{D_\lambda u_\mu [u^\mu,u_\nu]}}} & {\color{gray}\underline{\bin{\gamma^5\gamma^\nu\lrd_\lambda}\tra{D_\mu u^\lambda [u^\mu,u^\nu]}}}
\end{align}

$D N^2{\chi}^2$:
\begin{align}
    & \bin{\gamma^5\gamma^\mu} \lra{\chi_-D_\mu\chi_+} & \bin{\gamma^5\gamma^\mu\chi_+} D_\mu\lra{\chi_-} \notag \\
    & \bin{\gamma^5\gamma^\mu\chi_-} D_\mu\lra{\chi_+} & \bin{\gamma^5\gamma^\mu}\lra{\chi_-}D_\mu\lra{\chi_+} \notag \\
    & \bin{\gamma^\mu[\chi_+,D_\mu\chi_+]} & \bin{\gamma^\mu[\chi_-,D_\mu\chi_-]}
\end{align}

$N^2u{\chi}^2$:
\begin{align}
    & \bin{\gamma^5\gamma^\mu u_\mu}\lra{\chi_-}\lra{\chi_-} & \bin{\gamma^5\gamma^\mu u_\mu}\lra{\chi_+}\lra{\chi_+} \notag \\
    & \bin{\gamma^5\gamma^\mu}\tra{\chi_- u_\mu}\tra{\chi_-} & \bin{\gamma^5\gamma^\mu}\tra{\chi_+ u_\mu}\tra{\chi_+} \notag \\
    & \bin{\gamma^\mu[\chi_-,u_\mu]}\tra{\chi_+} & \bin{\gamma^\mu[\chi_+,u_\mu]}\tra{\chi_-} \notag \\
    & \bin{\gamma^\mu}\tra{u_\mu[\chi_+,\chi_-]} & \bin{\gamma^5\gamma^\mu u_\mu}\tra{\chi_-\chi_-} \notag \\
    & \bin{\gamma^5\gamma^\mu\chi_-}\tra{u_\mu\chi_-} & \bin{\gamma^5\gamma^\mu u_\mu}\tra{\chi_+\chi_+} \notag \\
    & \bin{\gamma^5\gamma^\mu\chi_+}\tra{u_\mu\chi_+}
 \end{align}

$N^2u^3{\chi}$:
\begin{align}
    & \bin{\gamma^\nu u^\mu}\tra{\chi_-[u_\mu,u_\nu]} & \bin{\gamma^\nu[u_\nu,\chi_-]}\tra{u_\mu u^\nu} \notag \\
    & \bin{\gamma^\nu[u_\mu,\chi_-]}\tra{u_\nu u^\mu} & \bin{\gamma_\rho\chi_+}\tra{u_\lambda [u_\mu,u_\nu]}\epsilon^{\mu\nu\rho\lambda} \notag \\
    & \bin{\gamma^5\gamma^\nu}\tra{u_\mu u^\mu}\tra{u_\nu\chi_+} & \bin{\gamma^5\gamma^\nu}\tra{u_\mu u_\nu}\tra{u^\mu\chi_+} \notag \\
    & \bin{\gamma_\rho}\tra{u_\lambda [u_\mu,u_\nu]}\tra{\chi_+}\epsilon^{\mu\nu\rho\lambda} & \bin{\gamma^5\gamma^\nu u_\nu}\tra{u_\mu u^\mu}\tra{\chi_+} \notag \\
    & \bin{\gamma^5\gamma^\nu u^\mu}\tra{u_\mu u_\nu}\tra{\chi_+} 
\end{align}

$DN^2u^2{\chi}$:
\begin{align}
    & (\overline{N}\gamma^\mu N)\langle \chi_+[D_\mu u_\nu,u^\nu]\rangle & (\overline{N}\gamma^\mu N)\langle D_\nu \chi_+[u_\mu,u^\nu]\rangle \notag \\
    & \epsilon_{\mu\nu\rho\lambda}(\overline{N}\gamma^5\gamma^\mu u^\rho N)\langle D^\nu \chi_+u^\lambda\rangle & (\overline{N}\gamma^\mu[D_\mu u_\nu,u^\nu] N)\langle \chi_+\rangle \notag \\
    & (\overline{N}\gamma^\mu[u_\mu,u_\nu] N)D^\nu\langle \chi_+\rangle & \epsilon_{\mu\nu\rho\lambda}(\overline{N}\gamma^5\gamma^\mu u^\rho u^\lambda N)D^\nu\langle\chi_+\rangle \notag \\
    & \epsilon_{\mu\nu\rho\lambda}(\overline{N}\gamma^\mu N)\langle D^\nu \chi_-u^\rho u^\lambda\rangle & (\overline{N}\gamma^5\gamma^\mu D_\mu u_\nu N)\langle  \chi_-u^\nu\rangle \notag \\
    & (\overline{N}\gamma^5\gamma^\mu u^\nu N)\langle\chi_-D_\mu u_\nu\rangle & (\overline{N}\gamma^5\gamma^\mu u^\nu N)\langle D_\nu\chi_- u^\mu\rangle \notag \\
    & (\overline{N}\gamma^5\gamma^\mu u^\mu N)\langle D_\nu\chi_-u^\nu \rangle & (\overline{N}\gamma^5\gamma^\mu \chi_-N)\langle D_\mu u_\nu u^\nu\rangle \notag \\
    & (\overline{N}\gamma^5\gamma^\mu D^\nu\chi_-N)\langle u_\mu u_\nu\rangle & (\overline{N}\gamma^5\gamma^\mu N)\langle D_\mu u_\nu u^\nu\rangle\langle \chi_-\rangle \notag \\
    & (\overline{N}\gamma^5\gamma^\mu N)\langle u_\mu u_\nu\rangle D^\nu\langle \chi_-\rangle \notag \\
    & {\color{gray}\underline{\bin{\gamma^5\gamma^\mu[u_\mu u_\nu]\lrd^\nu }\tra{\chi_-}}} & {\color{gray}\underline{\bin{\gamma^\nu \lrd^\mu}\tra{u_\mu u_\nu}\tra{\chi_+}}} \notag \\
    & {\color{gray}\underline{\bin{\gamma^5\gamma^\nu\lrd^\mu}\tra{\chi_-[u_\mu,u_\nu]}}}
\end{align}

$fN^2u\chi$:
\begin{align}
    & \bin{\gamma^\nu}\tra{u^\mu\tilde{f}_+{}_{\mu\nu}}\tra{\chi_+} & \bin{\gamma^5\gamma^\nu [u_\mu,f_+{}_{\mu\nu}]}\tra{\chi_+} \notag \\
    & \bin{\gamma^5\gamma^\nu}\tra{u^\mu\tilde{f}_-{}_{\mu\nu}}\tra{\chi_+} & \bin{\gamma^\nu [u_\mu,f_-{}_{\mu\nu}]}\tra{\chi_+} \notag \\
    & \bin{\gamma^\nu}\tra{u^\mu f_+{}_{\mu\nu}}\tra{\chi_-} & \bin{\gamma^5\gamma^\nu [u_\mu,\tilde{f}_+{}_{\mu\nu}]}\tra{\chi_-} \notag \\
    & \bin{\gamma^5\gamma^\nu}\tra{u^\mu f_-{}_{\mu\nu}}\tra{\chi_-} & \bin{\gamma^\nu [u_\mu,\tilde{f}_-{}_{\mu\nu}]}\tra{\chi_-} \notag \\
    & \bin{\gamma^5\gamma^\nu\chi_-}\tra{f_-{}_{\mu\nu} u^\mu} & \bin{\gamma^5\gamma^\nu f_-{}_{\mu\nu}}\tra{\chi_- u^\mu} \notag \\
    & \bin{\gamma^5\gamma^\nu u^\mu}\tra{f_-{}_{\mu\nu} \chi_-} & \bin{\gamma^\nu}\tra{\tilde{f}_-{}_{\mu\nu}[u_\mu,\chi_-]} \notag \\
    & \bin{\gamma^\nu\chi_-}\tra{f_+{}_{\mu\nu} u^\mu} & \bin{\gamma^\nu f_+{}_{\mu\nu}}\tra{\chi_- u^\mu} \notag \\
    & \bin{\gamma^\nu u^\mu}\tra{f_+{}_{\mu\nu} \chi_-} & \bin{\gamma^5\gamma^\nu}\tra{\tilde{f}_+{}_{\mu\nu}[u_\mu,\chi_-]} \notag \\
    & \bin{\gamma^5\gamma^\nu\chi_+}\tra{\tilde{f}_-{}_{\mu\nu} u^\mu} & \bin{\gamma^5\gamma^\nu \tilde{f}_-{}_{\mu\nu}}\tra{\chi_+ u^\mu} \notag \\
    & \bin{\gamma^5\gamma^\nu u^\mu}\tra{\tilde{f}_-{}_{\mu\nu} \chi_+} & \bin{\gamma^\nu}\tra{f_-{}_{\mu\nu}[u_\mu,\chi_+]} \notag \\
    & \bin{\gamma^\nu\chi_+}\tra{\tilde{f}_+{}_{\mu\nu} u^\mu} & \bin{\gamma^\nu \tilde{f}_+{}_{\mu\nu}}\tra{\chi_+ u^\mu} \notag \\
    & \bin{\gamma^\nu u^\mu}\tra{\tilde{f}_+{}_{\mu\nu} \chi_+} & \bin{\gamma^5\gamma^\nu}\tra{f_+{}_{\mu\nu}[u_\mu,\chi_+]} \notag \\
\end{align}

$D^2N^2u\chi$:
\begin{align}
    & \bin{\gamma^5\gamma^\nu}\lra{u^\nu D_\nu D_\nu \chi_+} & \bin{\gamma^5\gamma^\nu}\lra{u_\nu D^2\chi_+} \notag \\
    & \bin{\gamma^\mu[u^\nu,D_\nu D_\mu\chi_-]} & \bin{\gamma^\mu[u_\mu,D^2\chi_-]} \notag \\
    & \bin{\gamma^5\gamma^\mu u^\nu}D_\nu D_\mu\lra{\chi_+} & \bin{\gamma^5\gamma^\nu}D^2\lra{\chi_+} 
\end{align}

{\color{gray}
\begin{align}
    & \underline{\bin{\gamma^5\gamma^\mu\lrd^\nu[u_\mu,D_\nu\chi_+]}} & \underline{\bin{\gamma^\mu\lrd^\nu}\lra{u_\mu D_\nu\chi_-}} \notag \\
    & \underline{\bin{\gamma^\mu\lrd^\nu u_\mu}D_\nu\lra{\chi_-}}
\end{align}

}



$DN^2f\chi$:
\begin{align}
    & (\overline{N}\gamma^\mu N)\langle f_{+\mu\nu}D^\nu \chi_+\rangle & (\overline{N}\gamma^5\gamma^\mu [\Tilde{f}_{+\mu\nu},D^\nu \chi_+]N) \notag \\
    & (\overline{N}\gamma^\mu N)\langle D^\nu f_{+\mu\nu} \chi_+\rangle & (\overline{N}\gamma^5\gamma^\mu [f_{+\mu\nu},D^\nu \chi_-]N) \notag \\
    & (\overline{N}\gamma^\mu N)\langle\Tilde{f}_{+\mu\nu}D^\nu \chi_-\rangle & (\overline{N}\gamma^5\gamma^\mu [D^\nu f_{+\mu\nu} ,\chi_-]N) \notag \\
    & (\overline{N}\gamma^\mu [f_{-\mu\nu},D^\nu \chi_-]N) & (\overline{N}\gamma^5\gamma^\mu N)\langle\Tilde{f}_{-\mu\nu}D^\nu \chi_-\rangle \notag \\
    & (\overline{N}\gamma^\mu [D^\nu f_{-\mu\nu} ,\chi_-]N) & (\overline{N}\gamma^5\gamma^\mu N)\langle f_{-\mu\nu}D^\nu \chi_+\rangle \notag \\
    & (\overline{N}\gamma^\mu [\Tilde{f}_{-\mu\nu},D^\nu \chi_+]N) & (\overline{N}\gamma^5\gamma^\mu N)\langle D^\nu f_{-\mu\nu} \chi_+\rangle \notag \\
    & (\overline{N}\gamma^\mu f_{+\mu\nu}N)D^\nu\langle\chi_+\rangle & (\overline{N}\gamma^\mu \Tilde{f}_{+\mu\nu}N)D^\nu\langle\chi_-\rangle \notag \\
    & (\overline{N}\gamma^5\gamma^\mu f_{-\mu\nu}N)D^\nu\langle\chi_+\rangle & (\overline{N}\gamma^5\gamma^\mu \Tilde{f}_{-\mu\nu}N)D^\nu\langle\chi_-\rangle \notag \\
    & (\overline{N}\gamma^\mu D^\nu f_{+\mu\nu}N)\langle\chi_+\rangle & (\overline{N}\gamma^5\gamma^\mu D^\nu f_{-\mu\nu}N)\langle\chi_+\rangle 
\end{align}

$N^2u^5$:
\begin{align}
    & \bin{\gamma_\eta} \tra{u_\mu [u_\nu,u_\lambda]}\tra{u_\rho u^\mu}\epsilon^{\eta\lambda\nu\rho} & \bin{\gamma^5\gamma^\lambda u_\mu}\tra{u_\nu u^\mu}\tra{u_\lambda u^\nu} \notag \\
    & \bin{\gamma^5\gamma^\lambda u_\mu}\tra{u_\nu u^\nu}\tra{u_\lambda u^\mu} & \bin{\gamma^5\gamma^\mu u_\mu}\tra{u_\nu u^\nu}\tra{u_\lambda u^\lambda} \notag \\
    & \bin{\gamma^5\gamma^\mu u_\mu}\tra{u_\nu u_\lambda}\tra{u^\nu u^\lambda}
\end{align}

$DN^2u^4$:
\begin{align}
    & \bin{\gamma^\nu u^\lambda}\lra{D_\lambda u_\mu [u^\mu,u_\nu]} & \bin{\gamma^\lambda u^\nu}\lra{D_\lambda u_\mu [u^\mu,u_\nu]} \notag \\
    & \bin{\gamma^\lambda [u^\mu,u^\nu]}\lra{D_\lambda u_\mu, u_\nu} & \bin{\gamma^\lambda[u_\nu,u_\lambda]}\lra{u^\mu D_\mu u^\nu} \notag \\
    & \bin{\gamma^\lambda[D_\lambda u_\nu,u^\nu]}\lra{u_\mu u^\mu} & \bin{\gamma^\lambda[D^\mu u^\nu,u_\lambda]}\lra{u_\mu u_\nu} \notag \\
    & \bin{\gamma^5\gamma_\eta}\lra{u_\nu D_\mu u_\rho}\lra{u_\lambda u^\mu}\epsilon^{\eta\lambda\rho\nu} 
\end{align}

$N^2u^3f$:
\begin{align}
    & \bin{\gamma^\lambda u_\mu} \tra{f_-{}_{\lambda\nu}[u^\mu,u^\nu]} & \bin{\gamma^\mu u_\mu} \tra{f_-{}_{\lambda\nu}[u^\lambda,u^\nu]} \notag \\
    & \bin{\gamma^\nu u^\mu}\tra{f_-{}_{\lambda\mu}[u_\nu,u^\lambda]} & \bin{\gamma^\lambda [u^\mu,u^\nu]}\tra{f_-{}_{\lambda\nu}u_\mu} \notag \\
    & \bin{\gamma^\mu [u_\nu,u_\lambda]}\tra{f_-{}^{\nu\lambda} u_\mu} & \bin{\gamma^\lambda [f_-{}_{\lambda\nu},u^\mu]}\tra{u_\mu u^\nu} \notag \\
    & \bin{\gamma^5\gamma^\lambda}\tra{\tilde{f}_-{}_{\lambda\nu} u_\mu}\tra{u^\mu u^\nu} & \bin{\gamma^5\gamma^\lambda}\tra{\tilde{f}_-{}_{\lambda\nu} u^\nu}\tra{u_\mu u^\mu} \notag \\
    & \bin{\gamma^5\gamma^\mu}\tra{\tilde{f}_-{}^{\lambda\nu} u_\nu}\tra{u_\lambda u^\mu} & \bin{\gamma^\mu}\tra{\tilde{f}_+{}^{\lambda\nu} u_\nu}\tra{u_\lambda u^\mu}  \notag \\
    & \bin{\gamma^5\gamma^\lambda u_\mu} \tra{f_+{}_{\lambda\nu}[u^\mu,u^\nu]} & \bin{\gamma^5\gamma^\mu u_\mu} \tra{f_+{}_{\lambda\nu}[u^\lambda,u^\nu]} \notag \\
    & \bin{\gamma^5\gamma^\nu u^\mu}\tra{f_+{}_{\lambda\mu}[u_\nu,u^\lambda]} & \bin{\gamma^5\gamma^\lambda [u^\mu,u^\nu]}\tra{f_+{}_{\lambda\nu}u_\mu} \notag \\
    & \bin{\gamma^5\gamma^\mu [u_\nu,u_\lambda]}\tra{f_+{}^{\nu\lambda} u_\mu} & \bin{\gamma^5\gamma^\lambda [f_+{}_{\lambda\nu},u^\mu]}\tra{u_\mu u^\nu} \notag \\
    & \bin{\gamma^\lambda}\tra{\tilde{f}_+{}_{\lambda\nu} u_\mu}\tra{u^\mu u^\nu} & \bin{\gamma^\lambda}\tra{\tilde{f}_+{}_{\lambda\nu} u^\nu}\tra{u_\mu u^\mu} 
\end{align}

$D^2N^2u^3$:
\begin{align}
    & \bin{\gamma^5\gamma^\lambda D_\lambda D_\mu u^\nu}\lra{u_\nu u^\mu} & \bin{\gamma^5\gamma^\lambda u_\mu}\lra{u_\nu D^\mu u^\nu} \notag \\
    & \bin{\gamma^5\gamma^\nu D_\lambda u_\mu}\lra{u_\nu D^\lambda u^\mu} & \bin{\gamma^5\gamma^\lambda u_\mu}\lra{u_\nu D_\lambda D^\mu u^\nu} \notag \\
    & \bin{\gamma^5\gamma^\mu u_\mu}\lra{D_\lambda u_\nu D^\lambda u^\nu} & \epsilon_{\mu\nu\rho\lambda}(\overline{N}\gamma^\mu N)\langle u^\nu[D^\sigma u^\rho,D_\sigma u^\lambda]\rangle
\end{align}

$DN^2u^2f$:
\begin{align}
    & \bin{\gamma^\lambda u^\mu}\lra{f_+{}_{\nu\lambda}D_\mu u^\nu} & \bin{\gamma^\mu u^\lambda } \lra{f_+{}_{\nu\lambda}D_\mu u^\nu} \notag \\
    & \bin{\gamma^\lambda D^\nu u^\mu}\lra{f_+{}_{\nu\lambda}u_\mu} & \bin{\gamma^\mu D^\lambda u_\mu}\lra{f_+{}_{\nu\lambda}u^\nu} \notag \\
    & \bin{\gamma^\lambda u^\nu}\lra{D_\mu f_+{}_{\lambda\nu}u^\mu} & \bin{\gamma^\lambda u^\mu}\lra{D^\nu f_+{}_{\nu\lambda} u_\mu} \notag \\
    & \bin{\gamma^\lambda u^\mu}\lra{D_\mu f_+{}_{\nu\lambda} u^\nu} & \bin{\gamma^\mu u^\lambda}\lra{D^\nu f_+{}_{\nu\lambda}u_\mu} \notag \\
    & \bin{\gamma^\mu u_\mu}\lra{D^\nu f_+{}_{\nu\lambda}u^\lambda} & \bin{\gamma^\lambda f_+{}_{\lambda\nu}}\lra{D_\mu u^\nu u^\mu} \notag \\
    & \bin{\gamma^\mu f_+{}_{\nu\lambda}}\lra{D_\mu u_\nu u_\lambda} & \bin{\gamma^\lambda D_\mu f_+{}_{\nu\lambda}}\lra{u^\mu u^\nu} \notag \\
    & \bin{\gamma^\lambda D^\nu f_+{}_{\nu\lambda}}\lra{u_\mu u^\mu} & \bin{\gamma^\mu D^\nu f_+{}_{\nu\lambda}}\lra{u_\lambda u_\mu} \notag \\
    & \bin{\gamma^5\gamma^\lambda}\lra{\tilde{f}_+{}_{\nu\lambda}[D_\mu u^\nu,u^\mu]} & \bin{\gamma^5\gamma^\mu}\lra{\tilde{f}_+{}_{\nu\lambda}[D_\mu u^\nu,u^\lambda]} \notag \\
    & \bin{\gamma^5\gamma^\lambda u^\mu}\lra{f_-{}_{\nu\lambda}D_\mu u^\nu} & \bin{\gamma^5\gamma^\mu u^\lambda } \lra{f_-{}_{\nu\lambda}D_\mu u^\nu} \notag \\
    & \bin{\gamma^5\gamma^\lambda D^\nu u^\mu}\lra{f_-{}_{\nu\lambda}u_\mu} & \bin{\gamma^5\gamma^\mu D^\lambda u_\mu}\lra{f_-{}_{\nu\lambda}u^\nu} \notag \\
    & \bin{\gamma^5\gamma^\lambda u^\nu}\lra{D_\mu f_-{}_{\lambda\nu}u^\mu} & \bin{\gamma^5\gamma^\lambda u^\mu}\lra{D^\nu f_-{}_{\nu\lambda} u_\mu} \notag \\
    & \bin{\gamma^5\gamma^\lambda u^\mu}\lra{D_\mu f_-{}_{\nu\lambda} u^\nu} & \bin{\gamma^5\gamma^\mu u^\lambda}\lra{D^\nu f_-{}_{\nu\lambda}u_\mu} \notag \\
    & \bin{\gamma^5\gamma^\mu u_\mu}\lra{D^\nu f_-{}_{\nu\lambda}u^\lambda} & \bin{\gamma^5\gamma^\lambda f_-{}_{\lambda\nu}}\lra{D_\mu u^\nu u^\mu} \notag \\
    & \bin{\gamma^5\gamma^\mu f_-{}_{\nu\lambda}}\lra{D_\mu u_\nu u_\lambda} & \bin{\gamma^5\gamma^\lambda D_\mu f_-{}_{\nu\lambda}}\lra{u^\mu u^\nu} \notag \\
    & \bin{\gamma^5\gamma^\lambda D^\nu f_-{}_{\nu\lambda}}\lra{u_\mu u^\mu} & \bin{\gamma^5\gamma^\mu D^\nu f_-{}_{\nu\lambda}}\lra{u_\lambda u_\mu} \notag \\
    & \bin{\gamma^\lambda}\lra{\tilde{f}_-{}_{\nu\lambda}[D_\mu u^\nu,u^\mu]} & \bin{\gamma^\mu}\lra{\tilde{f}_-{}_{\nu\lambda}[D_\mu u^\nu,u^\lambda]} 
\end{align}

{\color{gray}
\begin{align}
    & \underline{\bin{\gamma^5\gamma^\lambda\lrd_\mu u^\nu}\lra{\tilde{f}_+{}_{\nu\lambda} u^\mu}} & \underline{\bin{\gamma^5\gamma^\lambda\lrd_\mu u^\mu}\lra{\tilde{f}_+{}_{\nu\lambda} u^\nu}} \notag \\
    & \underline{\bin{\gamma^5\gamma^\lambda \lrd^\mu\tilde{f}_+{}_{\nu\lambda}}\lra{u_\mu u^\nu}} & \underline{\bin{\gamma^\lambda\lrd^\mu}\lra{f_+{}_{\nu\lambda}[u_\mu,u^\nu]}} \notag \\
    & \underline{\bin{\gamma^\lambda\lrd_\mu u^\nu}\lra{\tilde{f}_-{}_{\nu\lambda} u^\mu}} & \underline{\bin{\gamma^\lambda\lrd_\mu u^\mu}\lra{\tilde{f}_-{}_{\nu\lambda} u^\nu}} \notag \\
    & \underline{\bin{\gamma^\lambda \lrd^\mu\tilde{f}_-{}_{\nu\lambda}}\lra{u_\mu u^\nu}} & \underline{\bin{\gamma^5\gamma^\lambda\lrd^\mu}\lra{f_-{}_{\nu\lambda}[u_\mu,u^\nu]}}
\end{align}
}

$N^2uf^2$:

\begin{align}
    & \bin{\gamma^\lambda}\tra{u_\mu [f_-{}^{\nu\mu},\tilde{f}_-{}_{\nu\lambda}]} & \bin{\gamma^5\gamma^\mu u_\mu}\tra{f_-{}^{\nu\lambda}f_-{}_{\nu\lambda}} \notag \\
    & \bin{\gamma^5\gamma^\mu f_-{}^{\nu\lambda}}\tra{u_\mu f_-{}_{\nu\lambda}} 
    & \bin{\gamma^5\gamma^\lambda u_\mu}\tra{f_-{}_{\nu\lambda} f_-^{\nu\mu}} \notag \\
    & \bin{\gamma^5\gamma^\lambda f_-{}_{\nu\lambda}} \tra{u_\mu f_-{}^{\nu\mu}} & \bin{\gamma^5\gamma^\lambda f_-{}^{\nu\mu}}\tra{u_\mu f_-{}_{\nu\lambda}} \notag \\
    & \bin{\gamma^\lambda} \tra{u_\mu [\tilde{f}_-{}^{\nu\mu},f_-{}_{\nu\lambda}]} & \bin{\gamma^\lambda} \tra{u_\mu [\tilde{f}_+{}^{\nu\mu},f_+{}_{\nu\lambda}]} \notag \\
    & \bin{\gamma^\lambda}\tra{u_\mu [f_+{}^{\nu\mu},\tilde{f}_+{}_{\nu\lambda}]} & \bin{\gamma^5\gamma^\mu u_\mu}\tra{f_+{}^{\nu\lambda}f_+{}_{\nu\lambda}} \notag \\
    & \bin{\gamma^5\gamma^\mu f_+{}^{\nu\lambda}}\tra{u_\mu f_+{}_{\nu\lambda}} 
    & \bin{\gamma^5\gamma^\lambda u_\mu}\tra{f_+{}_{\nu\lambda} f_+^{\nu\mu}} \notag \\
    & \bin{\gamma^5\gamma^\lambda f_+{}_{\nu\lambda}} \tra{u_\mu f_+{}^{\nu\mu}} & \bin{\gamma^5\gamma^\lambda f_+{}^{\nu\mu}}\tra{u_\mu f_+{}_{\nu\lambda}}
\end{align}

\begin{align}
    & \bin{\gamma^\mu}\tra{u_\mu [f_-{}_{\nu\lambda},f_+{}^{\nu\lambda}]} & \bin{\gamma^\lambda}\tra{u_\mu [f_-{}^{\nu\mu},f_+{}_{\nu\lambda}]} \notag \\
    & \bin{\gamma^\lambda}\tra{u_\mu [\tilde{f}_-{}^{\nu\mu},\tilde{f}_+{}_{\nu\lambda}]} & \bin{\gamma^\mu f_-{}_{\nu\lambda}}\tra{f_+{}^{\nu\lambda} u_\mu} \notag \\
    & \bin{\gamma^\mu f_+{}_{\nu\lambda}}\tra{f_-{}^{\nu\lambda} u_\mu} & \bin{\gamma^\mu u_\mu}\tra{f_+{}^{\nu\lambda} f_-{}_{\nu\lambda}} \notag \\
    & \bin{\gamma^\lambda f_-{}^{\nu\mu}}\tra{f_+{}_{\nu\lambda}u_\mu} & \bin{\gamma^\lambda f_+{}^{\nu\mu}}\tra{f_-{}_{\nu\lambda}u_\mu} \notag \\
    & \bin{\gamma^\lambda u_\mu}\tra{f_-{}^{\nu\mu}f_+{}_{\nu\lambda}} & \bin{\gamma^\lambda \tilde{f}_-{}^{\nu\mu}}\tra{\tilde{f}_+{}_{\nu\lambda}u_\mu} \notag \\
    & \bin{\gamma^\lambda \tilde{f}_+{}^{\nu\mu}}\tra{\tilde{f}_-{}_{\nu\lambda}u_\mu} & \bin{\gamma^\lambda u_\mu}\tra{\tilde{f}_-{}^{\nu\mu}\tilde{f}_+{}_{\nu\lambda}}
\end{align}

$D^2N^2uf$:
\begin{align}
    & (\overline{N}\gamma^5\gamma^\mu [D^\nu f_{+\mu\lambda},D_\nu u_\lambda]N) & (\overline{N}\gamma^\mu N)\langle D^\nu \Tilde{f}_{+\mu\lambda} D_\nu u_\lambda\rangle \notag \\
    & (\overline{N}\gamma^5\gamma^\mu [D^\nu f_{+\nu\lambda}, D_\mu u_\lambda]N) & (\overline{N}\gamma^5\gamma^\mu [D^2 f_{+\mu\nu} ,u_\nu]N) \notag \\
    & (\overline{N}\gamma^5\gamma^\mu[D^\nu D^\lambda f_{+\mu\lambda}, u_\nu]N) & (\overline{N}\gamma^5\gamma^\mu N)\langle D^\nu \Tilde{f}_{-\mu\lambda} D_\nu u_\lambda\rangle \notag \\
    & (\overline{N}\gamma^\mu [D^\nu f_{-\mu\lambda},D_\nu u_\lambda]N) & (\overline{N}\gamma^\mu [D^\nu f_{-\nu\lambda}, D_\mu u_\lambda]N) \notag \\
    & (\overline{N}\gamma^\mu [D^2 f_{-\mu\nu} ,u_\nu]N) & (\overline{N}\gamma^\mu[D^\nu D^\lambda f_{-\mu\lambda}, u_\nu]N)
\end{align}

{\color{gray}
\begin{align}
    & \underline{\bin{\gamma^5\gamma^\mu \lrd^\nu}\lra{f_+{}_{\mu\lambda} D^\lambda u_\nu}} & \underline{\bin{\gamma^\mu \lrd^\nu [\Tilde{f}_+{}_{\mu\lambda} ,D^\lambda u_\nu]}} \notag \\
    & \underline{\bin{\gamma^5\gamma^\mu \lrd^\nu}\lra{D^\lambda f_+{}_{\mu\lambda}u_\nu}} & \underline{\bin{\gamma^\mu \lrd^\nu}\lra{f_-{}_{\mu\lambda} D^\lambda u_\nu}} \notag \\
    & \underline{\bin{\gamma^5\gamma^\mu \lrd^\nu [\Tilde{f}_-{}_{\mu\lambda} ,D^\lambda u_\nu]}} & \underline{\bin{\gamma^\mu \lrd^\nu}\lra{D^\lambda f_-{}_{\mu\lambda}u_\nu]}} 
\end{align}
}

$DN^2f^2$:
\begin{align}
    & (\overline{N}\gamma^\mu[D_\mu f_{+\nu\lambda},f_{+\nu\lambda}]N) & (\overline{N}\gamma^\mu [D_\nu f_{+\mu\lambda},f_{+\nu\lambda}]N) \notag \\
    & (\overline{N}\gamma^5\gamma^\mu N)\langle D_\nu f_{+\mu\lambda}\Tilde{f}_{+\nu\lambda}\rangle & (\overline{N}\gamma^\mu[D_\mu f_{-\nu\lambda},f_{-\nu\lambda}]N) \notag \\
    & (\overline{N}\gamma^\mu [D_\nu f_{-\mu\lambda},f_{-\nu\lambda}]N) & (\overline{N}\gamma^5\gamma^\mu N)\langle D_\nu f_{-\mu\lambda}\Tilde{f}_{-\nu\lambda}\rangle \notag \\
    & (\overline{N}\gamma^\mu N)\langle D_\mu f_{+\nu\lambda}\Tilde{f}_{-\nu\lambda}\rangle & (\overline{N}\gamma^\mu N)\langle D_\nu f_{-\mu\lambda}\Tilde{f}_{+\nu\lambda}\rangle \notag \\
    & (\overline{N}\gamma^\mu N)\langle \Tilde{f}_{-\mu\lambda}D_\nu f_{+\nu\lambda}\rangle & (\overline{N}\gamma^5\gamma^\mu[D_\mu f_{+\nu\lambda},f_{-\nu\lambda}]N) \notag \\
    & (\overline{N}\gamma^5\gamma^\mu [D_\nu f_{-\mu\lambda},f_{+\nu\lambda}]N) & (\overline{N}\gamma^5\gamma^\mu [f_{-\mu\lambda},D_\nu f_{+\nu\lambda}]N) 
\end{align}

{\color{gray}
\begin{align}
    & \underline{\bin{\gamma^\mu\lrd_\nu}\lra{f_+{}_{\nu\lambda}f_+^{\lambda\mu}}} & \underline{\bin{\gamma^5\gamma^\mu\lrd_\nu[\Tilde{f}_+{}_{\nu\lambda},f_+^{\lambda\mu}]}} \notag \\
    & \underline{\bin{\gamma^\mu\lrd_\nu}\lra{f_-{}_{\nu\lambda}f_-^{\lambda\mu}}} & \underline{\bin{\gamma^5\gamma^\mu\lrd_\nu[\Tilde{f}_-{}_{\nu\lambda},f_-^{\lambda\mu}]}} \notag \\
    & \underline{\bin{\gamma^\mu\lrd_\nu [\tilde{f}_+{}_{\nu\lambda},f_-^\lambda{}_\mu]}} & \underline{\bin{\gamma^5\gamma^\mu\lrd_\nu}\lra{f_-{}_{\nu\lambda}f_+^{\lambda\mu}}} 
\end{align}
}

$D^3N^2f$:
\begin{align}
    & \bin{\gamma^\nu (D^\mu D^2 f_+{}_{\mu\nu})} & \bin{\gamma^5\gamma^\nu (D^\mu D^2 f_-{}_{\mu\nu})}
\end{align}


\section{Multi-nucleon Sector}
\label{sec:NNoperators}

Since the interactions involving two or more nucleons in the sectors $\mathcal{L}_{NN},\mathcal{L}_{\pi NN},\mathcal{L}_{NNN},\dots$ are not perturbative, the power-counting formula in Eq.~\eqref{eq:nuw} is applicable not for the amplitudes directly but for the effective potential, which is contributed from the Feynman diagrams without the pure nucleon inner states, called the one-nucleon irreducible diagrams. However, the Weinberg power-counting formula in Eq.~\eqref{eq:nuw} still needs modification since it may give a negative value when $A\geq 3$. This can be fixed by the addition of $(3A-6)$ to the formula in Eq.~\eqref{eq:nuw}, which leaves the $A=2$ case unaltered, and one obtain the modified power-counting formula~\cite{Meissner:2014lgi,Epelbaum:2008ga,Hammer:2019poc,Machleidt:2011zz} 
\begin{equation}
\label{eq:power}
    \nu=-4+2A+2L+\sum_{i=1}^V\Delta_i\,.
\end{equation}
According to the power-counting formula above, there are several remarks,
\begin{itemize}
    \item when the number of the external nucleons is fixed, at a certain dimension $\nu$ of the effective potential, as the number of the loops increases, the number and the interaction indices of the involving vertices decrease in principle, thus at every order, the contributing diagrams to the effective potential are finite, and consequently, the contributing interactions are finite,
    \item for each effective operator of certain interaction index $\Delta$ in the $\mathcal{L}_{NN}$ sector, it contributes to the effective potential not only at order $\nu=\Delta$ via the tree-level contact diagram but also at all the order $\nu>\Delta$ via tree-level or loop-level meson-exchange diagram. Such resummation property characterizes the non-perturbative nature of the nucleon self-interactions,
    \item while for the effective operators of $\mathcal{L}_{\pi NN}$ sector, the minimum order $\nu$ they contribute is larger than their interaction index $\Delta$ since they can only contribute to the effective potential via the meson-exchange diagram,
    \item the contribution of the $\mathcal{L}_{NNN}$ sector is also suppressed because $A=3$ gives a positive shift $2$ in the power counting formula in Eq.~\eqref{eq:power} compared to the $\mathcal{L}_{NN}$ sector.
\end{itemize}

According to the remarks above, we can list all the contributing interactions in terms of their interaction indices $\Delta$ and the minimum order $\nu$ of the effective potential they contribute. In Tab.~\ref{tab:types}, the contributing interaction types in the sectors $\mathcal{L}_{NN}\,,\mathcal{L}_{NNN}$, and $\mathcal{L}_{\pi NN}$ are presented up to the effective potential order $\nu=6$, organized by a two-element array $(\nu\,,\Delta)$, where $\nu$ is the minimum order they contribute.

\begin{table}[ht]
    \centering
    \begin{tabular}{|c|c|c|c|c|c|c|c|}
\hline
\diagbox{$\Delta$}{$\nu$} & 0 & 1 & 2 & 3 & 4 & 5 & 6 \\
\hline
0 & $N^4$ & & & & & & \\
\hline
1 & & $N^4D$ & & $N^4u\,,N^6$ & & & \\
\hline
2 & & & $N^4D^2$ & & $N^4uD\,,N^6D$ & & $N^4u^2$\\
\hline
3 & & & & $N^4D^3$ & & $N^4uD^2\,,N^6D^2$ & \\
\hline
4 & & & & & $N^4D^4$ & & $N^4uD^3\,,N^6D^3$ \\
\hline
5 & & & & & & $N^4D^5$ & \\
\hline
6 & & & & & & & $N^4D^6$ \\
\hline
    \end{tabular}
    \caption{The vertices types are organized in terms by a two-element array $(\nu,\Delta)$, where the horizontal coordinate $\nu$ denotes the lowest-order potential that vertex type can contribute, and the vertical coordinate denotes its interaction index.}
    \label{tab:types}
\end{table}

In this section, we present the effective operators in Tab.~\ref{tab:types} organized by their interaction indices rather than the power $\nu$, because of the non-perturbative property of the effective potential. In addition, the number of the complete and independent $C\,, P$-even operators is cross-checked by the Hilbert series, whose result is 
\begin{align}
    \mathcal{H}_{\text{C-even}}^{\text{P-even}} &= 5N^4 & \mathit{\Delta=0} \notag \\
    & + N^4D + 3N^4u + 5N^6 & \mathit{\Delta=1} \notag \\
    & + 9 N^4D^2 + 7N^4uD + 16N^4u^2 + 10N^6D & \mathit{\Delta=2} \notag \\
    & + 2 N^4D^3 + 18 N^4uD^2 + 39 N^6D^2 & \mathit{\Delta=3} \notag \\
    & + 13 N^4D^4 + 32 N^4uD^3 + 85 N^6D^3 & \mathit{\Delta =4} \notag \\
    & + 3 N^4D^5 & \mathit{\Delta=5} \notag \\
    & + 17 N^4D^6\,, & \mathit{\Delta=6} 
\end{align}
and the explicit operator forms will be presented in the subsequent subsections. The operators presented here are of relativistic form and follow the conventions in Sec.~\ref{sec:NMoperators}. In particular, for the operators without pion field, it belongs to the pion-less EFT ($\not\pi$-EFT), in the relativistic form. Expanding to the non-relativistic form using the heavy nucleon expansion defined above is straightforward. 

In Ref.~\cite{Girlanda:2010ya, Girlanda:2010zz} the authors constructed a complete but nonminimal set of the 2-nucleon sector $\mathcal{L}_{NN}$. Apart from the five operators without derivatives, all other ones are redundant. Besides, the authors constructed a set of operators in the 3-nucleon sector $\mathcal{L}_{NNN}$ in Ref.~\cite{Nasoni:2023adf}, which are also redundant.

\subsection{2-Nucleon Contact Interaction Operators}
\label{subsec:nninteraction}

\begin{center}
    \large\textbf{1. $\Delta=0$}
\end{center}

$N^4$:
\begin{equation}
    \begin{array}{ll}
        \bin{}\bin{} & {\color{gray}\bin{\gamma^5}\bin{\gamma^5}} \\
        \bin{\sigma^{\mu\nu}}\bin{\sigma_{\mu\nu}} & \bin{\gamma^\mu}\bin{\gamma_\mu} \\
        \bin{\gamma^5\gamma^\mu}\bin{\gamma^5\gamma_\mu} & 
    \end{array}
\end{equation}

\begin{center}
    \large\textbf{2. $\Delta=1$}
\end{center}
$N^4D$:
\begin{equation}
    \begin{array}{ll}
       \bin{\gamma^\mu}\bin{\lrd_\mu} &  
    \end{array}
\end{equation}

\begin{center}
    \large\textbf{3. $\Delta=2$}
\end{center}

$N^4D^2$:
\begin{equation}
    \begin{array}{ll}
        \bin{}D^2\bin{} & {\color{gray}\bin{\gamma^5}D^2\bin{\gamma^5}} \\
        \bin{\sigma_{\nu\lambda}}D^2\bin{\sigma^{\nu\lambda}} & \bin{\sigma_{\nu\lambda}\lrd_\mu}\bin{\sigma^{\nu\lambda}\lrd^\mu} \\
        \bin{\sigma_{\nu\lambda}}D_\mu D^\nu\bin{\sigma^{\mu\lambda}} & \bin{\gamma^\nu}D^2\bin{\gamma_\nu} \\
        \bin{\gamma_\nu\lrd_\mu}\bin{\gamma^\nu\lrd^\mu} & \bin{\gamma^5\gamma^\nu}D^2\bin{\gamma^5\gamma_\nu} \\
        \bin{\gamma^5\gamma_\nu\lrd_\mu}\bin{\gamma^5\gamma^\nu\lrd^\mu} & \\
    \end{array}
\end{equation}

\begin{center}
    \large\textbf{4. $\Delta=3$}
\end{center}

$N^4D^3$:
\begin{equation}
    \begin{array}{ll}
        D^2\bin{\gamma^\mu}\bin{\lrd_\mu} & \bin{\gamma^\mu\lrd^\nu}\bin{\lrd_\mu\lrd_\nu} 
    \end{array}
\end{equation}

\begin{center}
    \large\textbf{5. $\Delta=4$}
\end{center}
$N^4D^4$:
\begin{equation}
    \begin{array}{ll}
        \bin{}D^4\bin{} & {\color{gray}\bin{\gamma^5}D^4\bin{\gamma^5}} \\
        \bin{\sigma_{\mu\rho}}D^4\bin{\sigma^{\mu\rho}} & \bin{\sigma_{\nu\rho}}D^2\bin{\sigma^{\mu\rho}\lrd^\nu\lrd_\mu} \\
        \bin{\sigma_{\nu\rho}\lrd_\mu}D^2\bin{\sigma^{\mu\rho}\lrd^\nu} & \bin{\sigma_{\mu\nu}\lrd_\lambda}D^\mu\bin{\lrd^\nu\lrd^\lambda} \\
        \bin{\sigma_{\lambda\kappa}}D^2D_\mu\bin{\gamma^5\lrd_\nu}\epsilon^{\lambda\kappa\mu\nu} & \bin{\gamma^\mu}D^4\bin{\gamma_\mu} \\
        \bin{\gamma^5\gamma^\mu}D^4\bin{\gamma^5\gamma_\mu} & \bin{\gamma^\mu\lrd^\nu}D^2\bin{\gamma_\nu\lrd_\mu} \\
        \bin{\gamma^5\gamma^\mu\lrd^\nu}D^2\bin{\gamma^5\gamma_\nu\lrd_\mu} & \bin{\gamma^\mu\lrd^\nu\lrd^\lambda}\bin{\gamma_\mu\lrd_\nu\lrd_\lambda} \\
        \bin{\gamma^5\gamma^\mu\lrd^\nu\lrd^\lambda}\bin{\gamma^5\gamma_\mu\lrd_\nu\lrd_\lambda} & 
    \end{array}
\end{equation}

\begin{center}
    \large\textbf{6. $\Delta=5$}
\end{center}
$N^4D^5$:
\begin{equation}
    \begin{array}{ll}
        \bin{\gamma^\mu}D^4\bin{\lrd_\mu} & \bin{\gamma^\mu\lrd^\nu}D^2\bin{\lrd_\mu\lrd_\nu} \\
        \bin{\lrd_\nu}D^2D_\mu\bin{\gamma^5\gamma_\rho \lrd_\lambda}\epsilon^{\rho\lambda\mu\nu} & 
    \end{array}
\end{equation}

\begin{center}
    \large\textbf{7. $\Delta=6$}
\end{center}
$N^4D^6$:
\begin{equation}
\begin{array}{ll}
    \bin{}D^6\bin{} & {\color{gray}\bin{\gamma^5}D^6\bin{\gamma^5}} \\
    \bin{\lrd^\mu}D^4\bin{\lrd_\mu} & {\color{gray}\bin{\gamma^5\lrd^\mu}D^4\bin{\gamma^5\lrd_\mu}} \\
    \bin{\lrd^\mu\lrd^\nu}D^2\bin{\lrd_\mu\lrd_\nu} & {\color{gray}\bin{\gamma^5\lrd^\mu\lrd^\nu}D^2\bin{\gamma^5\lrd_\mu\lrd_\nu}} \\
    \bin{\lrd^\mu\lrd^\nu\lrd^\lambda}\bin{\lrd_\mu\lrd_\nu\lrd_\lambda} & {\color{gray}\bin{\gamma^5\lrd^\mu\lrd^\nu\lrd^\lambda}\bin{\gamma^5\lrd_\mu\lrd_\nu\lrd_\lambda}} \\
    \bin{\sigma_{\rho\kappa}\lrd_\mu\lrd_\nu\lrd_\lambda}\bin{\sigma^{\rho\kappa}\lrd^\mu\lrd^\nu\lrd^\lambda} & \bin{\sigma_{\mu\nu}}D^4\bin{\sigma^{\mu\nu}} \\
    \bin{\sigma_{\rho\kappa}\lrd_\mu\lrd_\nu}D^2\bin{\sigma^{\rho\kappa}\lrd^\mu\lrd^\nu} & \bin{\gamma^\mu}D^6\bin{\gamma_\mu} \\
    \bin{\gamma^5\gamma^\mu}D^6\bin{\gamma^5\gamma_\mu} & \bin{\gamma^\mu\lrd^\nu}D^4\bin{\gamma_\mu\lrd_\nu} \\
    \bin{\gamma^5\gamma^\mu\lrd^\nu}D^4\bin{\gamma^5\gamma_\mu\lrd_\nu} & \bin{\gamma^\mu\lrd^\nu\lrd^\lambda}D^2\bin{\gamma_\mu\lrd_\nu\lrd_\lambda} \\
    \bin{\gamma^5\gamma^\mu\lrd^\nu\lrd^\lambda}D^2\bin{\gamma^5\gamma_\mu\lrd_\nu\lrd_\lambda} & 
\end{array}
\end{equation}

\subsection{2 Nucleon-Meson Interaction Operators}
\label{subsec:nminteraction}
\begin{center}
    \large\textbf{1. $\Delta=1$}
\end{center}
$N^4u$:
\begin{equation}
    \begin{array}{ll}
{\color{gray}\bin{\gamma^\mu u_\mu}\bin{\gamma^5}} & \bin{\gamma^5\gamma^\mu u_\mu}\bin{}\\
\bin{\sigma_{\rho\lambda}}\bin{\gamma_\mu u_\nu}\epsilon^{\rho\lambda\mu\nu} & 
    \end{array}
\end{equation}

\begin{center}
    \large\textbf{2. $\Delta=2$}
\end{center}
$N^4uD$:
\begin{equation}
    \begin{array}{ll}
{\color{gray}\bin{\gamma^5}D_\mu\bin{u^\mu}} & {\color{gray}\bin{}D_\mu\bin{\gamma^5u^\mu}} \\
\bin{\gamma^5\gamma^\mu u_\mu\lrd^\nu}\bin{\gamma_\nu} & \bin{\gamma^\mu u_\mu\lrd^\nu}\bin{\gamma^5\gamma_\nu} \\
\bin{\gamma^\mu u^\nu}\bin{\gamma^5\gamma_\mu\lrd_\nu} & \bin{\gamma^5\gamma^\mu u^\nu}\bin{\gamma_\mu\lrd_\nu} \\
\bin{\sigma_{\rho\lambda}D_\mu u_\nu}\bin{{\sigma^\nu}_\kappa}\epsilon^{\rho\lambda\mu\kappa} & 
    \end{array}
\end{equation}

$N^4u^2$:
\begin{equation}
    \begin{array}{ll}
\bin{u_\mu u^\mu}\bin{} & \bin{u_\mu}\bin{u^\mu} \\
\bin{}\bin{}\tra{u_\mu u^\mu} & {\color{gray}\bin{\gamma^5u_\mu u^\mu}\bin{\gamma^5}} \\
{\color{gray}\bin{\gamma^5u_\mu}\bin{\gamma^5u^\mu}} & {\color{gray}\bin{\gamma^5}\bin{\gamma^5}\tra{u_\mu u^\mu}} \\
\bin{\gamma^\mu u^\nu}\bin{\gamma_\mu u_\nu} & \bin{\gamma^\mu}\bin{\gamma_\mu}\tra{u^\nu u_\nu} \\
\bin{\gamma^5\gamma^\mu u^\nu}\bin{\gamma^5\gamma_\mu u_\nu} & \bin{\gamma^5\gamma^\mu}\bin{\gamma^5\gamma_\mu}\tra{u^\nu u_\nu} \\
\bin{\gamma^\mu u^\nu}\bin{\gamma_\nu u_\mu} & \bin{\gamma^\mu}\bin{\gamma^\nu}\tra{u_\nu u_\mu} \\
\bin{\gamma^5\gamma^\mu u^\nu}\bin{\gamma^5\gamma_\nu u_\mu} & \bin{\gamma^5\gamma^\mu}\bin{\gamma^5\gamma^\nu}\tra{u_\mu u_\nu} \\
\bin{\gamma^\mu u_\mu u_\nu}\bin{\gamma^\nu} & \bin{\gamma^5\gamma^\mu u_\mu u_\nu}\bin{\gamma^5\gamma^\nu}
    \end{array}
\end{equation}

\begin{center}
    \large\textbf{3. $\Delta=3$}
\end{center}
$N^4uD^2$:
\begin{equation}
    \begin{array}{ll}
{\color{gray}\bin{\gamma^\mu u^\nu}D_\nu\bin{\gamma^5\lrd_\mu}} & {\color{gray}\bin{\gamma^\mu }D_\nu\bin{\gamma^5u^\nu\lrd_\mu}} \\
\bin{\gamma^5\gamma^\mu u_\mu}D^2\bin{} & \bin{\gamma^5\gamma^\mu u_\mu \lrd^\nu}\bin{\lrd_\nu} \\
\bin{\gamma^5\gamma^\mu u^\nu\lrd_\nu}\bin{\lrd_\mu} & \bin{\gamma^5\gamma^\mu }D^2\bin{u_\mu} \\
\bin{\sigma_{\mu\nu}}D^\mu\bin{\gamma^5\gamma^\rho u_\rho\lrd^\nu} & \bin{\sigma_{\mu\nu}\lrd^\nu}D^\mu\bin{\gamma^5\gamma^\rho u_\rho} \\
\bin{\sigma_{\mu\nu}u_\rho}D^\mu\bin{\gamma^5\gamma^\rho \lrd^\nu} & \bin{\sigma_{\alpha\beta}}D^2\bin{\gamma_\mu u_\rho}\epsilon^{\alpha\beta\mu\rho} \\ 
\bin{\sigma_{\alpha\beta}\lrd^\nu}\bin{\gamma_\mu u_\rho\lrd_\nu}\epsilon^{\alpha\beta\mu\rho} & \bin{\sigma_{\alpha\beta}u_\rho}D^2\bin{\gamma_\mu }\epsilon^{\alpha\beta\mu\rho} \\
\bin{\sigma_{\mu\nu}}D^\mu D^\rho\bin{\gamma^5\gamma_\rho u^\nu} & \bin{\sigma_{\mu\nu}\lrd^\rho} D^\mu\bin{\gamma^5\gamma_\rho u^\nu} \\
\bin{\sigma_{\mu\nu}u^\nu}D^\mu D^\rho\bin{\gamma^5\gamma_\rho} & \bin{\gamma^\mu D_\mu u_\nu}\bin{\gamma^5\lrd^\nu} \\
\bin{\gamma^5\gamma^\mu D_\mu u_\nu}D^\nu\bin{} & \bin{\gamma^\mu}\bin{\gamma^5 D_\mu u_\nu \lrd^\nu} 
    \end{array}
\end{equation}

\begin{center}
    \large\textbf{4. $\Delta=4$}
\end{center}
$N^4uD^3$:
\begin{equation}
    \begin{array}{ll}
{\color{gray}\bin{\gamma^5 u_\mu}D^2 D^\mu\bin{}} & {\color{gray}\bin{\gamma^5}D^2D^\mu\bin{u_\mu}} \\
{\color{gray}\bin{u_\mu}D^2D^\mu\bin{\gamma^5}} & {\color{gray}\bin{}D^2D^\mu\bin{\gamma^5u_\mu}} \\
{\color{gray}\bin{\gamma^5u_\mu\lrd^\nu}D^\mu\bin{\lrd_\nu}} & {\color{gray}\bin{\gamma^5\lrd^\nu}D^\mu\bin{u_\mu\lrd_\nu}} \\
{\color{gray}\bin{u_\mu\lrd^\nu}D^\mu\bin{\gamma^5\lrd_\nu}} & \bin{\gamma^5\gamma^\mu u_\nu}D^2\bin{\gamma_\nu\lrd^\nu} \\
\bin{\gamma^5\gamma^\mu}D^2\bin{\gamma_\mu u^\nu\lrd_\nu} & \bin{\gamma^\mu u_\nu}D^2\bin{\gamma^5\gamma_\mu\lrd_\nu} \\
\bin{\gamma^\mu}D^2\bin{\gamma^5\gamma_\mu u^\nu\lrd_\nu} & \bin{\gamma^5\gamma^\mu u_\nu\lrd^\rho}\bin{\gamma_\mu\lrd_\rho\lrd^\nu} \\
\bin{\gamma^5\gamma^\mu\lrd^\rho}\bin{\gamma_\mu u^\nu\lrd_\rho\lrd_\nu} & \bin{\gamma^\mu u_\nu\lrd^\rho}\bin{\gamma^5\gamma_\mu\lrd_\rho\lrd^\nu} \\
\bin{\gamma^\mu\lrd^\rho}\bin{\gamma^5\gamma_\mu u^\nu\lrd_\rho\lrd_\nu} & \bin{\gamma^5\gamma^\mu u_\nu\lrd^\nu}D^2\bin{\gamma_\mu} \\
\bin{\gamma^5\gamma^\mu\lrd^\nu}D^2\bin{\gamma_\mu u_\nu} & \bin{\gamma^\mu u_\nu\lrd^\nu}D^2\bin{\gamma^5\gamma_\mu} \\
\bin{\gamma^\mu\lrd^\nu}D^2\bin{\gamma^5\gamma_\mu u_\nu} & \bin{\gamma^5\gamma^\mu u_\mu\lrd^\nu}D^2\bin{\gamma_\nu} \\
\bin{\gamma^5\gamma^\mu\lrd^\nu}D^2\bin{\gamma_\nu u_\mu} & \bin{\gamma^\mu u_\mu\lrd^\nu}D^2\bin{\gamma^5\gamma_\nu} \\
\bin{\gamma^\mu\lrd^\nu}D^2\bin{\gamma^5\gamma_\nu u_\mu} & \bin{\gamma^5\gamma^\mu u_\mu\lrd^\nu\lrd^\rho}\bin{\gamma_\nu\lrd_\rho} \\
\bin{\gamma^5\gamma^\mu\lrd^\nu\lrd^\rho}\bin{\gamma_\nu u_\mu\lrd_\rho} & \bin{\sigma_{\mu\nu} u^\mu}D^2\bin{\gamma^5\lrd^\nu} \\
\bin{\sigma_{\mu\nu}}D^2\bin{\gamma^5 u^\mu\lrd^\nu} & \bin{\sigma_{\mu\nu}u^\mu\lrd^\rho}\bin{\gamma^5\lrd_\rho\lrd^\nu} \\
\bin{\sigma_{\mu\nu}\lrd^\rho}\bin{\gamma^5u^\mu\lrd_\rho\lrd^\nu} & \bin{\sigma_{\alpha\beta}u_\mu}D^2\bin{\lrd_\nu}\epsilon^{\alpha\beta\mu\nu} \\
\bin{\sigma_{\alpha\beta}}D^2\bin{u_\mu\lrd_\nu}\epsilon^{\alpha\beta\mu\nu} & \bin{\sigma_{\alpha\beta}\lrd^\rho}\bin{u_\mu\lrd_\rho\lrd_\nu}\epsilon^{\alpha\beta\mu\nu}
    \end{array}
\end{equation}

\subsection{3-Nucleon Contact Interaction Operators}
\label{subsec:nnninteraction}
\begin{center}
    \large\textbf{1. $\Delta=1$}
\end{center}

$N^6$:
\begin{equation}
    \begin{array}{ll}
\bin{}\bin{}\bin{} & {\color{gray}\bin{}\bin{\gamma^5}\bin{\gamma^5}} \\
\bin{}\bin{\gamma_\mu}\bin{\gamma^\mu} & \bin{}\bin{\gamma^5\gamma^\mu}\bin{\gamma^5\gamma_\mu} \\
\bin{}\bin{\sigma_{\mu\nu}}\bin{\sigma^{\mu\nu}} & 
    \end{array}
\end{equation}

\begin{center}
    \large\textbf{2. $\Delta=2$}
\end{center}
$N^6D$:
\begin{equation}
    \begin{array}{ll}
\bin{\gamma^\mu}\bin{\lrd_\mu}\bin{} & {\color{gray}\bin{\gamma^\mu}\bin{\gamma^5\lrd_\mu}\bin{\gamma^5}} \\
{\color{gray}\bin{\gamma^5\gamma^\mu} D_\mu\bin{\gamma^5}\bin{}} & {\color{gray}\bin{\gamma^5\gamma^\mu}\bin{\gamma^5\lrd^\nu}\bin{\sigma_{\mu\nu}}} \\
\bin{\gamma^\mu}\bin{\gamma^\nu\lrd_\mu}\bin{\gamma_\nu} & \bin{\gamma^5\gamma^\mu}\bin{\gamma^5\gamma^\nu\lrd_\mu}\bin{\gamma_\nu} \\
\bin{\gamma^5\gamma^\mu}\bin{\gamma^\nu\lrd_\mu}\bin{\gamma^5\gamma_\nu} & \bin{\gamma^\mu}\bin{\gamma^5\gamma^\nu\lrd_\mu}\bin{\gamma^5\gamma_\nu} \\
{\color{gray}\bin{\sigma_{\mu\nu}}\bin{\gamma^5\gamma_\mu}\bin{\gamma^5\lrd^\nu}} & {\color{gray}\bin{\sigma_{\mu\nu}}\bin{\gamma^5\gamma_\mu\lrd^\nu}\bin{\gamma^5}}
    \end{array}
\end{equation}

\begin{center}
    \large\textbf{3. $\Delta=3$}
\end{center}
$N^6D^2$:
\begin{equation}
    \begin{array}{ll}
\bin{}\bin{}D^2\bin{} & \bin{}\bin{\lrd_\mu}\bin{\lrd^\mu} \\
\bin{}\bin{\sigma_{\mu\nu}}D^2\bin{\sigma^{\mu\nu}} & \bin{}\bin{\sigma_{\mu\nu}\lrd^\lambda}\bin{\sigma^{\mu\nu}\lrd_\lambda} \\
\bin{\lrd^\lambda}\bin{\sigma_{\mu\nu}}\bin{\sigma^{\mu\nu}\lrd_\lambda} & {\color{gray}\bin{\gamma^5}\bin{\gamma^5}D^2\bin{}} \\
{\color{gray}\bin{}\bin{\gamma^5}D^2\bin{\gamma^5}} & {\color{gray}\bin{}\bin{\gamma^5\lrd_\mu}\bin{\gamma^5\lrd^\mu}} \\
{\color{gray}\bin{\gamma^5}\bin{\lrd_\mu}\bin{\gamma^5\lrd^\mu}} & {\color{gray}\bin{\gamma^5}\bin{\sigma_{\rho\kappa}}D^2\bin{\sigma_{\nu\lambda}}\epsilon^{\rho\kappa\nu\lambda}} \\
{\color{gray}D^2\bin{\gamma^5}\bin{\sigma_{\rho\kappa}}\bin{\sigma_{\nu\lambda}}\epsilon^{\rho\kappa\nu\lambda}} & {\color{gray}\bin{\gamma^5}\bin{\sigma_{\rho\kappa}\lrd^\mu}\bin{\sigma_{\nu\lambda}\lrd_\mu}\epsilon^{\rho\kappa\nu\lambda}} \\
{\color{gray}\bin{\gamma^5\lrd^\mu}\bin{\sigma_{\rho\kappa}}\bin{\sigma_{\nu\lambda}\lrd_\mu}\epsilon^{\rho\kappa\nu\lambda}} & \bin{}\bin{\gamma^\mu}D^2\bin{\gamma_\mu} \\
\bin{}\bin{\gamma^\mu\lrd^\nu}\bin{\gamma_\mu\lrd_\nu} & \bin{\lrd^\nu}\bin{\gamma^\mu}\bin{\gamma_\mu\lrd_\nu} \\
\bin{}\bin{\gamma^\mu\lrd^\nu}\bin{\gamma_\nu\lrd_\mu} & \bin{\lrd^\nu}\bin{\gamma^\mu}\bin{\gamma_\nu\lrd_\mu} \\
\bin{}\bin{\gamma^5\gamma^\mu\lrd^\nu}\bin{\gamma^5\gamma_\mu\lrd_\nu} & \bin{}\bin{\gamma^5\gamma^\mu\lrd^\nu}\bin{\gamma^5\gamma_\nu\lrd_\mu} \\
{\color{gray}\bin{\gamma^5}\bin{\gamma^\mu\lrd^\nu}D_\mu\bin{\gamma^5\gamma_\nu}} & {\color{gray}\bin{\gamma^5}D^\nu\bin{\gamma^\mu}\bin{\gamma^5\gamma_\nu\lrd_\mu}} \\
{\color{gray}\bin{\gamma^5\lrd^\nu}\bin{\gamma^5\gamma^\mu}D_\nu\bin{\gamma_\mu}} & {\color{gray}\bin{\gamma^5\lrd^\nu}\bin{\gamma^\mu}D_\nu\bin{\gamma^5\gamma_\mu}} \\
\bin{\lrd^\nu}\bin{\gamma^5\gamma^\mu}\bin{\gamma^5\gamma_\mu\lrd_\nu} & {\color{gray}D^\nu\bin{\gamma^5}\bin{\gamma^5\gamma^\mu}\bin{\gamma_\mu\lrd_\nu}} \\
{\color{gray}D^\nu\bin{\gamma^5}\bin{\gamma^\mu}\bin{\gamma^5\gamma_\mu\lrd_\nu}} & D^\nu\bin{}\bin{\gamma^5\gamma^\mu}D_\nu\bin{\gamma^5\gamma_\mu} \\
{\color{gray}\bin{\gamma^5\lrd^\nu}\bin{\gamma^5\gamma^\mu}D_\mu\bin{\gamma_\nu}} & {\color{gray}\bin{\gamma^5\lrd^\nu}\bin{\gamma^\mu}D_\mu\bin{\gamma^5\gamma_\nu}} \\
\bin{\lrd^\nu}\bin{\gamma^5\gamma^\mu}\bin{\gamma^5\gamma_\nu\lrd_\mu} & {\color{gray}D^\nu\bin{\gamma^5}\bin{\gamma^5\gamma^\mu}\bin{\gamma_\nu\lrd_\mu}} \\
{\color{gray}D^\nu\bin{\gamma^5}\bin{\gamma^\mu}\bin{\gamma^5\gamma_\nu\lrd_\mu}} & D^\nu\bin{}\bin{\gamma^5\gamma^\mu}D_\mu\bin{\gamma^5\gamma_\nu} \\
\bin{\sigma^{\mu\nu}}D^\lambda\bin{\gamma_\mu}\bin{\gamma_\nu\lrd_\lambda} & \bin{\sigma^{\mu\nu}}D^\lambda\bin{\gamma^5\gamma_\mu}\bin{\gamma^5\gamma_\nu\lrd_\lambda} \\
\bin{\sigma_{\rho\kappa}}D^2\bin{\gamma^5\gamma_\mu}\bin{\gamma_\nu}\epsilon^{\rho\kappa\mu\nu} & \bin{\sigma_{\rho\kappa}}D^\lambda\bin{\gamma^5\gamma_\mu}D_\lambda\bin{\gamma_\nu}\epsilon^{\rho\kappa\mu\nu} \\
\bin{\sigma^{\mu\nu}}D^\lambda\bin{\gamma_\mu}\bin{\gamma_\lambda\lrd_\nu} & 
    \end{array}
\end{equation}

\begin{center}
    \large\textbf{4. $\Delta=4$}
\end{center}
\paragraph{$N^6D^3$:}
\begin{equation}
    \begin{array}{ll}
D^\mu D^\nu\bin{}\bin{}\bin{\gamma_\mu\lrd_\nu} & {\color{gray}D^\mu D^\nu\bin{\gamma^5}\bin{\gamma^5}\bin{\gamma_\mu\lrd_\nu}} \\
{\color{gray}D^\mu\bin{\gamma^5\lrd^\nu}\bin{}\bin{\gamma^5\gamma_\mu\lrd_\nu}} & {\color{gray}D^\mu\bin{\lrd^\nu}\bin{\gamma^5}\bin{\gamma^5\gamma_\mu\lrd_\nu}} \\
\bin{\lrd^\mu}\bin{\lrd^\nu}\bin{\gamma_\mu\lrd_\nu} & {\color{gray}\bin{\gamma^5\lrd^\mu}\bin{\gamma^5\lrd^\nu}\bin{\gamma_\mu\lrd_\nu}} \\
{\color{gray}\bin{\gamma^5\lrd^\mu}\bin{\lrd^\nu}D_\nu\bin{\gamma^5\gamma_\mu}} & {\color{gray}\bin{\lrd^\mu}\bin{\gamma^5\lrd^\nu}D_\nu\bin{\gamma^5\gamma_\mu}} \\
D^\mu D^\nu \bin{\sigma_{\rho\lambda}}\bin{\sigma^{\rho\lambda}}\bin{\gamma_\mu\lrd_\nu} & D^\mu\bin{\sigma_{\alpha\beta}\lrd^\nu}\bin{\sigma_{\rho\lambda}}\bin{\gamma^5\gamma_\mu\lrd_\nu}\epsilon^{\alpha\beta\rho\lambda} \\
D_\mu D^\nu\bin{\sigma_{\rho\lambda}}\bin{\sigma^{\mu\rho}}\bin{\gamma^\lambda\lrd_\nu} & D^\mu \bin{\sigma_{\alpha\beta}\lrd^\nu}\bin{\sigma_{\mu\rho}}\bin{\gamma^5\gamma_\lambda\lrd_\nu}\epsilon^{\alpha\beta\rho\lambda} \\
D_\mu \bin{\sigma_{\rho\lambda}\lrd^\nu}\bin{\sigma_{\alpha\beta}}\bin{\gamma^5\gamma^\lambda\lrd_\nu}\epsilon^{\alpha\beta\mu\rho} & D_\mu D^\nu\bin{\sigma_{\rho\lambda}}\bin{\sigma^{\mu\rho}\lrd_\nu}\bin{\gamma^\lambda} \\
D^\mu \bin{\sigma_{\alpha\beta}\lrd^\nu}\bin{\sigma_{\mu\rho}\lrd_\nu}\bin{\gamma^5\gamma_\lambda}\epsilon^{\alpha\beta\rho\lambda} & D_\mu \bin{\sigma_{\rho\lambda}\lrd^\nu}\bin{\sigma_{\alpha\beta}\lrd_\nu}\bin{\gamma^5\gamma^\lambda}\epsilon^{\alpha\beta\mu\rho} \\
\bin{\sigma_{\rho\lambda}}\bin{\sigma^{\mu\rho}\lrd^\nu}D_\mu D_\nu\bin{\gamma^\lambda} & \bin{\sigma_{\alpha\beta}}\bin{\sigma_{\mu\rho}\lrd^\nu}D^\mu\bin{\gamma^5\gamma_\lambda\lrd_\nu}\epsilon^{\alpha\beta\rho\lambda} \\
\bin{\sigma_{\rho\lambda}}\bin{\sigma_{\alpha\beta}\lrd^\nu}D_\mu\bin{\gamma^5\gamma^\lambda\lrd_\nu}\epsilon^{\alpha\beta\mu\rho} & \bin{\sigma_{\mu\nu}}\bin{\sigma_{\lambda\rho}\lrd^\mu}D^\lambda D^\nu\bin{\gamma^\rho} \\
\bin{\sigma_{\alpha\beta}}\bin{\sigma_{\lambda\rho}\lrd_\mu}D^\lambda\bin{\gamma^5\gamma_\rho\lrd_\nu}\epsilon^{\alpha\beta\mu\nu} & \bin{\sigma_{\mu\nu}}\bin{\sigma_{\alpha\beta}\lrd^\mu}D_\lambda\bin{\gamma^5\gamma_\rho\lrd^\nu}\epsilon^{\alpha\beta\lambda\rho} \\
D^\mu D^\nu\bin{\sigma_{\mu\rho}}\bin{}D_\nu\bin{\gamma^\rho} & D_\mu {\color{gray}D^\nu\bin{\sigma_{\alpha\beta}}\bin{\gamma^5}D_\nu\bin{\gamma_\rho}\epsilon^{\alpha\beta\rho\mu}} \\
D_\mu D^\nu\bin{\sigma_{\alpha\beta}}\bin{}\bin{\gamma^5\gamma_\rho\lrd_\nu}\epsilon^{\alpha\beta\rho\mu} & D^\mu {\color{gray}D^\nu\bin{\sigma_{\mu\rho}}\bin{\gamma^5}\bin{\gamma^5\gamma^\rho\lrd_\nu}} \\
\bin{\sigma_{\mu\rho}\lrd^\nu}\bin{}D^\mu\bin{\gamma^\rho\lrd_\nu} & {\color{gray}\bin{\sigma_{\mu\rho}\lrd^\nu}\bin{\gamma^5}\bin{\gamma^5\gamma^\rho\lrd^\mu\lrd_\nu}} \\
\bin{\lrd^\mu\lrd^\nu}\bin{}\bin{\gamma_\mu\lrd_\nu} & {\color{gray}\bin{\gamma^5\lrd^\mu\lrd^\nu}\bin{\gamma^5}\bin{\gamma_\mu\lrd_\nu}} \\
\bin{\lrd^\mu\lrd^\nu}\bin{\lrd_\nu}\bin{\gamma_\mu} & {\color{gray}\bin{\gamma^5\lrd^\mu\lrd^\nu}\bin{\gamma^5\lrd_\nu}\bin{\gamma_\mu}} \\
{\color{gray}D^\mu\bin{\gamma^5\lrd^\nu}\bin{\lrd_\nu}\bin{\gamma^5\gamma_\mu}} & {\color{gray}D^\mu\bin{\lrd^\nu}\bin{\gamma^5\lrd_\nu}\bin{\gamma^5\gamma_\mu}} \\
\bin{\sigma_{\nu\lambda}\lrd^\mu}\bin{}D^\lambda\bin{\gamma_\mu\lrd^\nu} & {\color{gray}\bin{\sigma_{\alpha\beta}\lrd^\mu}\bin{\gamma^5}D_\lambda\bin{\gamma_\mu\lrd_\nu}\epsilon^{\alpha\beta\nu\lambda}} \\
\bin{\sigma_{\nu\lambda}}\bin{\lrd^\mu}D^\lambda\bin{\gamma_\mu\lrd^\nu} & {\color{gray}\bin{\sigma_{\alpha\beta}}\bin{\gamma^5\lrd^\mu}D_\lambda\bin{\gamma_\mu\lrd_\nu}\epsilon^{\alpha\beta\nu\lambda}} \\
\bin{\sigma_{\nu\lambda}}D^\lambda\bin{\lrd^\mu}\bin{\gamma_\mu\lrd_\nu} & {\color{gray}\bin{\sigma_{\alpha\beta}}D_\lambda\bin{\gamma^5\lrd^\mu}\bin{\gamma_\mu\lrd_\nu}\epsilon^{\alpha\beta\nu\lambda}} \\
\bin{\sigma_{\alpha\beta}}\bin{\lrd^\mu\lrd_\lambda}\bin{\gamma^5\gamma_\mu\lrd_\nu}\epsilon^{\alpha\beta\nu\lambda} & {\color{gray}\bin{\sigma_{\nu\lambda}}\bin{\gamma^5\lrd^\nu\lrd^\lambda}\bin{\gamma^5\gamma_\mu\lrd_\nu}} \\
D^\mu\bin{\sigma_{\mu\nu}}\bin{\lrd^\lambda}\bin{\gamma_\lambda\lrd^\nu} & {\color{gray}D_\mu\bin{\sigma_{\alpha\beta}}\bin{\gamma^5\lrd^\lambda}\bin{\gamma_\lambda\lrd_\nu}\epsilon^{\alpha\beta\mu\nu}} \\
\bin{\sigma_{\alpha\beta}\lrd_\mu}\bin{\lrd^\lambda}\bin{\gamma^5\gamma_\lambda\lrd_\nu}\epsilon^{\alpha\beta\mu\nu} & {\color{gray}\bin{\sigma_{\mu\nu}\lrd^\mu}\bin{\gamma^5\lrd^\lambda}\bin{\gamma^5\gamma_\lambda\lrd^\nu}} \\
\bin{\sigma_{\nu\rho}}D_\lambda D^\mu\bin{\sigma^{\lambda\rho}}\bin{\gamma_\mu\lrd^\nu} & \bin{\sigma_{\mu\nu}}D^\mu\bin{\sigma_{\lambda\rho}}D^\lambda\bin{\gamma^\rho\lrd^\nu} \\
\bin{\sigma_{\alpha\beta}}D_\mu\bin{\sigma_{\lambda\rho}}\bin{\gamma^5\gamma^\rho\lrd^\lambda\lrd_\nu}\epsilon^{\mu\nu\alpha\beta} & \bin{\sigma_{\mu\rho}\lrd^\nu}\bin{\sigma^{\rho\lambda}\lrd_\nu}\bin{\gamma_\lambda\lrd^\mu} \\
\bin{\sigma_{\alpha\beta}\lrd^\nu}\bin{\sigma_{\rho\lambda}\lrd_\nu}D_\mu\bin{\gamma^5\gamma^\lambda}\epsilon^{\alpha\beta\mu\rho} & \bin{\sigma_{\mu\rho}\lrd^\nu}\bin{\sigma_{\alpha\beta}\lrd_\nu}D^\mu\bin{\gamma^5\gamma_\lambda}\epsilon^{\alpha\beta\rho\lambda} \\
\bin{\sigma_{\lambda\mu}\lrd^\mu}\bin{\sigma_{\nu\rho}\lrd^\nu}\bin{\gamma^\rho\lrd^\lambda} & \bin{\sigma_{\alpha\beta}\lrd_\mu}\bin{\sigma_{\nu\rho}\lrd^\nu}D_\lambda\bin{\gamma^5\gamma^\rho}\epsilon^{\alpha\beta\lambda\mu} \\
\bin{\sigma_{\lambda\mu}\lrd^\mu}\bin{\sigma_{\alpha\beta}\lrd_\nu}D^\lambda\bin{\gamma^5\gamma_\rho}\epsilon^{\alpha\beta\nu\rho} & D^\nu\bin{}\bin{\lrd^\mu}D_\nu\bin{\gamma_\mu} \\
{\color{gray}D^\nu\bin{\gamma^5}\bin{\gamma^5\lrd^\mu}D_\nu\bin{\gamma_\mu}} & {\color{gray}D^\nu\bin{\gamma^5}D^\mu\bin{}D_\nu\bin{\gamma^5\gamma_\mu}} \\
{\color{gray}D^\nu\bin{}D^\mu\bin{\gamma^5}D_\nu\bin{\gamma^5\gamma_\mu}} & D^2\bin{}\bin{\lrd^\mu}\bin{\gamma_\mu} \\
{\color{gray}D^2\bin{\gamma^5}\bin{\gamma^5\lrd^\mu}\bin{\gamma_\mu}} & {\color{gray}D^2\bin{\gamma^5}D^\mu\bin{}\bin{\gamma^5\gamma_\mu}} \\
{\color{gray}D^2\bin{}D^\mu\bin{\gamma^5}\bin{\gamma^5\gamma_\mu}} & \bin{\gamma^\mu}D^\nu\bin{\gamma^\rho}D_\nu\bin{\gamma_\rho\lrd_\mu} \\
\bin{\gamma^5\gamma^\mu}D^\nu\bin{\gamma^5\gamma^\rho}D_\nu\bin{\gamma_\rho\lrd_\mu} & \bin{\gamma^5\gamma^\mu}D^\nu\bin{\gamma^\rho}D_\nu\bin{\gamma^5\gamma_\rho\lrd_\mu} \\
\bin{\gamma^\mu}D^\nu\bin{\gamma^5\gamma^\rho}D_\nu\bin{\gamma^5\gamma_\rho\lrd_\mu} & \bin{\gamma^\mu}\bin{\gamma^\rho\lrd^\nu}\bin{\gamma_\rho\lrd_\nu\lrd_\mu} \\
\bin{\gamma^5\gamma^\mu}\bin{\gamma^5\gamma^\rho\lrd^\nu}\bin{\gamma_\rho\lrd_\nu\lrd_\mu} & \bin{\gamma^5\gamma^\mu}\bin{\gamma^\rho\lrd^\nu}\bin{\gamma^5\gamma_\rho\lrd_\nu\lrd_\mu} \\
\bin{\gamma^\mu}\bin{\gamma^5\gamma^\rho\lrd^\nu}\bin{\gamma^5\gamma_\rho\lrd_\nu\lrd_\mu} & D^\nu\bin{\gamma^\mu}\bin{\gamma^\rho}D_\nu\bin{\gamma_\rho\lrd_\mu} \\
D^\nu\bin{\gamma^5\gamma^\mu}\bin{\gamma^5\gamma^\rho}D_\nu\bin{\gamma_\rho\lrd_\mu} & D^\nu\bin{\gamma^5\gamma^\mu}\bin{\gamma^\rho}D_\nu\bin{\gamma^5\gamma_\rho\lrd_\mu} \\
D^\nu\bin{\gamma^\mu}\bin{\gamma^5\gamma^\rho}D_\nu\bin{\gamma^5\gamma_\rho\lrd_\mu} & \bin{\gamma^\mu}\bin{\gamma^\rho}D^2\bin{\gamma_\rho\lrd_\mu} \\
\bin{\gamma^5\gamma^\mu}\bin{\gamma^\rho}D^2\bin{\gamma^5\gamma_\rho\lrd_\mu} & \bin{\gamma^\mu}\bin{\gamma^5\gamma^\rho}D^2\bin{\gamma^5\gamma_\rho\lrd_\mu} \\
D^2\bin{\gamma^\mu}\bin{\gamma^\rho}\bin{\gamma_\rho\lrd_\mu} & D^2\bin{\gamma^5\gamma^\mu}\bin{\gamma^5\gamma^\rho}\bin{\gamma_\rho\lrd_\mu} \\
D^2\bin{\gamma^5\gamma^\mu}\bin{\gamma^\rho}\bin{\gamma^5\gamma_\rho\lrd_\mu} & 
     \end{array}
\end{equation}

\section{Beyond External Sources: Leptonic Chiral Lagrangian}
\label{sec:weakoperators}


The interactions of the nucleons and the leptons are of importance in the low-energy probe of the new physics effects, for example, the beta decay~\cite{Cirigliano:2002ng,Cirigliano:2022hob,Cirigliano:2023fnz,Cirigliano:2024rfk}, the neutrinoless double beta decay~\cite{Savage:1998yh,Prezeau:2003xn,Graesser:2016bpz,Cirigliano:2017ymo,Cirigliano:2017djv,Cirigliano:2017tvr,Pastore:2017ofx,Cirigliano:2018yza,Cirigliano:2019vdj,Dekens:2024eae}, the coherent neutrino-nucleus scattering~\cite{Altmannshofer:2018xyo,Hoferichter:2020osn,Du:2021idh,Breso-Pla:2023tnz,Kouzakov:2024xnq}, and so on. In this section, we shall extend the ChEFT to include such interactions by the spurion method, in which the leptons are added as its new degrees of freedom, and some spurions are needed. 

\subsection{The Spurion Method of The ChEFT}

In Sec.~\ref{sec:ccwz} we have introduced the traditional external source method of the ChEFT. Some external sources $\chi\,,\chi^\dagger\,,l_\mu\,,r_\mu$ of the QCD Lagrangian are presented.
Considering the interactions between the hadrons and the leptons, these external sources are identified with the lepton currents as~\cite{Bishara:2016hek} 
\begin{align}
    \chi &= T_\chi (C_1\overline{e}e + C_2\overline{e}i\gamma^5 e) \,,\label{eq:ext_1_int}\\
    \chi^\dagger &= T_\chi^\dagger(C_1\overline{e}e + C_2\overline{e}i\gamma^5 e) \\
    l^\mu &= T_l(C_3\overline{e}\gamma^\mu e + C_4 \overline{e}\gamma^5\gamma^\mu e)  \,, \\
    r^\mu &= T_r (C_5\overline{e}\gamma^\mu e + C_6 \overline{e}\gamma^5\gamma^\mu e) \,, \label{eq:ext_2_int}
\end{align}
where $T$'s are the Wilson coefficients associated with the quark bilinears in Eq.~\eqref{eq:QCD_ext} in the $SU(2)_V$ space, and $C_i$'s are the coefficients of the leptonic currents to indicate their independence. As discussed before, these external sources are embedded in the ChEFT building blocks that
\begin{align}
    \chi_\pm &= \xi^{-1} \chi \xi^{-1} \pm \xi \chi^\dagger \xi\,, \label{eq:ext_cheft_1}\\
    u_\mu &= i\left[\xi^{-1}(\partial_\mu-il_\mu)\xi-\xi(\partial_\mu-ir_\mu)\xi^{-1}\right]\,.\label{eq:ext_cheft_2}
\end{align}
Considering the higher-dimension operators, on the one hand, more leptonic currents of the scalar or vector form appear, which extend the external sources above, and make the leptonic currents of different orders mix in the ChEFT operators. For example, the bilinear $(\overline{e}\lrd^\mu e)$ is though of vector form, it is of higher order compared to the $(\overline{e}\gamma^\mu e)$ according to the power-counting of the QCD Lagrangian, thus their contributions to the ChEFT should be distinguishable rather than unified in a single external source $l_\mu$ or $r_\mu$.
on the other hand, the leptonic currents of other forms beyond the ones in Eq.~\eqref{eq:ext_1_int} to Eq.~\eqref{eq:ext_2_int} could appear, which means these external sources are not adequate. For example, the tensor bilinears $(\overline{e}\sigma^{\mu\nu}e)\,,(\overline{e}\gamma^\mu \lrd^\nu e)$, and so on can appear in the high-dimension QCD Lagrangian, and can not be included in the external sources in Eq.~\eqref{eq:ext_1_int} to Eq.~\eqref{eq:ext_2_int}.

Instead of considering the complicated external sources, we adopt the spurion method. We regard the leptons as the building blocks of the ChEFT. At the same time, we introduce two new building blocks, which are of different dressing forms of the $SU(2)_V$ Wilson coefficients in the QCD Lagrangian,
\begin{align}
    \Sigma_{\pm} &= u^\dagger T u^\dagger \pm u T^\dagger u \,,\notag\\
    Q_{\pm}&=u^\dagger T u\pm u T^\dagger u^\dagger\,,
\end{align}
which is formally covariant under the $SU(2)_V$ symmetry, 
\begin{equation}
    \Sigma_\pm \rightarrow h \Sigma_\pm h^{-1}\,,\quad Q_\pm \rightarrow h Q_\pm h^\dagger \,,\quad h \in SU(2)_V\,,
\end{equation}
and are referred to as two spurions.
In summary, we list the building blocks of the ChEFT in the spurion method in Tab.~\ref{tab:building_blocks_spurion}. In particular, the $CP$ properties are also presented.

The parameterizations in terms of the spurions and the external sources of the ChEFT are equivalent. The external sources in Eq.~\eqref{eq:ext_cheft_1} and Eq.~\eqref{eq:ext_cheft_2} can be expressed by the spurions that
\begin{align}
    \chi_+ &\sim \Sigma_+ (C_1\overline{e}e + C_2\overline{e}i\gamma^5 e)\,, \\
    \chi_- &\sim \Sigma_- (C_1\overline{e}e + C_2\overline{e}i\gamma^5 e)\,, \\
    u_\mu &\sim i\left(\xi^{-1}\partial_\mu\xi-\xi\partial_\mu\xi^{-1}\right) + \frac{1}{2} Q_+ \left[(C_3+C_5)\overline{e}\gamma^\mu e + (C_4+C_6)\overline{e}\gamma^5\gamma^\mu e\right] \notag \\
    &+ \frac{1}{2}Q_-\left[(C_3-C_5)\overline{e}\gamma^\mu e + (C_4-C_6)\overline{e}\gamma^5\gamma^\mu e\right]\,.
\end{align}
Thus the spurions and the leptonic currents are extracted from the ChEFT building blocks. In particular, such an extraction makes the meson building block $u_\mu$ take the form without the external sources in Eq.~\eqref{eq:chptu}.

\begin{table}[t]
\renewcommand{\arraystretch}{1.5}
    \centering
    \begin{tabular}{|c|c|c|c|}
\hline
building block & $SU(2)_V$ & C & P \\
\hline
\multicolumn{4}{|c|}{meson} \\
\hline
$u_\mu$ & $\mathbf{3}$ & + & - \\
\hline
\multicolumn{4}{|c|}{spurions} \\
\hline
$\Sigma_\pm$ & $\mathbf{3}$ & + & $\pm$ \\
\hline
$Q_\pm$ & $\mathbf{3}$ & $\pm$ & $\pm$ \\
\hline
\multicolumn{4}{|c|}{nucleon} \\
\hline
$N$ & $\mathbf{2}$ & \multicolumn{2}{c|}{Tab.~\ref{tab:buildingblocks}} \\
\hline
\multicolumn{4}{|c|}{leptons} \\
\hline
$e_L\,,e_R\,,\nu_L$ & $\mathbf{1}$ & \multicolumn{2}{c|}{Tab.~\ref{tab:buildingblocks}} \\
\hline
    \end{tabular}
    \caption{The building blocks and their $SU(2)_V$ representations of the ChEFT in the spurion method. The $CP$ properties have been presented, and the spurions $\Sigma_\pm\,,Q_\pm$ are of definite $CP$ eigenvalues,}
    \label{tab:building_blocks_spurion}
\end{table}

\subsubsection*{The Young Tensor Method of Spurions}

The spurion is special in both the Lorentz and the internal structures,
\begin{itemize}
    \item In the Lorentz sector, the spurions can not be applied by the derivatives, since they are actually frozen degrees of freedom.
    \item In the internal sector, the self-contractions of the spurions are not independent, since they contribute just an overall constant. 
\end{itemize}
These properties require extra management in the Young tensor method discussed in Sec.~\ref{sec:young}. A similar manipulation in the Higgs effective field theory has been discussed~\cite{Sun:2022ssa,Sun:2022snw}, and we summarize the key points here.  

To ensure the derivatives do not apply to the spurions, we treat the operator types as if there is no spurion in the Lorentz sector. Because the spurions are scalar fields, dropping them does not affect the Lorentz structures. In the internal sector, we can extract the spurions, which should be symmetrical in the outer product since they can not contract with each other. For example, the spurion $\Sigma_\pm$ is of the adjoint representation of the $SU(2)_V$,
\begin{equation}
    \Sigma_\pm \sim \ydiagram{2}\,,
\end{equation}
and the symmetrical outer product of $n$ $\Sigma_\pm$ is a single-row Young diagram
\begin{center}
\begin{tikzpicture}
\draw (0pt,0pt) rectangle (14pt,14pt);
\draw (14pt,0pt) rectangle (28pt,14pt);
\draw [loosely dotted] (30pt,7pt)--(40pt,7pt);
\draw (42pt,0pt) rectangle (56pt,14pt);
\draw (56pt,0pt) rectangle (70pt,14pt);
\draw [|<-] (0pt,21pt) -- (28pt,21pt);
\draw [|<-] (70pt,21pt) -- (42pt,21pt);
\node (2n) at (35pt,21pt) {$2n$};
\end{tikzpicture}
\end{center}
As for the other fields, their outer product should be of the form
\begin{center}
    \begin{tikzpicture}
\draw (0pt,0pt) rectangle (14pt,14pt);
\draw (14pt,0pt) rectangle (28pt,14pt);
\draw [loosely dotted] (30pt,7pt)--(40pt,7pt);
\draw (42pt,0pt) rectangle (56pt,14pt);
\draw (0pt,14pt) rectangle (14pt,28pt);
\draw (14pt,14pt) rectangle (28pt,28pt);
\draw [loosely dotted] (30pt,21pt)--(40pt,21pt);
\draw (42pt,14pt) rectangle (56pt,28pt);
\draw (56pt,14pt) rectangle (70pt,28pt);
\draw (70pt,14pt) rectangle (84pt,28pt);
\draw [loosely dotted] (86pt,21pt)--(96pt,21pt);
\draw (98pt,14pt) rectangle (112pt,28pt);
\draw (112pt,14pt) rectangle (126pt,28pt);
\draw [|<-] (56pt,35pt) -- (84pt,35pt);
\draw [|<-] (126pt,35pt) -- (98pt,35pt);
\node (2j) at (91pt,35pt) {$2n$};
    \end{tikzpicture}
\end{center}
so that they can form a trivial representation. 

Utilizing the improvements, we can construct the leptonic operators in the sector $\mathcal{L}_{\text{lepton}}$ by the Young tensor method. Subsequently, we will present the effective operators of the nucleon-lepton sector, the meson-lepton sector, and the lepton number violating operators contributing to the neutrinoless double beta decay.

\subsection{Nucleon-Lepton Interactions}

In this subsection, we follow the power-counting scheme similar to the $\mathcal{L}_{\pi N}$ sector, where the spurions are of $p^2$ order, and present the effective operators up to $p^4$ order.


The LO operators are of order $p^2$:
\begin{align}
\mathcal{O}(p^2): & \notag \\
    & \begin{array}{ll}
\bin{}(\overline{e}e)\,, & \bin{}(\overline{e}\gamma^5 e)\,, \\
\bin{\Sigma_\pm}(\overline{e}e)\,, & \bin{\Sigma_\pm} (\overline{e}\gamma^5 e)\,, \\
\bin{\gamma^5}(\overline{e}e)\,, & \bin{\gamma^5}(\overline{e}\gamma^5 e)\,, \\
\bin{\gamma^5\Sigma_\pm}(\overline{e}e)\,, & \bin{\gamma^5\Sigma_\pm} (\overline{e}\gamma^5 e)\,, \\
\bin{\Sigma_\pm}(\overline{e}\nu_L)\,, & \bin{\Sigma_\pm}(\overline{\nu}_L e)\,, \\
\bin{\gamma^5\Sigma_\pm}(\overline{e}\nu_L)\,, & \bin{\gamma^5\Sigma_\pm}(\overline{\nu}_L e)\,, \\
\bin{\gamma^\mu}(\overline{e}\gamma_\mu e)\,, & \bin{\gamma^\mu}(\overline{e}\gamma^5\gamma_\mu e)\,, \\
\bin{\gamma^\mu Q_\pm}(\overline{e}\gamma_\mu e)\,, & \bin{\gamma^\mu Q_\pm}(\overline{e}\gamma^5\gamma_\mu e)\,, \\
\bin{\gamma^5\gamma^\mu}(\overline{e}\gamma_\mu e)\,, & \bin{\gamma^5\gamma^\mu}(\overline{e}\gamma^5\gamma_\mu e)\,, \\
\bin{\gamma^5\gamma^\mu Q_\pm}(\overline{e}\gamma_\mu e)\,, & \bin{\gamma^5\gamma^\mu Q_\pm}(\overline{e}\gamma^5\gamma_\mu e)\,, \\
\bin{\gamma^\mu}(\overline{\nu}_L\gamma_\mu \nu_L)\,, & \bin{\gamma^5\gamma^\mu}(\overline{\nu}_L\gamma_\mu \nu_L)\,, \\
\bin{\gamma^\mu Q_\pm}(\overline{\nu}_L\gamma_\mu \nu_L)\,, & \bin{\gamma^5\gamma^\mu Q_\pm}(\overline{\nu}_L\gamma_\mu \nu_L)\,, \\
\bin{\gamma^\mu Q_\pm}(\overline{e}\gamma_\mu \nu_L)\,, & \bin{\gamma^\mu Q_\pm}(\overline{\nu}_L\gamma_\mu e)\,, \\
\bin{\gamma^5\gamma^\mu Q_\pm}(\overline{e}\gamma_\mu \nu_L)\,, & \bin{\gamma^5\gamma^\mu Q_\pm}(\overline{\nu}_L\gamma_\mu e)\,, \\
\bin{\sigma^{\mu\nu}}(\overline{e}\sigma_{\mu\nu} e)\,, & \bin{\sigma^{\mu\nu}}(\overline{e}\gamma^5\sigma_{\mu\nu} e)\,, \\
\bin{\sigma^{\mu\nu}\Sigma_\pm}(\overline{e}\sigma_{\mu\nu} e)\,, & \bin{\sigma^{\mu\nu}\Sigma_\pm}(\overline{e}\gamma^5\sigma_{\mu\nu} e)\,, \\
\bin{\sigma^{\mu\nu}\Sigma_\pm}(\overline{e}\sigma_{\mu\nu} \nu_L)\,, & \bin{\sigma^{\mu\nu}\Sigma_\pm}(\overline{e}\gamma^5\sigma_{\mu\nu} \nu_L)\,, \\
\bin{\sigma^{\mu\nu}}(\overline{\nu}_L\sigma_{\mu\nu} e)\,, & \bin{\sigma^{\mu\nu}}(\overline{\nu}_L\gamma^5\sigma_{\mu\nu} e)\,, \\
\bin{\sigma^{\mu\nu}\Sigma_\pm}(\overline{\nu}_L\sigma_{\mu\nu} e)\,, & \bin{\sigma^{\mu\nu}\Sigma_\pm}(\overline{\nu}_L\gamma^5\sigma_{\mu\nu} e)\,, 
    \end{array} \\
    & \begin{array}{ll}
\bin{\lrd^\mu}(\overline{e}\gamma_\mu e)\,, & \bin{\lrd^\mu}(\overline{e}\gamma^5\gamma_\mu e)\,, \\
\bin{\Sigma_\pm\lrd^\mu}(\overline{e}\gamma_\mu e)\,, & \bin{\Sigma_\pm\lrd^\mu}(\overline{e}\gamma^5\gamma_\mu e)\,, \\
\bin{\gamma^5\lrd^\mu}(\overline{e}\gamma_\mu e)\,, & \bin{\gamma^5\lrd^\mu}(\overline{e}\gamma^5\gamma_\mu e)\,, \\
\bin{\gamma^5\Sigma_\pm\lrd^\mu}(\overline{e}\gamma_\mu e)\,, & \bin{\gamma^5\Sigma_\pm\lrd^\mu}(\overline{e}\gamma^5\gamma_\mu e)\,, \\
\bin{\lrd^\mu}(\overline{\nu}_L\gamma_\mu \nu_L)\,, & \bin{\gamma^5\lrd^\mu}(\overline{\nu}_L\gamma_\mu \nu_L)\,, \\
\bin{\Sigma_\pm\lrd^\mu}(\overline{\nu}_L\gamma_\mu \nu_L)\,, & \bin{\gamma^5\Sigma_\pm\lrd^\mu}(\overline{\nu}_L\gamma_\mu \nu_L)\,, \\
\bin{Q_\pm\lrd^\mu}(\overline{e}\gamma_\mu \nu_L)\,, & \bin{Q_\pm\lrd^\mu}(\overline{\nu}_L\gamma_\mu e)\,.
    \end{array}
\end{align}

The $p^3$ order operators containing 1 derivative applying on the leptons are 
\begin{align}
    \mathcal{O}(p^3): & \notag \\
    & \begin{array}{ll}
\bin{\gamma^\mu}(\overline{e}\lrd_\mu e)\,, & \bin{\gamma^\mu}(\overline{e}\gamma^5\lrd_\mu e)\,, \\
\bin{\gamma^\mu Q_\pm}(\overline{e}\lrd_\mu e)\,, & \bin{\gamma^\mu Q_\pm}(\overline{e}\gamma^5\lrd_\mu e)\,, \\
\bin{\gamma^5\gamma^\mu}(\overline{e}\lrd_\mu e)\,, & \bin{\gamma^5\gamma^\mu}(\overline{e}\gamma^5\lrd_\mu e)\,, \\
\bin{\gamma^5\gamma^\mu Q_\pm}(\overline{e}\lrd_\mu e)\,, & \bin{\gamma^5\gamma^\mu Q_\pm}(\overline{e}\gamma^5\lrd_\mu e)\,, \\
\bin{\gamma^\mu Q_\pm}(\overline{e}\lrd_\mu \nu_L)\,, & \bin{\gamma^\mu Q_\pm}(\overline{\nu}_L\lrd_\mu e)\,, \\
\bin{\gamma^5\gamma^\mu Q_\pm}(\overline{e}\lrd_\mu \nu_L)\,, & \bin{\gamma^5\gamma^\mu Q_\pm}(\overline{\nu}_L\lrd_\mu e)\,, \\
\bin{\gamma^\mu\lrd^\nu}(\overline{e}\gamma_\mu\lrd_\nu e)\,, & \bin{\gamma^\mu\lrd^\nu}(\overline{e}\gamma^5\gamma_\mu\lrd_\nu e)\,, \\
\bin{\gamma^\mu Q_\pm\lrd^\nu}(\overline{e}\gamma_\mu\lrd_\nu e)\,, & \bin{\gamma^\mu Q_\pm\lrd^\nu}(\overline{e}\gamma^5\gamma_\mu\lrd_\nu e)\,, \\
\bin{\gamma^5\gamma^\mu\lrd^\nu}(\overline{e}\gamma_\mu\lrd_\nu e)\,, & \bin{\gamma^5\gamma^\mu\lrd^\nu}(\overline{e}\gamma^5\gamma_\mu\lrd_\nu e)\,, \\
\bin{\gamma^5\gamma^\mu Q_\pm\lrd^\nu}(\overline{e}\gamma_\mu\lrd_\nu e)\,, & \bin{\gamma^5\gamma^\mu Q_\pm\lrd^\nu}(\overline{e}\gamma^5\gamma_\mu\lrd_\nu e)\,, \\
\bin{\gamma^\mu\lrd^\nu}(\overline{\nu}_L\gamma_\mu \lrd_\nu \nu_L)\,, & \bin{\gamma^5\gamma^\mu\lrd^\nu}(\overline{\nu}_L\gamma_\mu \lrd_\nu \nu_L)\,,  \\
\bin{\gamma^\mu Q_\pm \lrd^\nu}(\overline{\nu}_L\gamma_\mu \lrd_\nu \nu_L)\,, & \bin{\gamma^5\gamma^\mu Q_\pm \lrd^\nu}(\overline{\nu}_L\gamma_\mu \lrd_\nu \nu_L)\,, \\
\bin{\gamma^\mu Q_\pm \lrd^\nu}(\overline{e}\gamma_\mu \lrd_\nu \nu_L)\,, & \bin{\gamma^\mu Q_\pm \lrd^\nu}(\overline{\nu}_L\gamma_\mu \lrd_\nu e)\,, \\
\bin{\gamma^5\gamma^\mu Q_\pm \lrd^\nu}(\overline{e}\gamma_\mu \lrd_\nu \nu_L)\,, & \bin{\gamma^5\gamma^\mu Q_\pm \lrd^\nu}(\overline{\nu}_L\gamma_\mu \lrd_\nu e)\,, \\
\bin{\sigma^{\mu\nu}\lrd^\rho}(\overline{e}\sigma_{\mu\nu}\lrd_\rho e)\,, & \bin{\sigma^{\mu\nu}\lrd^\rho}(\overline{e}\gamma^5\sigma_{\mu\nu}\lrd_\rho e)\,, \\
\bin{\sigma^{\mu\nu}\Sigma_\pm\lrd^\rho}(\overline{e}\sigma_{\mu\nu}\lrd_\rho e)\,, & \bin{\sigma^{\mu\nu}\Sigma_\pm\lrd^\rho}(\overline{e}\gamma^5\sigma_{\mu\nu}\lrd_\rho e)\,, \\
\bin{\sigma^{\mu\nu}\Sigma_\pm\lrd^\rho}(\overline{e}\sigma_{\mu\nu}\lrd_\rho \nu_L)\,, & \bin{\sigma^{\mu\nu}\Sigma_\pm\lrd^\rho}(\overline{e}\gamma^5\sigma_{\mu\nu}\lrd_\rho \nu_L)\,, \\
\bin{\sigma^{\mu\nu}\lrd^\rho}(\overline{\nu}_L\sigma_{\mu\nu}\lrd_\rho e)\,, & \bin{\sigma^{\mu\nu}\lrd^\rho}(\overline{\nu}_L\gamma^5\sigma_{\mu\nu}\lrd_\rho e)\,, \\
\bin{\sigma^{\mu\nu}\Sigma_\pm\lrd^\rho}(\overline{\nu}_L\sigma_{\mu\nu}\lrd_\rho e)\,, & \bin{\sigma^{\mu\nu}\Sigma_\pm\lrd^\rho}(\overline{\nu}_L\gamma^5\sigma_{\mu\nu}\lrd_\rho e) \,.
    \end{array}
\end{align}
Furthermore, the $p^4$ order operators containing 2 derivatives applying on the leptons are 
\begin{align}
    \mathcal{O}(p^4): & \notag \\
    & \begin{array}{ll}
\bin{}D^2(\overline{e}e)\,, & \bin{}D^2(\overline{e}\gamma^5 e)\,, \\
\bin{\Sigma_\pm}D^2(\overline{e}e)\,, & \bin{\Sigma_\pm} D^2(\overline{e}\gamma^5 e)\,, \\
\bin{\gamma^5}D^2(\overline{e}e)\,, & \bin{\gamma^5}D^2(\overline{e}\gamma^5 e)\,, \\
\bin{\gamma^5\Sigma_\pm}D^2(\overline{e}e)\,, & \bin{\gamma^5\Sigma_\pm} D^2(\overline{e}\gamma^5 e)\,, \\
\bin{\Sigma_\pm}D^2(\overline{e}\nu_L)\,, & \bin{\Sigma_\pm}D^2(\overline{\nu}_L e)\,, \\
\bin{\gamma^5\Sigma_\pm}D^2(\overline{e}\nu_L)\,, & \bin{\gamma^5\Sigma_\pm}D^2(\overline{\nu}_L e)\,, \\
\bin{\gamma^\mu}D^2(\overline{e}\gamma_\mu e)\,, & \bin{\gamma^\mu}D^2(\overline{e}\gamma^5\gamma_\mu e)\,, \\
\bin{\gamma^\mu Q_\pm}D^2(\overline{e}\gamma_\mu e)\,, & \bin{\gamma^\mu Q_\pm}D^2(\overline{e}\gamma^5\gamma_\mu e)\,, \\
\bin{\gamma^5\gamma^\mu}D^2(\overline{e}\gamma_\mu e)\,, & \bin{\gamma^5\gamma^\mu}D^2(\overline{e}\gamma^5\gamma_\mu e)\,, \\
\bin{\gamma^5\gamma^\mu Q_\pm}D^2(\overline{e}\gamma_\mu e)\,, & \bin{\gamma^5\gamma^\mu Q_\pm}D^2(\overline{e}\gamma^5\gamma_\mu e)\,, \\
\bin{\gamma^\mu}D^2(\overline{\nu}_L\gamma_\mu \nu_L)\,, & \bin{\gamma^5\gamma^\mu}D^2(\overline{\nu}_L\gamma_\mu \nu_L)\,, \\
\bin{\gamma^\mu Q_\pm}D^2(\overline{\nu}_L\gamma_\mu \nu_L)\,, & \bin{\gamma^5\gamma^\mu Q_\pm}D^2(\overline{\nu}_L\gamma_\mu \nu_L)\,, \\
\bin{\gamma^\mu Q_\pm}D^2(\overline{e}\gamma_\mu \nu_L)\,, & \bin{\gamma^\mu Q_\pm}D^2(\overline{\nu}_L\gamma_\mu e)\,, \\
\bin{\gamma^5\gamma^\mu Q_\pm}D^2(\overline{e}\gamma_\mu \nu_L)\,, & \bin{\gamma^5\gamma^\mu Q_\pm}D^2(\overline{\nu}_L\gamma_\mu e)\,, \\
\bin{\sigma^{\mu\nu}}D_\mu(\overline{e}\lrd_\nu e)\,, & \bin{\sigma^{\mu\nu}}D_\mu(\overline{e}\gamma^5\lrd_\nu e)\,, \\
\bin{\sigma^{\mu\nu}\Sigma_\pm}D_\mu(\overline{e}\lrd_\nu e)\,, & \bin{\sigma^{\mu\nu}\Sigma_\pm}D_\mu(\overline{e}\gamma^5\lrd_\nu e)\,, \\
\bin{\gamma^5\sigma^{\mu\nu}}D_\mu(\overline{e}\lrd_\nu e)\,, & \bin{\gamma^5\sigma^{\mu\nu}}D_\mu(\overline{e}\gamma^5\lrd_\nu e)\,, \\
\bin{\gamma^5\sigma^{\mu\nu}\Sigma_\pm}D_\mu(\overline{e}\lrd_\nu e)\,, & \bin{\gamma^5\sigma^{\mu\nu}\Sigma_\pm}D_\mu(\overline{e}\gamma^5\lrd_\nu e)\,, \\
\bin{\sigma^{\mu\nu} \Sigma_\pm}D_\mu(\overline{e}\lrd_\nu \nu_L)\,, & \bin{\sigma^{\mu\nu} \Sigma_\pm}D_\mu(\overline{\nu}_L\lrd_\nu e)\,.
    \end{array}
\end{align}

\subsection{Meson-Lepton Interactions}
\label{sec:mli}

In this subsection, we shall present the lepton number conserving effective operators up to $p^4$.

The LO operators are of $p^2$ order, containing one spurion, one $u_\mu$, and two leptons,
\begin{align}
    \mathcal{O}(p^3): & \notag \\
    & \begin{array}{ll}
\lra{Q_\pm u_\mu}(\overline{e} \gamma^\mu e)\,,  & \lra{Q_\pm u_\mu}(\overline{e} \gamma^5\gamma^\mu e)\,, \\
\lra{Q_\pm u_\mu}(\overline{\nu}_L \gamma^\mu \nu_L)\,,  & \lra{Q_\pm u_\mu}(\overline{e} \gamma^\mu \nu_L)\,, \\
\lra{Q_\pm u_\mu}(\overline{\nu}_L \gamma^\mu e)\,.
    \end{array}
\end{align}

The $\mathcal{O}(p^4)$ operators are the ones containing more derivatives or $u_\mu$,
\begin{align}
    \mathcal{O}(p^4): & \notag \\
    & \begin{array}{ll}
\tra{\Sigma_\pm u_\mu} (\overline{e} \lrd^\mu e)\,, &  \tra{\Sigma_\pm u_\mu} (\overline{e} \gamma^5\lrd^\mu e)\,, \\
\tra{\Sigma_\pm u_\mu} (\overline{e} \lrd^\mu \nu_L)\,, & \tra{\Sigma_\pm u_\mu} (\overline{\nu}_L \lrd^\mu e)\,, \\
\tra{\Sigma_\pm [u_\mu,u_\nu]}(\overline{e}\sigma^{\mu\nu} e)\,, & \tra{\Sigma_\pm [u_\mu,u_\nu]}(\overline{e}\gamma^5\sigma^{\mu\nu} e)\,, \\
\tra{\Sigma_\pm [u_\mu,u_\nu]}(\overline{e}\sigma^{\mu\nu}\nu_L)\,, & \tra{\Sigma_\pm [u_\mu,u_\nu]}(\overline{\nu}_L\sigma^{\mu\nu}e)\,.
    \end{array}
\end{align}

\subsection{Neutrinoless Double Beta Decay}

In addition to the lepton number conserving interactions, the neutrinoless double beta decay ($0\nu\beta\beta$) is an important process violating the lepton number~\cite{Savage:1998yh,Prezeau:2003xn,Graesser:2016bpz,Cirigliano:2017ymo,Cirigliano:2017djv,Cirigliano:2017tvr,Pastore:2017ofx,Cirigliano:2018yza,Cirigliano:2019vdj,Dekens:2024eae}, during which the two neutrons transfer to two protons with two electrons emitting. Although only the upper limits of its decay rate have been set up to now, non-zero observations are expected in the next-generation experiments.

\begin{figure}
    \centering
\includegraphics[scale=0.5,trim=10 15 10 15]{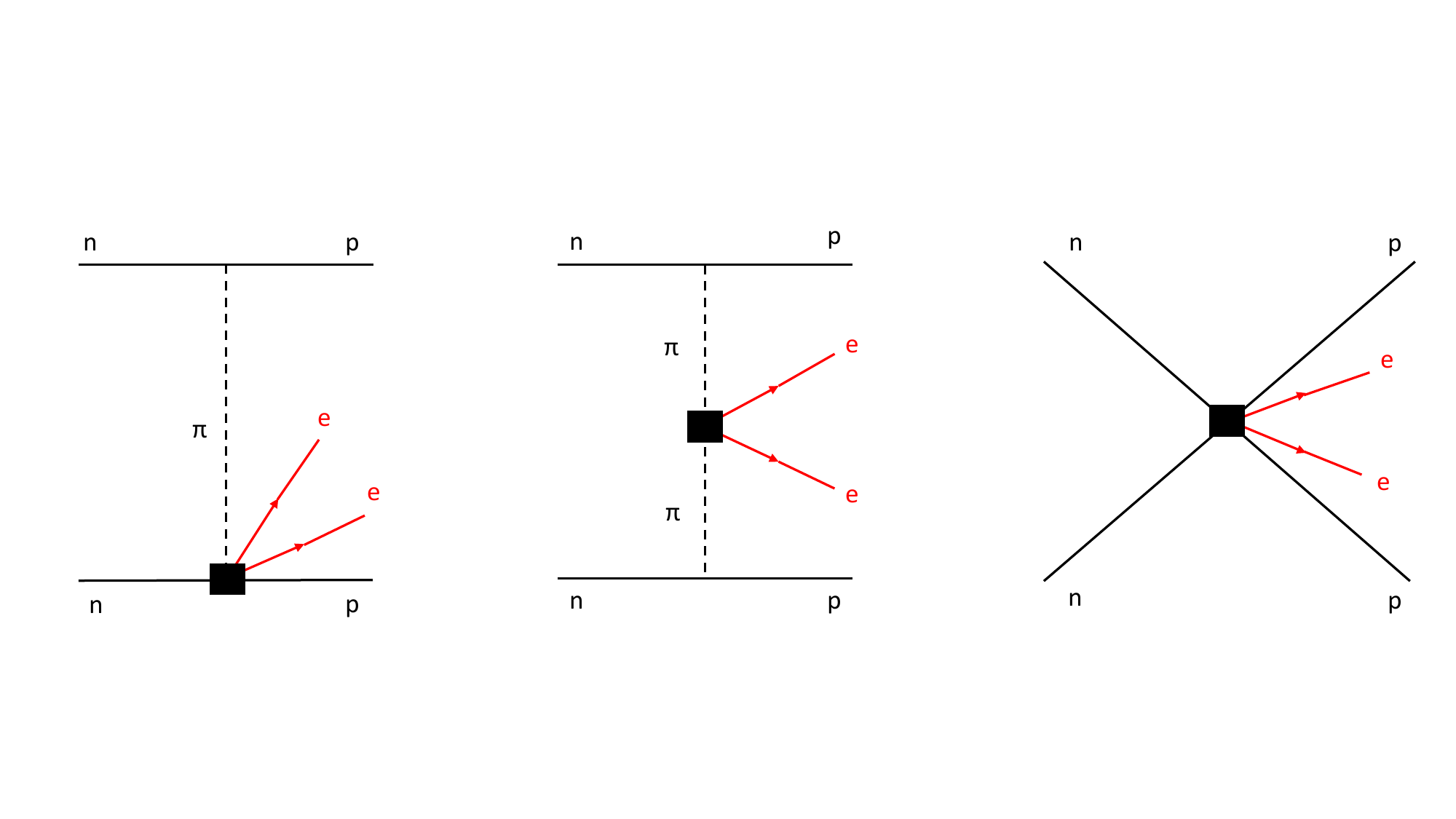}
    \caption{The tree-level diagrams contributing to the $0\nu\beta\beta$.}
    \label{tab:fig1}
\end{figure}

At the nuclear level, the $0\nu\beta\beta$ is contributed from 3 diagrams at the tree level in Fig.~\ref{tab:fig1}. The effective operators of these diagrams can be constructed in our framework. In particular, these operators involve two spurions. In the following, we presented the LO operators of each vertex.

In the first diagram in Fig.~\ref{tab:fig1}, the lepton number violating vertex involves 2 nucleons, 2 electrons, and a meson. Thus the contributing effective operators to this vertex at leading order are
\begin{align}
    \begin{array}{ll}
\bin{\Sigma_\pm}\lra{u_\mu Q_\pm}(\overline{e}^c \gamma^\mu e)\,, & \bin{\Sigma_\pm}\lra{u_\mu Q_\pm}(\overline{e}^c \gamma^5\gamma^\mu e)\,, \\
\bin{\gamma^5\Sigma_\pm}\lra{u_\mu Q_\pm}(\overline{e}^c \gamma^\mu e)\,, & \bin{\gamma^5\Sigma_\pm}\lra{u_\mu Q_\pm}(\overline{e}^c \gamma^5\gamma^\mu e)\,, \\
\bin{\gamma^\mu Q_\pm}\lra{u_\mu Q_\pm}(\overline{e}^c e)\,, & \bin{\gamma^\mu Q_\pm}\lra{u_\mu Q_\pm}(\overline{e}^c\gamma^5 e)\,, \\
\bin{\gamma^5\gamma^\mu Q_\pm}\lra{u_\mu Q_\pm}(\overline{e}^c e)\,, & \bin{\gamma^5\gamma^\mu Q_\pm}\lra{u_\mu Q_\pm}(\overline{e}^c\gamma^5 e)\,, \\
\bin{\gamma^\mu Q_\pm}\lra{u^\nu Q_\pm} (\overline{e}^c\sigma_{\mu\nu}e)\,, & \bin{\gamma^\mu Q_\pm}\lra{u^\nu Q_\pm} (\overline{e}^c\gamma^5\sigma_{\mu\nu}e)\,, \\
\bin{\gamma^5\gamma^\mu Q_\pm}\lra{u^\nu Q_\pm} (\overline{e}^c\sigma_{\mu\nu}e)\,, & \bin{\gamma^5\gamma^\mu Q_\pm}\lra{u^\nu Q_\pm} (\overline{e}^c\gamma^5\sigma_{\mu\nu}e)\,, \\
\bin{\sigma^{\mu\nu}\Sigma_\pm} \lra{u_\mu Q_\pm} (\overline{e}^c \gamma_\nu e)\,, & \bin{\sigma^{\mu\nu}\Sigma_\pm} \lra{u_\mu Q_\pm} (\overline{e}^c \gamma^5\gamma_\nu e)\,, \\
\bin{\gamma^5\sigma^{\mu\nu}\Sigma_\pm} \lra{u_\mu Q_\pm} (\overline{e}^c \gamma_\nu e)\,, & \bin{\gamma^5\sigma^{\mu\nu}\Sigma_\pm} \lra{u_\mu Q_\pm} (\overline{e}^c \gamma^5\gamma_\nu e)\,.
    \end{array}
\end{align}

In the second diagram in Fig.~\ref{tab:fig1}, the lepton number violating vertex involves 2 electrons and two mesons, to which the contributing effective operators at LO are
\begin{equation}
    \begin{array}{ll}
\lra{Q_\pm u_\mu}\lra{Q_\pm u_\mu}(\overline{e}^c e)\,, & \lra{Q_\pm u_\mu}\lra{Q_\pm u_\mu}(\overline{e}^c\gamma^5 e)\,.
    \end{array}
\end{equation}

The last diagram is the contract six-fermion interactions, the LO effective operators contributing to it are
\begin{equation}
    \begin{array}{ll}
\bin{\Sigma_\pm}\bin{\Sigma_\pm}(\overline{e}^c e)\,, & \bin{\Sigma_\pm}\bin{\Sigma_\pm}(\overline{e}^c \gamma^5 e)\,, \\
\bin{\gamma^5\Sigma_\pm}\bin{\gamma^5\Sigma_\pm}(\overline{e}^c e)\,, & \bin{\gamma^5\Sigma_\pm}\bin{\gamma^5\Sigma_\pm}(\overline{e}^c \gamma^5 e)\,, \\
\bin{\gamma^5\Sigma_\pm}\bin{\Sigma_\pm}(\overline{e}^c e)\,, & \bin{\gamma^5\Sigma_\pm}\bin{\Sigma_\pm}(\overline{e}^c \gamma^5 e)\,, \\
\bin{\gamma^\mu Q_\pm}\bin{\gamma_\mu Q_\pm}(\overline{e}^c e)\,, & \bin{\gamma^\mu Q_\pm}\bin{\gamma_\mu Q_\pm}(\overline{e}^c \gamma^5 e)\,, \\
\bin{\gamma^5\gamma^\mu Q_\pm}\bin{\gamma^5\gamma_\mu Q_\pm}(\overline{e}^c e)\,, & \bin{\gamma^5\gamma^\mu Q_\pm}\bin{\gamma^5\gamma_\mu Q_\pm}(\overline{e}^c \gamma^5 e)\,, \\
\bin{\gamma^5\gamma^\mu Q_\pm}\bin{\gamma_\mu Q_\pm}(\overline{e}^c e)\,, & \bin{\gamma^5\gamma^\mu Q_\pm}\bin{\gamma_\mu Q_\pm}(\overline{e}^c \gamma^5 e)\,, \\
\bin{\sigma^{\mu\nu} \Sigma_\pm}\bin{\sigma_{\mu\nu}\Sigma_\pm}(\overline{e}^c e)\,, & \bin{\sigma^{\mu\nu} \Sigma_\pm}\bin{\sigma_{\mu\nu}\Sigma_\pm}(\overline{e}^c\gamma^5 e)\,, \\
\bin{\gamma^5\sigma^{\mu\nu}\Sigma_\pm}\bin{\sigma_{\mu\nu}\Sigma_\pm}(\overline{e}^c e)\,, & \bin{\gamma^5\sigma^{\mu\nu} \Sigma_\pm}\bin{\sigma_{\mu\nu}\Sigma_\pm}(\overline{e}^c\gamma^5 e)\,, \\
\bin{\Sigma_\pm}\bin{\gamma^\mu Q_\pm}(\overline{e}^c \gamma_\mu e)\,, & \bin{\Sigma_\pm}\bin{\gamma^\mu Q_\pm}(\overline{e}^c \gamma^5\gamma_\mu e)\,, \\
\bin{\gamma^5\Sigma_\pm}\bin{\gamma^5\gamma^\mu Q_\pm}(\overline{e}^c \gamma_\mu e)\,, & \bin{\gamma^5\Sigma_\pm}\bin{\gamma^5\gamma^\mu Q_\pm}(\overline{e}^c \gamma^5\gamma_\mu e)\,, \\
\bin{\gamma^5\Sigma_\pm}\bin{\gamma^\mu Q_\pm}(\overline{e}^c \gamma_\mu e)\,, & \bin{\gamma^5\Sigma_\pm}\bin{\gamma^\mu Q_\pm}(\overline{e}^c \gamma^5\gamma_\mu e)\,, \\
\bin{\Sigma_\pm}\bin{\gamma^5\gamma^\mu Q_\pm}(\overline{e}^c \gamma_\mu e)\,, & \bin{\Sigma_\pm}\bin{\gamma^5\gamma^\mu Q_\pm}(\overline{e}^c \gamma^5\gamma_\mu e)\,, \\
\bin{\sigma_{\mu\nu}\Sigma_\pm}\bin{\gamma^\nu Q_\pm}(\overline{e}^c \gamma^\mu e)\,, & \bin{\sigma_{\mu\nu}\Sigma_\pm}\bin{\gamma^\nu Q_\pm}(\overline{e}^c \gamma^5\gamma^\mu e)\,, \\
\bin{\gamma^5\sigma_{\mu\nu}\Sigma_\pm}\bin{\gamma^5\gamma^\nu Q_\pm}(\overline{e}^c \gamma^\mu e)\,, & \bin{\gamma^5\sigma_{\mu\nu}\Sigma_\pm}\bin{\gamma^5\gamma^\nu Q_\pm}(\overline{e}^c \gamma^5\gamma^\mu e)\,, \\
\bin{\gamma^5\sigma_{\mu\nu}\Sigma_\pm}\bin{\gamma^\nu Q_\pm}(\overline{e}^c \gamma^\mu e)\,, & \bin{\gamma^5\sigma_{\mu\nu}\Sigma_\pm}\bin{\gamma^\nu Q_\pm}(\overline{e}^c \gamma^5\gamma^\mu e)\,, \\
\bin{\sigma_{\mu\nu}\Sigma_\pm}\bin{\gamma^5\gamma^\nu Q_\pm}(\overline{e}^c \gamma^\mu e)\,, & \bin{\sigma_{\mu\nu}\Sigma_\pm}\bin{\gamma^5\gamma^\nu Q_\pm}(\overline{e}^c \gamma^5\gamma^\mu e)\,.
    \end{array}
\end{equation}

\section{Conclusion}
\label{sec:conclu}

The chiral effective field theory (ChEFT) is an effective field theory describing the strong and the electroweak interactions among the nucleons, the meson, and the leptons. Originated from the spontaneous symmetry breaking $SU(2)_L\times SU(2)_R \rightarrow SU(2)_V$ caused by the quark condensation, the effective Lagrangian can be constructed by the CCWZ coset formalism, which takes the general form that
\begin{equation}
\mathcal{L}_\text{ChEFT} = \mathcal{L}_\pi+\mathcal{L}_{\pi N}+\mathcal{L}_{NN}+\mathcal{L}_{\pi NN} + \mathcal{L}_{NNN}+\mathcal{L}_{\text{leptonic}}\dots\,,     
\end{equation}
with several different sectors. The pure meson sector $\mathcal{L}_\pi$ and the meson-nucleon sector $\mathcal{L}_{\pi N}$ compose the ChPT Lagrangian, and the multi-nucleon sectors, including the 2-nucleon sector $\mathcal{L}_{NN}$, the 2-nucleon-meson sector $\mathcal{L}_{\pi NN}$, the 3-nucleon sector $\mathcal{L}_{NNN}$, and so on, contain the effective interactions of more than two nucleons describing the nuclear forces. The leptonic sector $\mathcal{L}_{\text{leptonic}}$ is responsible for the electroweak interactions among the hadrons and the leptons, which are embedded in the external sources to be taken into the ChEFT conventionally. Nevertheless, the external source method has shortages when constructing the high-dimension operators. In principle, more and more external sources with complicated Lorentz structures should be taken into account for the high-dimension operators. Besides, a specific external source unifies many different leptonic currents of different power counting, which raises ambiguity in the ChEFT Lagrangian.

In this paper, we propose a systematic description of the strong and electroweak interactions of the hadrons and the leptons in the ChEFT framework. To construct the complete and independent operators of the ChEFT systematically, we utilize the Hilbert series method and the Young tensor method. The Hilbert series is used to count the numbers of the independent operators with the parity and the charge conjugation symmetries considered. This method has been applied to the pure meson sector $\mathcal{L}_\pi$ previously, and in this paper, we extend it to the fermions and apply it to all the remaining sectors of the ChEFT for the first time. The Young tensor method is general and has been applied to many EFTs widely. In this paper, we apply it to the ChEFT with certain developments to manage the NGBs and the external sources.

Traditionally, the external source method is used to take the leptonic currents into account, which, however, encounters problems when constructing the high-dimension operators. To avoid the shortages of the external source method mentioned above, we adopt the spurion method in the leptonic sector $\mathcal{L}_{\text{leptonic}}$. 
In the spurion method, the leptons are regarded as the building blocks of the ChEFT, and the $SU(2)_V$ symmetry breaking effects are compensated by additional spurions, $\Sigma_\pm\,,Q_\pm$. 

Within this framework, we present the effective operators of the different sectors of the ChEFT. 
The operators and their maximal dimensions given in this paper are summarized in Tab.~\ref{tab:organization}.
In particular, we present the minimal set of the 2-nucleon operators, the 3-nucleon operators, and the 2-nucleon-meson operators up to $p^6$ for the first time. 
For the leptonic sectors, we present the operators with the leptons up to $p^4$ by the spurion method, in which the leading order operators in terms of the spurions contributing to the neutrinoless double beta decay are also presented in addition.

\begin{table}[htbp]
    \centering
    \renewcommand{\arraystretch}{1.8}
    \begin{tabular}{|c|c|c|c|c|c|}
\hline
sector & $\mathcal{L}_\pi$ & $\mathcal{L}_{\pi N}$ & $\mathcal{L}_{NN}+\mathcal{L}_{NNN}$  & $\mathcal{L}_{\pi NN}$ & $\mathcal{L}_{\text{leptonic}}$\\
\hline
operator order & $p^8$ & $p^5$ & $p^6$ & $p^6$ & $p^4$ \\
\hline
section & Sec.~\ref{sec:operators} & Sec.~\ref{sec:NMoperators} & Sec.~\ref{sec:NNoperators} & Sec.~\ref{sec:NNoperators} & Sec.~\ref{sec:weakoperators} \\
\hline
    \end{tabular}
    \caption{The effective operators and their maximal orders in this paper.}
    \label{tab:organization}
\end{table}

The ChEFT framework developed in this paper is systematic to parameterize the various low-energy strong and electroweak processes, which, together with the effective operators presented in this paper, are of phenomenological importance in the low-energy experiments of the precision frontier. For example, the neutrinoless double beta decay highlighted in this paper is an important process closely related to high-energy new physics, and it can be characterized systematically in the ChEFT. The proposition of the ChEFT in this paper can be extended to the $SU(3)_V$ case to take the light baryon octet into account. Besides, the spurion method is general and can used in the theories with different light fields rather than the leptons, such as the axion, light dark matter, and so on.

\section*{Acknowledgement}
We thank Li-Sheng Geng, Feng-kun Guo, Bingwei Long for their useful discussions and Ulf-G. Meißner for the valuable comments on the draft. 
This work is supported by the National Science Foundation of China under Grants No. 12347105, No. 12375099, and No. 12047503, and the National Key Research and Development Program of China Grant No. 2020YFC2201501, No. 2021YFA0718304.

\appendix
\section{Hilbert Series of the Meson-Nucleon Sector}
\label{sec:app1}

In this appendix, we present the Hilbert series of the $C$-even and $P$-even operators of the $\mathcal{L}_{\pi N}$ sector  as follows,
\begin{align}
    \mathcal{O}(p^2): & \notag \\
    \mathcal{H}^{\text{P-even}}_{\text{C-even}} &= f_+ N^2+N^2 \left\langle \Sigma _-\right\rangle +N^2 \left\langle \Sigma _+\right\rangle +N^2 \Sigma _-+N^2 \Sigma _++2 N^2 u^2 \\
    \mathcal{O}(p^3): & \notag \\
    \mathcal{H}^{\text{P-even}}_{\text{C-even}} &= D f_- N^2+D f_+ N^2+2 D N^2 u^2+2 f_- N^2 u+2 f_+ N^2 u+3 N^2 u^3+N^2 u \left\langle \Sigma _+\right\rangle \notag \\
    & +N^2 \Sigma _- u+N^2 \Sigma _+ u \\
    \mathcal{O}(p^4): & \notag \\
    \mathcal{H}^{\text{P-even}}_{\text{C-even}} &= 3 D^2 N^2 u^2+4 D N^2 u^3+4 N^2 u^4+D^2 N^2 \left\langle \Sigma _-\right\rangle +D N^2 u \left\langle \Sigma _-\right\rangle +2 N^2 u^2 \left\langle \Sigma _-\right\rangle +N^2 \left\langle \Sigma _-\right\rangle {}^2 \notag \\
    & +D^2 N^2 \left\langle \Sigma _+\right\rangle + D N^2 u \left\langle \Sigma _+\right\rangle +2 N^2 u^2 \left\langle \Sigma _+\right\rangle +N^2 \left\langle \Sigma _-\right\rangle  \left\langle \Sigma _+\right\rangle +N^2 \left\langle \Sigma _+\right\rangle {}^2+5 D f_- N^2 u\notag \\
    & +3 f_- N^2 u^2 +3 f_-^2 N^2+D^2 N^2 \Sigma _-+2 D N^2 \Sigma _- u+3 N^2 \Sigma _- u^2+N^2 \Sigma _- \left\langle \Sigma _-\right\rangle +N^2 \Sigma _- \left\langle \Sigma _+\right\rangle \notag \\
    & +f_- N^2 \Sigma _-+N^2\Sigma _-^2+D^2 f_+ N^2 +5 D f_+ N^2 u+7 f_+ N^2 u^2+f_+ N^2 \left\langle \Sigma _-\right\rangle +f_+ N^2 \left\langle \Sigma _+\right\rangle \notag \\
    & +3 f_- f_+ N^2+f_+ N^2 \Sigma _-+3 f_+^2 N^2+D^2 N^2 \Sigma _++2 D N^2 \Sigma _+ u+3 N^2 \Sigma _+ u^2+N^2 \Sigma _+ \left\langle \Sigma _-\right\rangle \notag \\
    & +N^2 \Sigma _+ \left\langle \Sigma _+\right\rangle +f_- N^2 \Sigma _++N^2 \Sigma _- \Sigma _++f_+ N^2 \Sigma _++N^2\Sigma _+^2
\end{align}

\begin{align}
    \mathcal{O}(p^5): & \notag \\
    \mathcal{H}^{\text{P-even}}_{\text{C-even}} &= 3 D^3 N^2 u^2+9 D^2 N^2 u^3+10 D N^2 u^4+5 N^2 u^5+D^2 N^2 u \left\langle \Sigma _-\right\rangle +3 D N^2 u^2 \left\langle \Sigma _-\right\rangle +N^2 u \left\langle \Sigma _-\right\rangle {}^2 \notag \\
    & +2 D^2 N^2 u \left\langle \Sigma _+\right\rangle +3 D N^2 u^2 \left\langle \Sigma _+\right\rangle +3 N^2 u^3 \left\langle \Sigma _+\right\rangle +D N^2 \left\langle \Sigma _-\right\rangle  \left\langle \Sigma _+\right\rangle +N^2 u \left\langle \Sigma _+\right\rangle {}^2+D^3 f_- N^2\notag \\
    & +8 D^2 f_- N^2 u+20 D f_- N^2 u^2+9 f_- N^2 u^3+D f_- N^2 \left\langle \Sigma _-\right\rangle +2 f_- N^2 u \left\langle \Sigma _-\right\rangle +2 D f_- N^2 \left\langle \Sigma _+\right\rangle \notag \\
    & +2 f_- N^2 u \left\langle \Sigma _+\right\rangle +5 D f_-^2 N^2+7 f_-^2 N^2 u+3 D^2 N^2 \Sigma _- u+7 D N^2 \Sigma _- u^2+3 N^2 \Sigma _- u^3+N^2 \Sigma _- u \left\langle \Sigma _-\right\rangle \notag \\
    & +D N^2 \Sigma _- \left\langle \Sigma _+\right\rangle +N^2 \Sigma _- u \left\langle \Sigma _+\right\rangle +3 D f_- N^2 \Sigma _-+4 f_- N^2 \Sigma _- u+DN^2\Sigma _-^2+2N^2u\Sigma _-^2+D^3 f_+ N^2 \notag \\
    & +8 D^2 f_+ N^2 u+20 D f_+ N^2 u^2+9 f_+ N^2 u^3+D f_+ N^2 \left\langle \Sigma _-\right\rangle +2 f_+ N^2 u \left\langle \Sigma _-\right\rangle +2 D f_+ N^2 \left\langle \Sigma _+\right\rangle \notag \\
    & +2 f_+ N^2 u \left\langle \Sigma _+\right\rangle +8 D f_- f_+ N^2+12 f_- f_+ N^2 u+3 D f_+ N^2 \Sigma _- +4 f_+ N^2 \Sigma _- u+5 D f_+^2 N^2 \notag \\
    & +7 f_+^2 N^2 u+3 D^2 N^2 \Sigma _+ u+5 D N^2 \Sigma _+ u^2+3 N^2 \Sigma _+ u^3+D N^2 \Sigma _+ \left\langle \Sigma _-\right\rangle + N^2 \Sigma _+ u \left\langle \Sigma _-\right\rangle \notag \\
    & +N^2 \Sigma _+ u \left\langle \Sigma _+\right\rangle +3 D f_- N^2 \Sigma _++4 f_- N^2 \Sigma _+ u+D N^2 \Sigma _- \Sigma _+ +N^2 \Sigma _- \Sigma _+ u+3 D f_+ N^2 \Sigma _+ \notag \\
    & +4 f_+ N^2 \Sigma _+ u+DN^2\Sigma _+^2+2N^2u\Sigma _+^2
\end{align}

\begin{align}
    \mathcal{O}(p^6): & \notag \\
    \mathcal{H}^{\text{P-even}}_{\text{C-even}} &= 4 D^4 N^2 u^2+14 D^3 N^2 u^3+30 D^2 N^2 u^4+15 D N^2 u^5+7 N^2 u^6+D^4 N^2 \left\langle \Sigma _-\right\rangle +3 D^3 N^2 u \left\langle \Sigma _-\right\rangle \notag \\
    & +11 D^2 N^2 u^2 \left\langle \Sigma _-\right\rangle +7 D N^2 u^3 \left\langle \Sigma _-\right\rangle +4 N^2 u^4 \left\langle \Sigma _-\right\rangle +2 D^2 N^2 \left\langle \Sigma _-\right\rangle {}^2+D N^2 u \left\langle \Sigma _-\right\rangle {}^2 \notag \\
    & +2 N^2 u^2 \left\langle \Sigma _-\right\rangle {}^2+N^2 \left\langle \Sigma _-\right\rangle {}^3+D^4 N^2 \left\langle \Sigma _+\right\rangle +3 D^3 N^2 u \left\langle \Sigma _+\right\rangle +11 D^2 N^2 u^2 \left\langle \Sigma _+\right\rangle +7 D N^2 u^3 \left\langle \Sigma _+\right\rangle \notag \\
    & +4 N^2 u^4 \left\langle \Sigma _+\right\rangle +3 D^2 N^2 \left\langle \Sigma _-\right\rangle  \left\langle \Sigma _+\right\rangle +2 D N^2 u \left\langle \Sigma _-\right\rangle  \left\langle \Sigma _+\right\rangle +2 N^2 u^2 \left\langle \Sigma _-\right\rangle  \left\langle \Sigma _+\right\rangle +N^2 \left\langle \Sigma _-\right\rangle {}^2 \left\langle \Sigma _+\right\rangle \notag \notag \\
    & +2 D^2 N^2 \left\langle \Sigma _+\right\rangle {}^2+D N^2 u \left\langle \Sigma _+\right\rangle {}^2+2 N^2 u^2 \left\langle \Sigma _+\right\rangle {}^2+N^2 \left\langle \Sigma _-\right\rangle  \left\langle \Sigma _+\right\rangle {}^2+N^2 \left\langle \Sigma _+\right\rangle {}^3+13 D^3 f_- N^2 u\notag \\
    & +43 D^2 f_- N^2 u^2+54 D f_- N^2 u^3+10 f_- N^2 u^4+2 D^2 f_- N^2 \left\langle \Sigma _-\right\rangle +10 D f_- N^2 u \left\langle \Sigma _-\right\rangle +3 f_- N^2 u^2 \left\langle \Sigma _-\right\rangle \notag \\
    & +2 D^2 f_- N^2 \left\langle \Sigma _+\right\rangle +10 D f_- N^2 u \left\langle \Sigma _+\right\rangle +3 f_- N^2 u^2 \left\langle \Sigma _+\right\rangle +14 D^2 f_-^2 N^2+30 D f_-^2 N^2 u+25 f_-^2 N^2 u^2 \notag \\
    & +3 f_-^2 N^2 \left\langle \Sigma _-\right\rangle +3 f_-^2 N^2 \left\langle \Sigma _+\right\rangle +D^4 N^2 \Sigma _-+5 D^3 N^2 \Sigma _- u+19 D^2 N^2 \Sigma _- u^2\notag +17 D N^2 \Sigma _- u^3 \notag \\
    & +6 N^2 \Sigma _- u^4+3 D^2 N^2 \Sigma _- \left\langle \Sigma _-\right\rangle +4 D N^2 \Sigma _- u \left\langle \Sigma _-\right\rangle +3 N^2 \Sigma _- u^2 \left\langle \Sigma _-\right\rangle +N^2 \Sigma _- \left\langle \Sigma _-\right\rangle {}^2 \notag \\
    & +3 D^2 N^2 \Sigma _- \left\langle \Sigma _+\right\rangle +4 D N^2 \Sigma _- u \left\langle \Sigma _+\right\rangle +3 N^2 \Sigma _- u^2 \left\langle \Sigma _+\right\rangle +N^2 \Sigma _- \left\langle \Sigma _-\right\rangle  \left\langle \Sigma _+\right\rangle +N^2 \Sigma _- \left\langle \Sigma _+\right\rangle {}^2 \notag \\
    & +7 D^2 f_- N^2 \Sigma _- +18 D f_- N^2 \Sigma _- u+11 f_- N^2 \Sigma _- u^2+f_- N^2 \Sigma _- \left\langle \Sigma _-\right\rangle +f_- N^2 \Sigma _- \left\langle \Sigma _+\right\rangle +5 f_-^2 N^2 \Sigma _- \notag \\
    & +3D^2N^2\Sigma _-^2+4DN^2u\Sigma _-^2+4N^2u^2\Sigma _-^2+N^2\left\langle \Sigma _-\right\rangle \Sigma _-^2+N^2\left\langle \Sigma _+\right\rangle \Sigma _-^2+N^2\Sigma _-^3+D^4 f_+ N^2 \notag \\
    & +13 D^3 f_+ N^2 u+53 D^2 f_+ N^2 u^2+50 D f_+ N^2 u^3+20 f_+ N^2 u^4+5 D^2 f_+ N^2 \left\langle \Sigma _-\right\rangle +10 D f_+ N^2 u \left\langle \Sigma _-\right\rangle \notag \\
    & +7 f_+ N^2 u^2 \left\langle \Sigma _-\right\rangle +f_+ N^2 \left\langle \Sigma _-\right\rangle {}^2+5 D^2 f_+ N^2 \left\langle \Sigma _+\right\rangle +10 D f_+ N^2 u \left\langle \Sigma _+\right\rangle +7 f_+ N^2 u^2 \left\langle \Sigma _+\right\rangle  \notag \\
    & +f_+ N^2 \left\langle \Sigma _-\right\rangle  \left\langle \Sigma _+\right\rangle +f_+ N^2 \left\langle \Sigma _+\right\rangle {}^2+21 D^2 f_- f_+ N^2+65 D f_- f_+ N^2 u+32 f_- f_+ N^2 u^2 +3 f_- f_+ N^2 \left\langle \Sigma _-\right\rangle \notag \\
    & +3 f_- f_+ N^2 \left\langle \Sigma _+\right\rangle +12 f_-^2 f_+ N^2+7 D^2 f_+ N^2 \Sigma _-+22 D f_+ N^2 \Sigma _- u+11 f_+ N^2 \Sigma _- u^2+f_+ N^2 \Sigma _- \left\langle \Sigma _-\right\rangle \notag \\
    & +f_+ N^2 \Sigma _- \left\langle \Sigma _+\right\rangle +5 f_- f_+ N^2 \Sigma _-+2N^2\Sigma _-^2f_++14 D^2 f_+^2 N^2+30 D f_+^2 N^2 u+25 f_+^2 N^2 u^2 \notag \\
    & +3 f_+^2 N^2 \left\langle \Sigma _-\right\rangle +3 f_+^2 N^2 \left\langle \Sigma _+\right\rangle +4 f_- f_+^2 N^2+5 f_+^2 N^2 \Sigma _-+6 f_+^3 N^2+D^4 N^2 \Sigma _++5 D^3 N^2 \Sigma _+ u \notag \\
    & +19 D^2 N^2 \Sigma _+ u^2+17 D N^2 \Sigma _+ u^3+6 N^2 \Sigma _+ u^4+3 D^2 N^2 \Sigma _+ \left\langle \Sigma _-\right\rangle +4 D N^2 \Sigma _+ u \left\langle \Sigma _-\right\rangle \notag \\
    & +3 N^2 \Sigma _+ u^2 \left\langle \Sigma _-\right\rangle +N^2 \Sigma _+ \left\langle \Sigma _-\right\rangle {}^2+3 D^2 N^2 \Sigma _+ \left\langle \Sigma _+\right\rangle +4 D N^2 \Sigma _+ u \left\langle \Sigma _+\right\rangle +3 N^2 \Sigma _+ u^2 \left\langle \Sigma _+\right\rangle \notag \\
    & +N^2 \Sigma _+ \left\langle \Sigma _-\right\rangle  \left\langle \Sigma _+\right\rangle +N^2 \Sigma _+ \left\langle \Sigma _+\right\rangle {}^2+7 D^2 f_- N^2 \Sigma _++18 D f_- N^2 \Sigma _+ u+11 f_- N^2 \Sigma _+ u^2\notag \\
    & +f_- N^2 \Sigma _+ \left\langle \Sigma _-\right\rangle +f_- N^2 \Sigma _+ \left\langle \Sigma _+\right\rangle +5 f_-^2 N^2 \Sigma _++4 D^2 N^2 \Sigma _- \Sigma _++8 D N^2 \Sigma _- \Sigma _+ u+5 N^2 \Sigma _- \Sigma _+ u^2 \notag \\
    & +N^2 \Sigma _- \Sigma _+ \left\langle \Sigma _-\right\rangle +N^2 \Sigma _- \Sigma _+ \left\langle \Sigma _+\right\rangle +f_- N^2 \Sigma _- \Sigma _++2N^2\Sigma _-^2\Sigma _++7 D^2 f_+ N^2 \Sigma _++22 D f_+ N^2 \Sigma _+ u \notag \\
    & +11 f_+ N^2 \Sigma _+ u^2+f_+ N^2 \Sigma _+ \left\langle \Sigma _-\right\rangle +f_+ N^2 \Sigma _+ \left\langle \Sigma _+\right\rangle +5 f_- f_+ N^2 \Sigma _++3 f_+ N^2 \Sigma _- \Sigma _++5 f_+^2 N^2 \Sigma _+ \notag \\
    & +3D^2N^2\Sigma _+^2+4DN^2u\Sigma _+^2+4N^2u^2\Sigma _+^2+N^2\left\langle \Sigma _-\right\rangle \Sigma _+^2+N^2\left\langle \Sigma _+\right\rangle \Sigma _+^2+2N^2\Sigma _-\Sigma _+^2 \notag \\
    & +2N^2f_+\Sigma _+^2+N^2\Sigma _+^3
\end{align}

\begin{figure}
\centering
  \includegraphics[scale=0.5]{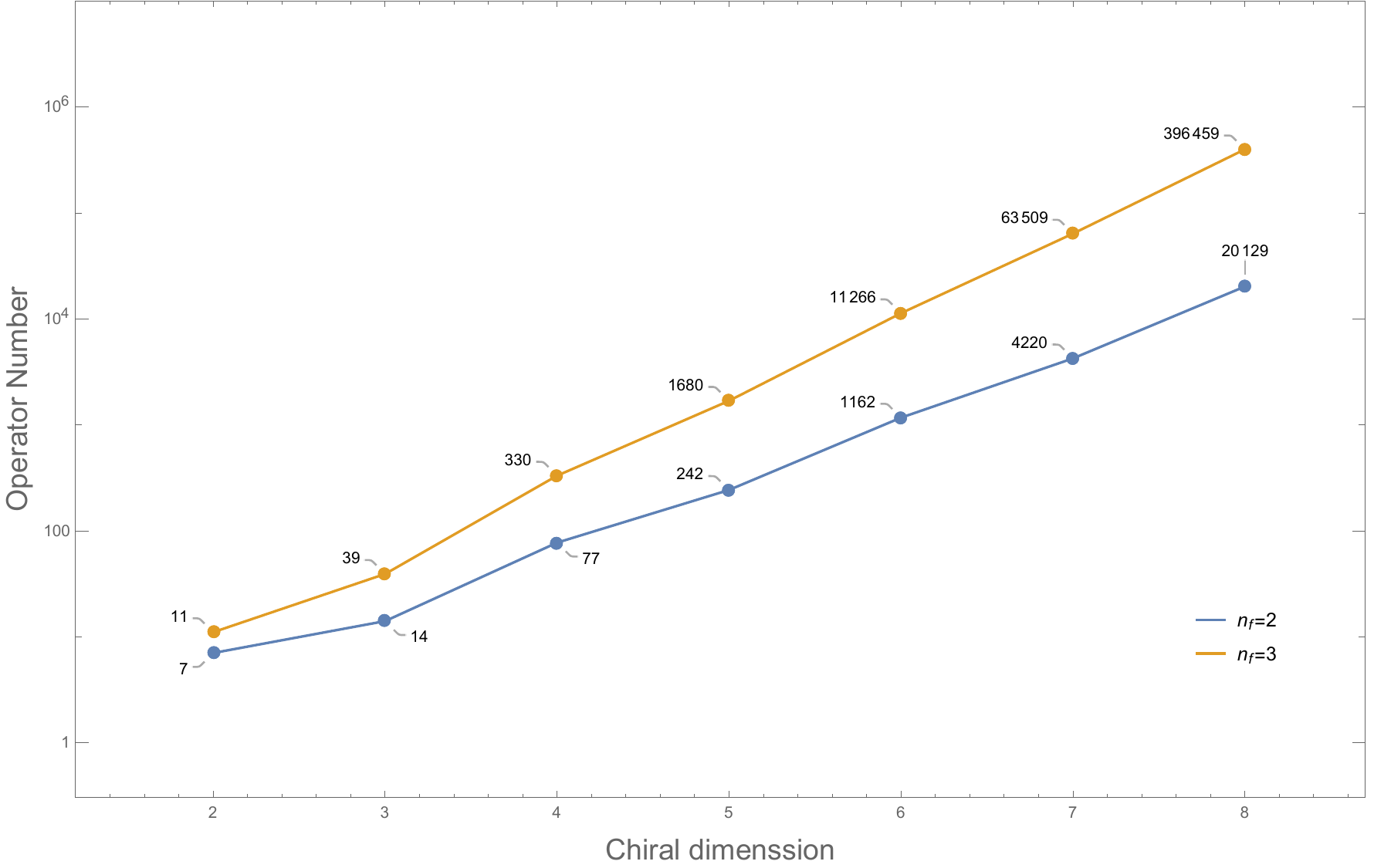}
  \caption{The growth of the number of the $C$-even and $P$-even operators in the $\mathcal{L}_{\pi N}$ sector.}
  \label{graph:pig4}
\end{figure}

\begin{figure}
\centering
  \includegraphics[scale=0.5]{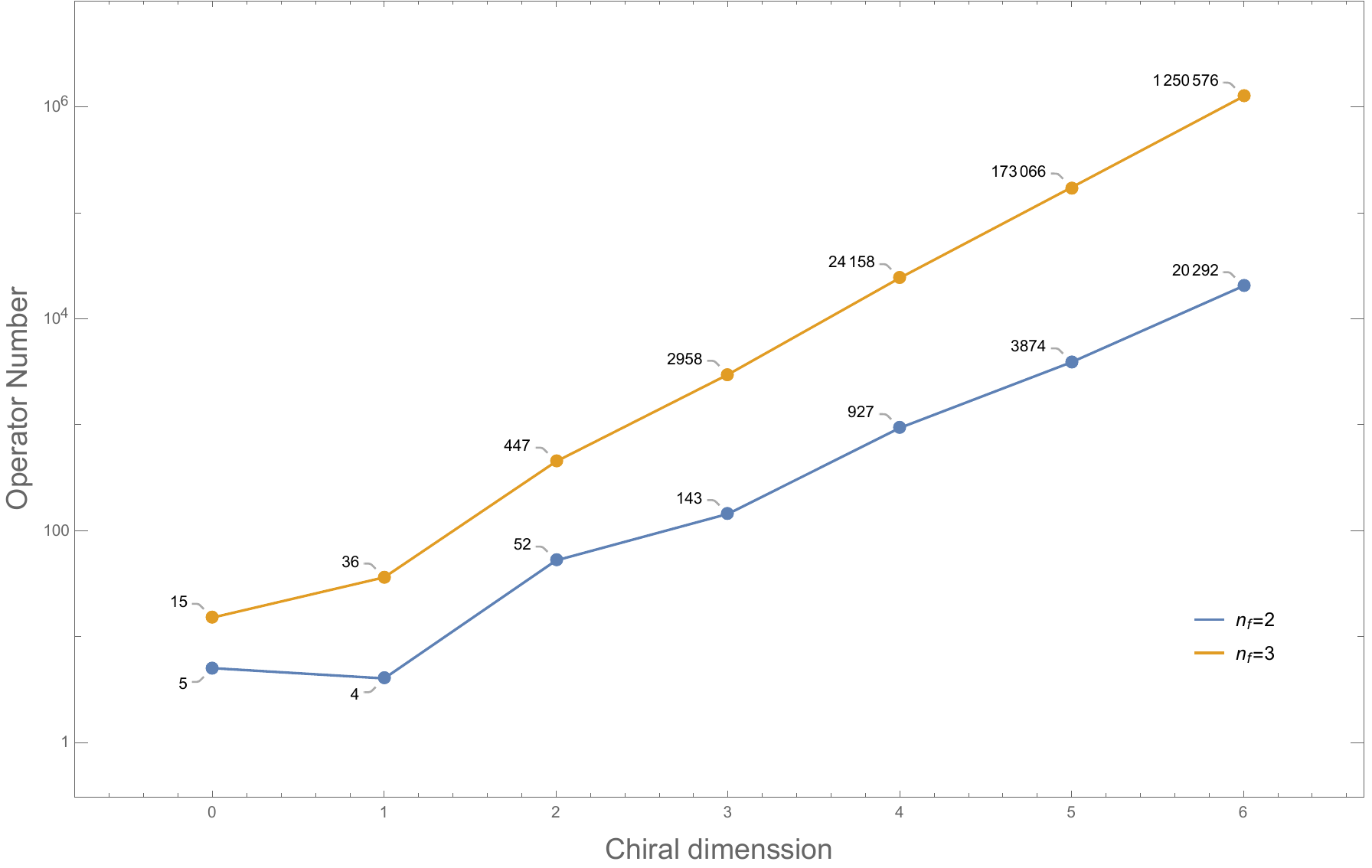}
  \caption{The growth of the number of the $C$-even and $P$-even operators in the $\mathcal{L}_{NN}$ sector.}
  \label{graph:pig42}
\end{figure}

In addition to the explicit Hilbert series presented above, we work out the growth of the numbers of the operators as the increase of the chiral dimension.
The growth of the $C\,, P$-even operators in the $\mathcal{L}_{\pi N}$ and $\mathcal{L}_{NN}$ sectors are presented in Fig.~\ref{graph:pig4} and Fig.~\ref{graph:pig42}, where both the $SU(2)$ and $SU(3)$ cases are shown.


\bibliography{ref}
\end{document}